\documentclass[11pt]{article}
\usepackage[centertags]{amsmath}
\usepackage{amsfonts}
\usepackage{amssymb}
\usepackage{amsthm}
\usepackage{boxedminipage}
\usepackage{mathptm}
\usepackage{psfig}
\usepackage{color}
\usepackage{pdfcolmk}	

\newcommand{\url}[1]{\footnote{\small \tt \textcolor{blue}{#1}}}

\begin{document}


\title{Simulation of networks of spiking neurons: \\ A review of tools and
strategies}

\author{Romain Brette$^1$, Michelle Rudolph$^2$, Ted Carnevale$^3$,
Michael Hines$^3$, David Beeman$^4$, \\
James M.\ Bower$^{5}$, Markus Diesmann$^{6,7}$, Abigail
Morrison$^{7}$, Philip H. Goodman$^{8}$, \\
Frederick C. Harris, Jr.$^{8}$, 
Milind Zirpe$^{8}$, Thomas Natschl\"ager$^9$, Dejan Pecevski$^{10}$, 
Bard Ermentrout$^{11}$, \\
Mikael Djurfeldt$^{12}$, Anders Lansner$^{12}$, Olivier Rochel$^{13}$, 
Thierry Vieville$^{14}$, Eilif Muller$^{15}$, \\ 
Andrew P.\ Davison$^2$, Sami El~Boustani$^2$ and 
Alain Destexhe$^2$\footnote{
%
%
Address for correspondence: Alain Destexhe, Unit\'e de Neurosciences 
Int\'egratives et Computationnelles (UNIC), CNRS (Bat 33), 1 
Avenue de la Terrasse, 91190 Gif-sur-Yvette, France 
(destexhe@iaf.cnrs-gif.fr)}. \\
\ \\
%
%
1: Ecole Normale Sup\'erieure, Paris, France \\
2: CNRS, Gif-sur-Yvette, France \\
3: Yale University, New Haven, CT, USA \\
4: University of Colorado, Boulder, CO, USA \\
5: University of Texas, San Antonio, TX, USA \\
6: University of Freiburg, Germany \\
7: RIKEN Brain Science Institute, Wako City, Japan \\
8: University of Nevada, Reno NV, USA \\
9: Software Competence Center Hagenberg, Hagenberg, Austria \\
10: Technical University of Graz, Austria \\
11: University of Pittsburgh, Pittsburgh, PA, USA \\
12: KTH, Sweden \\
13: University of Leeds, UK \\
14: INRIA, Nice, France \\
15: University of Heidelberg, Germany \\
}

\maketitle


\begin{abstract}

We review different aspects of the simulation of spiking neural
networks.  We start by reviewing the different types of simulation
strategies and algorithms that are currently implemented.  We next
review the precision of those simulation strategies, in particular
in cases where plasticity depends on the exact timing of the
spikes.  We overview different simulators and simulation
environments presently available (restricted to those freely
available, open source and documented).  For each simulation tool,
its advantages and pitfalls are reviewed, with an aim to allow the
reader to identify which simulator is appropriate for a given task.
Finally, we provide a series of benchmark simulations of different
types of networks of spiking neurons, including Hodgkin-Huxley
type, integrate-and-fire models, interacting with current-based or
conductance-based synapses, using clock-driven or event-driven
integration strategies.  The same set of models are implemented on
the different simulators, and the codes are made available.  The
ultimate goal of this review is to provide a resource to facilitate
identifying the appropriate integration strategy and simulation
tool to use for a given modeling problem related to spiking neural
networks.

\end{abstract}


\clearpage
\section*{Introduction}
\label{secIntroduction}

The growing experimental evidence that spike timing may be important
to explain neural computations has motivated the use of spiking
neuron models, rather than the traditional rate-based models.  At the
same time, a growing number of tools have appeared, allowing the
simulation of spiking neural networks.  Such tools offer the user to
obtain precise simulations of a given computational paradigm, as well
as publishable figures in a relatively short amount of time. 
However, the range of computational problems related to spiking
neurons is very large.
It requires in some cases to use
detailed biophysical representations of the neurons, for example
when intracellular electrophysiological measurements are to be
reproduced (e.g., see Destexhe and Sejnowski, 2001).
In this case, one uses conductance-based models, such as
the Hodgkin and Huxley (1952) type of models.  In other cases, one
does not need to realistically capture the spike generating
mechanisms, and simpler models, such as the integrate-and-fire (IF)
model are sufficient.  IF type models are also very fast to simulate,
and are particularly attractive for large-scale network simulations.  

There are two families of algorithms
for the simulation of neural networks:
synchronous or ``clock-driven'' algorithms, in which
all neurons are updated simultaneously at every tick
of a clock,
and asynchronous or ``event-driven'' algorithms, in
which neurons are updated only when they receive or
emit a spike (hybrid strategies also exist).
Synchronous algorithms can be easily coded and
apply to any model. Because spike times are typically
bound to a discrete time grid, the precision of the simulation
can be an issue.
Asynchronous algorithms have been developed mostly for
exact simulation, which is possible for simple models.
For very large networks, the simulation time for both methods
scale as the total number of spike transmissions, but each
strategy has its own assets and disadvantages.

In this paper, we start by providing an overview of different
simulation strategies, and outline to which extent the temporal
precision of spiking events impacts on neuronal dynamics of single as
well as small networks of IF neurons with plastic synapses.  Next, we
review the currently available simulators or simulation environments,
with an aim to focus only on publically-available and non-commercial
tools to simulate networks of spiking neurons.  For each type of
simulator, we describe the simulation strategy used, outline the type
of models which are most optimal, as well as provide concrete
examples.  The ultimate goal of this paper is to provide a resource
to enable the researcher to identify which strategy or simulator to
use for a given modeling problem related to spiking neural networks.


\section{Simulation strategies}
\label{secAlgorithms}

This discussion is restricted to
serial algorithms for brevity.
The specific sections of
NEST and SPLIT contain
additional information on
concepts for parallel computing.

There are two families of algorithms
for the simulation of neural networks:
synchronous or ``clock-driven'' algorithms, in which
all neurons are updated simultaneously at every tick
of a clock,
and asynchronous or ``event-driven'' algorithms, in
which neurons are updated only when they receive or
emit a spike.
These two approaches have some common features
that we will first describe
by expressing the problem of simulating
neural networks
in the formalism of hybrid systems, i.e.,
differential equations with discrete events
(spikes). In this framework some
common strategies for
efficient representation and simulation
appear.

Since we are going to compare algorithms in terms
of computational efficiency,
let us first ask ourselves the following question:
how much time can it possibly take for
a good algorithm to simulate a large network?
Suppose there are $N$ neurons whose average
firing rate is $F$ and average number of
synapses is $p$. If all spike transmissions are taken
into account, then a simulation lasting 1s
(biological time)
must process $N\times p\times F$
spike transmissions. The goal of efficient
algorithm design is to reach this minimal number
of operations
(of course, up to a constant
multiplicative factor).
If the simulation is not restricted to spike-mediated
interactions, e.g. if the model includes gap junctions
or dendro-dendritic interactions, then the optimal
number of operations can be much larger,
but in this review we chose not to address the problem
of graded interactions.

\subsection{A hybrid system formalism}
\label{hybrid}

Mathematically, neurons can be described as \emph{hybrid systems}:
the state of a neuron evolves continuously
according to some biophysical equations,
which are typically differential equations
(deterministic or stochastic,
ordinary or partial differential equations),
and spikes received through the synapses
trigger changes in some of the variables.
Thus the dynamics of a neuron can be
described as follows:
\begin{eqnarray*}
\frac{d\bf{X}}{dt}&=&f(\bf{X})\\
\bf{X} &\leftarrow& g_i({\bf X})\qquad\textrm{upon spike from synapse i}
\end{eqnarray*}
where $\bf{X}$ is a vector describing the state of the neuron.
In theory, taking into account the morphology of the neuron
would lead to partial differential equations; however, in
practice, one usually approximates the dendritic tree by
coupled isopotential compartments, which also leads to a
differential system with discrete events.
Spikes are emitted when some threshold condition
is satisfied, for instance $V_m\geq \theta$ for
integrate-and-fire models (where $V_m$ is the
membrane potential and would be the first component
of vector $\bf{X}$), and/or
$dV_m/dt \geq \theta$ for Hodgkin-Huxley type models.
This can be summarized by saying that a spike is emitted
whenever some condition $\bf{X}\in A$ is satisfied.
For integrate-and-fire models, the membrane potential,
which would be the first component of $\bf{X}$, is reset when a
spike is produced. The reset can be integrated into the hybrid system formalism
by considering for example that outgoing spikes act on $\bf{X}$ through
an additional (virtual) synapse: ${\bf X} \leftarrow g_0(\bf{X})$.

With this formalism, it appears clearly that
\emph{spike times need not be stored}
(except of course if transmission delays are included),
even though it would seem so from more phenomenological formulations.
For example, consider the following integrate-and-fire model
(described for example in G{\"u}tig and Sompolinsky (2006)):
\begin{displaymath}
V(t)=\sum_i\omega_i\sum_{t_i}K(t-t_i)+V_{\textrm{rest}}
\end{displaymath}
where $V(t)$ is the membrane potential,
$V_{\textrm{rest}}$ is the rest potential,
$\omega_i$ is the synaptic weight of synapse $i$,
$t_i$ are the timings of the spikes coming from
synapse $i$, and
$K(t-t_i)=\exp(-(t-t_i)/\tau)-\exp(-(t-t_i)/\tau_s)$
is the post-synaptic potential (PSP) contributed by each
incoming spike.
The model can
be restated as a two-variables differential system with discrete events
as follows:
\begin{eqnarray*}
\tau\frac{dV}{dt}&=&V_{\textrm{rest}}-V+J\\
\tau_s\frac{dJ}{dt}&=&-J\\
J&\leftarrow&J+\frac{\tau-\tau_s}{\tau}w_i\qquad\textrm{upon spike from synapse i}
\end{eqnarray*}

Virtually all post-synaptic potentials or currents
described in the literature
(e.g. $\alpha$-functions, bi-exponential functions)
can be expressed this way.
Several authors have described the transformation
from phenomenological expressions to the hybrid system
formalism for
synaptic conductances and currents (Destexhe et al, 1994a, 1994b,
Rotter and Diesmann, 1999,
Giugliano, 2000),
short-term synaptic depression (Giugliano et al, 1999),
and spike-timing-dependent plasticity (Song et al, 2000).
In many cases, the Spike Response Model (Gerstner and Kistler, 2002)
is also the integral expression of a hybrid system.
To derive the differential formulation of
a given post-synaptic current or conductance (PSC),
one way is to see the latter as the impulse response
of a linear time-invariant system
(which can be seen as a filter (Janke et al, 1998))
and use transformation
tools from signal processing theory such as
the Z-transform (Kohn and W{\"o}rg{\"o}tter, 1998; see also Sanchez, 
2001)
or the Laplace transform
(the Z-transform is the equivalent of the Laplace transform
in the digital time domain, i.e., for synchronous
algorithms).

\subsection{Using linearities for fast synaptic simulation}
\label{linearities}

In general, the number of state variables of a neuron
(length of vector $\bf{X}$) scales with the number of synapses,
since each synapse has its own dynamics. This fact constitutes
a major problem for efficient simulation of neural networks, both
in terms of memory consumption and computation time.
However,
several authors have observed that all synaptic variables
sharing the same linear dynamics can be reduced to a single one
(Wilson and Bower, 1989, Bernard et al, 1994, Lytton, 1996 and Song et al, 2000).
For example, the following set of differential equations,
describing an integrate-and-fire model with $n$ synapses
with exponential conductances:
\begin{eqnarray*}
C\frac{dV}{dt}&=&V_0-V+\sum_i g_i(t)(V-E_s)\\
\tau_s \frac{dg_1}{dt}&=&-g_1\\
&\ldots&\\
\tau_s \frac{dg_n}{dt}&=&-g_n\\
g_i &\leftarrow& g_i+w_i\qquad\textrm{upon spike arriving at synapse $i$}
\end{eqnarray*}
is mathematically equivalent to the following set of two
differential equations:
\begin{eqnarray*}
C\frac{dV}{dt}&=&V_0-V+g(t)(V-E_s)\\
\tau_s \frac{dg}{dt}&=&-g\\
g &\leftarrow& g+w_i\qquad\textrm{upon spike arriving at synapse $i$}
\end{eqnarray*}
where $g$ is the total synaptic conductance.
The same reduction applies to synapses with
higher dimensional dynamics, as long as it is
linear and the spike-triggered changes
($g_i \leftarrow g_i+w_i$) are additive and
do not depend on the state of the synapse
(e.g. the rule
$g_i \leftarrow g_i+w_i*f(g_i)$ would cause
a problem).
Some models of spike-timing dependent plasticity
(with linear interactions between pairs of spikes)
can also be simulated in this way (see
e.g. Abbott and Nelson (2000)).
However, some important biophysical models
are not linear and thus cannot benefit from this optimization,
in particular NMDA-mediated interactions and saturating synapses.

\subsection{Synchronous or ``clock-driven'' algorithms}
\label{secClock}

In a synchronous or ``clock-driven'' algorithm
(see pseudo-code in figure \ref{figclockdriven}),
the state variables of all neurons
(and possibly synapses) are updated at every
tick of a clock:
${\bf X}(t)\rightarrow {\bf X}(t+dt)$.
With non-linear differential equations, one
would use an integration method such as
Euler or Runge-Kutta (Press et al, 1993) or,
for Hodgkin-Huxley models, implicit methods
(Hines, 1984). Neurons with complex morphologies
are usually spatially discretized and
modelled as interacting compartments:
they are also described mathematically by coupled
differential equations, for which dedicated
integration methods have been developed
(for details see e.g. the specific
section of Neuron in this review).
If the differential equations are linear, then
the update operation
${\bf X}(t)\rightarrow {\bf X}(t+dt)$ is also linear,
which means updating the state variables
amounts simply to multiplying $\bf{X}$ by a matrix:
${\bf X}(t+dt)={\bf AX}(t)$ (Hirsch and Smale, 1974;
see also Rotter and Diesmann (1999) for an application
to neural networks),
which is very convenient in
vector-based scientific softwares
such as Matlab 
or Scilab. 
Then, after updating all variables,
the threshold condition is checked
for every neuron. Each neuron that
satisfies this condition
produces a spike which is transmitted
to its target neurons, updating
the corresponding variables
(${\bf X} \leftarrow g_i({\bf X})$).
For integrate-and-fire models,
the membrane potential of
every spiking neuron is reset.

\subsubsection*{Computational complexity}
The simulation time of such an algorithm
consists of two parts:
1) state updates and 2) propagation
of spikes.
Assuming the number of state variables for
the whole network scales
with the number of neurons $N$ in the network
(which is the case when the reduction
described in section \ref{linearities}
applies), the cost of the update phase
is of order $N$ for each step, so it is $O(N/dt)$ per second
of biological time
($dt$ is the duration of the time bin).
This component grows with the complexity of the neuron models
and the precision of the simulation.
Every second (biological time),
an average of $F\times N$ spikes are produced by the neurons
($F$ is the average firing rate), and
each of these needs to be propagated to $p$ target neurons.
Thus, the propagation phase consists in
$F\times N\times p$ spike propagations per second. These are essentially
additions of weights $w_i$ to state variables, and thus
are simple operations
whose cost does not grow with the complexity of the models.
Summing up, the total computational cost per second of
biological time is of order
\begin{eqnarray*}
\textrm{\it Update} & + & \textrm{\it Propagation}\\
c_U\times \frac{N}{dt} & + & c_P\times F\times N \times p\qquad(*)
\end{eqnarray*}
where $c_U$ is the cost of one update and $c_P$ is the
cost of one spike propagation; typically, $c_U$ is much
higher than $c_P$ but this is implementation-dependent.
Therefore, for very dense networks, the total is dominated
by the propagation phase and is linear in the number of
synapses, which is optimal. However, in practice
the first phase is negligible only when the following condition
is met:
\begin{displaymath}
\frac{c_P}{c_U}\times F\times p\times dt >> 1
\end{displaymath}
For example,
the average firing rate in the cortex might be as low
as $F = 1$ Hz (Olshausen and Field, 2005),
and assuming $p = 10,000$ synapses per neuron
and $dt = 0.1$ ms, we get $F\times p\times dt=1$.
In this case, considering that each operation in the update
phase is heavier than in the propagation phase (especially
for complex models), i.e., $c_P<c_U$,
the former is likely to dominate the total computational
cost. Thus, it appears that even in networks with realistic
connectivity, increases in precision (smaller $dt$,
see section \ref{secPrecision}) can be
detrimental to the efficiency of the simulation.

\subsubsection*{Delays}
For the sake of simplicity,
we ignored transmission delays in the description
above. However it is not very complicated to
include them in a synchronous
clock-driven algorithm.
The straightforward
way is to store
the future synaptic events
in a circular array.
Each element of the
array corresponds to a time bin and
contains a list of synaptic events that
are scheduled for that time
(see e.g. Morrison et al, 2005).
For example, if neuron $i$ sends a spike
to neuron $j$ with delay $d$
(in units of the time bin $dt$),
then the synaptic event
``$i\rightarrow j$''
is placed in the circular
array at position $p+d$, where
$p$ is the present position.
Circularity of the array means
the addition $p+d$ is modular
($(p+d) \mod n$, where $n$ is the size
of the array --- which corresponds to
the largest delay in the system).

What is the additional computational
cost of managing delays?
In fact, it is not very high and does not
depend on the duration of the time bin.
Since every synaptic event ($i\rightarrow j$)
is stored and retrieved exactly once,
the computational cost of
managing delays for one
second of biological time is
\begin{displaymath}
c_D\times F\times N\times p
\end{displaymath}
where $c_D$ is the cost of one
store and one retrieve operation in
the circular array (which is low).
In other words, managing delays
increases the cost of the propagation
phase in equation $(*)$
by a small multiplicative factor.

\subsubsection*{Exact clock-driven simulation}

The obvious drawback
of clock-driven algorithms as
described above is that
spike timings are aligned to a grid
(ticks of the clock), thus
the simulation is approximate even
when the differential equations are
computed exactly.
Other specific errors
come from the fact that threshold
conditions are checked only at the
ticks of the clock, implying that
some spikes might be missed
(see section \ref{secPrecision}).
However, in principle, it is possible
to simulate a network exactly
in a clock-driven fashion when
the minimum transmission delay
is larger than the time step.
It implies that the precise
timing of synaptic events is stored
in the circular array
(as described in Morrison et al, 2006).
Then within each time bin,
synaptic events for each neuron
are sorted and processed in
the right order, and when the
neuron spikes, the exact
spike timing is calculated.
Neurons can be processed independently
in this way only because the time bin
is smaller than the smallest transmission
delay
(neurons have no influence on each other
within one time bin).

Some sort of clock signals can also be used in
general event-driven algorithms without
the assumption of a minimum
positive delay. For example, one
efficient data structure used
in discrete event systems to store
events is a priority queue
known as ``calendar queue'' (Brown, 1988),
which is a dynamic
circular array of sorted lists.
Each ``day'' corresponds to a time bin,
as in a classical circular array,
and each event is placed in the calendar
at the corresponding day; all events
on a given day are sorted according to
their scheduling time. If the duration of
the day is correctly set, then insertions and
extractions of events take constant time on
average. Note that, in contrast with standard
clock-driven simulations,
the state variables are not updated
at ticks of the clock and the duration of the
days depends neither on the precision of the simulation
or on the transmission delays (it is rather linked
to the rate of events) --- in fact,
the management of the priority queue is separated
from the simulation itself.

Note however that in all these cases,
state variables need to be updated at
the time of every incoming spike
rather than at every tick of the clock
in order to simulate the network
exactly
(e.g. simple vector-based
updates ${\bf X}\leftarrow{\bf AX}$
are not possible),
so that the term \emph{event-driven}
may be a better description of these
algorithms
(the precise terminology may vary
between authors).

\subsubsection*{Noise in synchronous algorithms}

Noise can be introduced in synchronous
simulations by essentially two means:
1) adding random external spikes;
2) simulating a stochastic process.

Suppose a given neuron receives
$F$ random spikes per second, according
to a Poisson process. Then the
number of spikes in one time bin
follows a Poisson distribution with
mean $F\times dt$. Thus one can
simulate random external spike trains
by letting each tick of the clock
trigger a random number of synaptic updates.
If $F\times dt$ is low, the Poisson
distribution is almost a Bernouilli
distribution (i.e., there is one spike
with probability $F\times dt$).
It is straightforward to extend the
procedure to
inhomogeneous Poisson processes by
allowing $F$ to vary in time.
The additional computational cost
is proportional to $F_{\textrm{ext}}\times N$,
where $F_{\textrm{ext}}$ is the average
rate of external synaptic events for each neuron
and $N$ is the number of neurons.
Note that $F_{\textrm{ext}}$ can be quite
large since it represents the sum of firing
rates of all external neurons
(for example it would be $10,000$ Hz
for $10,000$ external synapses per neuron with
rate $1$ Hz).

To simulate a large number of external random
spikes, it can be advantageous to
simulate directly the total external synaptic input
as a stochastic process, e.g. white or colored
noise (Ornstein-Uhlenbeck).
Linear stochastic differential equations
are analytically solvable, therefore the
update
${\bf X}(t)\rightarrow {\bf X}(t+dt)$ can be
calculated exactly with matrix computations
(Arnold, 1974)
(${\bf X}(t+dt)$ is, conditionally to
${\bf X}(t)$, a normally distributed
random variable whose mean and covariance
matrix can be calculated as a function
of ${\bf X}(t)$).
Nonlinear stochastic differential equations
can be simulated using approximation
schemes, e.g. stochastic Runge-Kutta
(Honeycutt, 1992).

\subsection{Asynchronous or ``event-driven'' algorithms}
\label{secEvent}

Asynchronous or ``event-driven'' algorithms are not as widely used as
clock-driven ones because they are significantly more
complex to implement (see pseudo-code in figure \ref{figeventdriven2})
and less universal.
Their key advantages are a potential gain in speed due to not
calculating many small update steps for a neuron in which no event arrives
and that spike timings are computed exactly
(but see below for approximate event-driven algorithms);
in particular, spike timings are not
aligned to a time grid anymore (which is a source of potential
errors, see section \ref{secPrecision}).

The problem of simulating dynamical systems with discrete events
is a well established research topic in computer science
(Ferscha, 1996; Sloot et al, 1999; Fujimoto, 2000;
Zeigler et al, 2000;
see also Rochel and Martinez, 2003 and Mayrhofer et al, 2002),
with appropriate data structures and algorithms already available
to the computational neuroscience community.
We start by describing 
the simple case when synaptic interactions are instantaneous,
i.e., when spikes can be produced only at times of incoming
spikes (no latency); then we will turn to the most general
case.

\subsubsection*{Instantaneous synaptic interactions}

In an asynchronous or ``event-driven'' algorithm,
the simulation advances from one event to the next event.
Events can be spikes coming from neurons in the network or
external spikes (typically random spikes described by a
Poisson process). For models in which spikes can be produced by a neuron
only at times of incoming spikes,
event-driven simulation is relatively easy
(see pseudo-code in figure \ref{figeventdriven1}).
Timed events are stored in a queue (which is some sort
of sorted list).
One iteration consists in
1) extracting the next event;
2) updating the state of the
corresponding neuron
(i.e., calculating the state according
to the differential equation and
adding the synaptic weight)
3) checking if the neuron satisfies the
threshold condition, in which case
events are inserted in the queue for
each downstream neuron.

In the simple case of identical
transmission delays, the data structure
for the queue can be just a
FIFO queue (first in, first out), which has
fast implementations (Cormen et al, 2001).
When the delays
take values in a small discrete set, the
easiest way is to use one FIFO queue for
each delay value, as described
in Mattia and Del Giudice (2000). It is also more efficient
to use a separate FIFO queue for handling
random external events
(see paragraph about noise below).

In the
case of arbitrary delays, one needs a
more complex data structure.
In computer science, efficient data structures
to maintain an ordered list of time-stamped
events are grouped under the name
\emph{priority queues} (Cormen et al, 2001).
The topic of priority queues is dense and
well documented; examples are
binary heaps, Fibonacci heaps (Cormen et al, 2001),
calendar queues (Brown, 1988, Claverol et al, 2002) or
van Emde Boas trees (van Emde Boas et al, 1976;
see also Connolly et al (2003) in which various
priority queues are compared).
Using an efficient priority queue
is a crucial element of a good
event-driven algorithm.
It is even more crucial
when synaptic interactions are not instantaneous.

\subsubsection*{Non-instantaneous synaptic interactions}

For models in which spike times do not necessarily occur at times
of incoming spikes, event-driven simulation is more complex.
We first describe the basic algorithm with no delays and
no external events
(see pseudo-code in figure \ref{figeventdriven2}).
One iteration consists in
1) finding which neuron is the next one to spike;
2) updating this neuron;
3) propagating the spike, i.e.,
updating its target neurons.
The general way to do that
is to maintain a sorted list
of the future spike timings of all
neurons. These spike timings are only
provisory since any spike in the network
can modify all future spike timings.
However, the spike with lowest timing in
the list is certified. Therefore, the
following algorithm for one iteration
guarantees the correctness of the simulation
(see figure \ref{figeventdriven2}):
1) extract the spike with lowest timing
in the list; 2) update the state of
the corresponding neuron and recalculate
its future spike timing; 3) update
the state of its target neurons;
4) recalculate the future spike timings
of the target neurons.

For the sake of simplicity,
we ignored transmission delays in the description
above.
Including them in an event-driven algorithm is not as
straightforward as in a clock-driven algorithm, but
it is a minor complication. When a spike is produced
by a neuron, the future synaptic events are stored
in another priority queue in which the timings
of events are non-modifiable.
The first phase of the algorithm
(extracting the spike with lowest timing)
is replaced by extracting the next event,
which can be either a synaptic event or
a spike emission. One can use two separate
queues or a single one.
External events can be handled in the
same way.
Although delays introduce complications
in coding event-driven algorithms, they can in fact
simplify the management of the priority queue for
outgoing spikes. Indeed, the main difficulty in
simulating networks with non-instantaneous synaptic
interactions is that scheduled outgoing spikes can
be canceled, postponed or advanced by future incoming
spikes. If transmission delays are greater than
some positive value $\tau_{\min}$, then
all outgoing spikes scheduled in $[t,t+\tau_{\min}]$
($t$ being the present time)
are certified. Thus, algorithms can exploit the
structure of delays
to speed up the simulation (Lee and Farhat, 2001).

\subsubsection*{Computational complexity}

Putting aside the cost of handling external
events (which is minor),
we can subdivide the computational cost
of handling one outgoing spike as follows
(assuming $p$ is the average number of
synapses per neuron):
\begin{itemize}
\item extracting the event (in case of
non-instantaneous synaptic interactions)
\item updating the neuron and its targets:
$p+1$ updates;
\item inserting $p$ synaptic events in the
queue (in case of delays);
\item updating the spike times of
$p+1$ neurons (in case of
non-instantaneous synaptic interactions);
\item inserting or rescheduling $p+1$ events in the queue
(future spikes for non-instantaneous synaptic interactions).
\end{itemize}
Since there are $F\times N$ spikes per second
of biological time, the number of operations is approximately
proportional to $F\times N\times p$.
The total computational cost per
second of biological time can be written concisely as follows:
\begin{center}
\begin{tabular}{rcccll}
{\it Update} & + & {\it Spike} & + & {\it Queue} & \\
$(c_U$ & + & $c_S$ & + & $c_Q)$ & $\times F\times N \times p$
\end{tabular}
\end{center}
where $c_U$ is the cost of one update of the state variables,
$c_S$ is the cost of calculating the time of the next spike
(non-instantaneous synaptic interactions)
and $c_Q$ is the average cost of insertions and extractions
in the priority queue(s).
Thus, the simulation
time is linear in the number of synapses, which is optimal.
Nevertheless, we note that the operations involved are heavier
than in the propagation phase of clock-driven algorithms
(see previous section),
therefore the multiplicative factor is
likely to be larger.
We have also assumed that $c_Q$ is $O(1)$, i.e., that
the dequeue and enqueue operations can be done in
constant average time with the data structure chosen
for the priority queue. In the simple case of
instantaneous synaptic interactions
and homogeneous delays, one can use a simple FIFO queue
(First In, First Out), in which insertions and extractions
are very fast and take constant time. For the general
case, data structures for which dequeue and enqueue operations
take constant average time ($O(1)$) exist,
e.g. calendar queues (Brown, 1988, Claverol et al, 2002), however they are quite
complex, i.e., $c_Q$ is a large constant. In simpler
implementations of priority queues such as binary
heaps, the dequeue and enqueue operations
take $O(\log m)$ operations, where $m$ is the number of events
in the queue.
Overall, it appears that the crucial component in
general event-driven algorithms is the queue management.

\subsubsection*{What models can be simulated in an event-driven fashion?}

Event-driven algorithms implicitly assume that
we can calculate the state of a neuron at any given time, i.e.,
we have an explicit solution of the differential equations
(but see below for approximate event-driven simulation).
This would not be the case with e.g. Hodgkin-Huxley models.
Besides, when synaptic interactions are not instantaneous,
we also need a function that maps
the current state of the neuron to the timing of the next
spike (possibly
$+\infty$ if there is none).

So far, algorithms have been developed for
simple pulse-coupled integrate-and-fire models
(Watts, 1994, Claverol et al, 2002, Delorme and Thorpe, 2003)
and more complex ones such as some instances of
the Spike Response Model (Makino, 2003, Marian et al, 2002, Gerstner and Kistler, 2002)
(note that the SRM model can usually be restated in the
differential formalism of section \ref{hybrid}).
Recently, Rudolph and Destexhe (2006) introduced several
integrate-and-fire models with synaptic conductances
which are suitable for event-driven simulation.
Algorithms were also recently developed by Brette
to simulate exactly integrate-and-fire models with exponential
synaptic currents (Brette, 2006a) and conductances (Brette, 2006b),
and Tonnelier et al (2006, submitted) extended this work to the
quadratic model (Ermentrout and Kopell, 1986).
However, there are still efforts to be made to design suitable
algorithms for more complex models (for example
the two-variable integrate-and-fire
models of Izhikevich (2003) and Brette and Gerstner (2005)),
or to develop more realistic models that are suitable for
event-driven simulation.

\subsubsection*{Noise in event-driven algorithms}

As for synchronous algorithms, there are two ways
to introduce noise in a simulation: 1) adding random external
spikes; 2) simulating a stochastic process.

The former case is by far easier in asynchronous algorithms.
It simply amounts to adding a queue with external events, which
is usually easy to implement. For example, if external spikes
are generated according to a Poisson process with rate $F$,
the timing of the next event if random variable with exponential
distribution with $1/F$. If $n$ neurons receive external
spike trains given by independent Poisson processes with rate
$F$, then the time of the next event is exponentially distributed
with mean $1/(nF)$ and the label of the neuron receiving this
event is picked at random in $\{1,2,\ldots,n\}$.
Inhomogeneous Poisson processes can be simulated exactly in 
a similar way (Daley and Vere-Jones, 1988). If $r(t)$ is the
instantaneous rate of the Poisson process and is bounded
by $M$ ($r(t)\leq M$), then
one way to generate a spike train according to this
Poisson process in the interval $[0,T]$ is as follows:
generate a spike train in $[0,T]$ according to a homogeneous
Poisson process with rate $T*M$; for each spike at time $t_i$,
draw a random number $x_i$ from a uniform distribution in $[0,M]$;
select all spikes such that $x_i\leq r(t_i)$.

Simulating directly a stochastic process in asynchronous algorithms
is much harder because even for the simplest stochastic neuron models,
there is no closed analytical formula for the distribution
of the time to the next spike
(see e.g. Tuckwell (1988)).
It is however possible to use precalculated tables when the
dynamical systems are low dimensional (Reutimann et al, 2003) (i.e., not more than 2
dimensions).
Note that simulating noise in this way
introduces provisory events in the same way
as for non-instantaneous
synaptic interactions.

\subsubsection*{Approximate event-driven algorithms}

We have described asynchronous algorithms for simulating
neural networks exactly. For complex neuron models of the
Hodgkin-Huxley type, Lytton and Hines (2005) have
developed an asynchronous simulation
algorithm which consists in using for each neuron
an independent time step whose width is reduced
when the membrane potential approaches the action
potential threshold.


\section{Precision of different simulation strategies}
\label{secPrecision}

As shown in this paper, a steadily growing number of neural
simulation environments does endow computational neuroscience with
tools which, together with the steady improvement of computational
hardware, allow to simulate neural systems with increasing
complexity, ranging from detailed biophysical models of single cells
up to large-scale neural networks. Each of these simulation tools
pursues the quest for a compromise between efficiency in speed and
memory consumption, flexibility in the type of questions addressable,
and precision or exactness in the numerical treatment of the latter. 
In all cases, this quest leads to the implementation of a specific
strategy for numerical simulations which is found to be optimal given
the set of constraints set by the particular simulation tool. 
However, as shown recently (Hansel et al., 1998; Lee and Farhat,
2001; Morrison et al, 2006), quantitative results and their
qualitative interpretation strongly depend on the simulation strategy
utilized, and may vary across available simulation tools or for
different settings within the same simulator.  The specificity of
neuronal simulations is that spikes induce either a discontinuity in
the dynamics (integrate-and-fire models) or have very fast dynamics
(Hodgkin-Huxley type models). When using approximation methods, this
problem can be tackled by spike timing interpolation in the former
case (Hansel et al., 1998; Shelley and Tao, 2001) or integration with
adaptive time step in the latter case (Lytton and Hines, 2005). 
Specifically in networks of integrate-and-fire neurons, which to date
remain almost exclusively the basis for accessing dynamics of
large-scale neural populations (but see Section~\ref{split-sec}),
crucial differences in the appearance of synchronous activity
patterns were observed, depending on the temporal resolution of the
neural simulator or the integration method used.

In this section we address this question using one of the most simple
analytically solvable LIF neuron model, namely the classic leaky
integrate-and-fire neuron, described by the state equation 
\begin{equation}
\tau_m \, \frac{d m(t)}{dt} + m(t) = 0 \, ,
\end{equation}
where $\tau_m$~=~20~ms denotes the membrane time constant and $0 \leq m(t)
\leq 1$. Upon arrival of a synaptic event at time $t_0$, $m(t)$ is updated by
a constant $\Delta m$~=0.1 ($\Delta m$~=~0.0085 in network simulations) after
which it decays according to 
\begin{equation}
m(t) = m(t_0) \exp[ - \frac{t-t_0}{\tau_m} ] \, .
\end{equation}
If $m$ exceeds a threshold $m_{thres}$~=~1, the neuron fires and is afterwards
reset to a resting state $m_{rest}$~=~0 in which it stays for an absolute
refractory period $t_{ref}$~=~1~ms. The neurons were subject to non-plastic or
plastic synaptic interactions. In the latter case, spike timing dependent
synaptic plasticity was used according to a model by Song and Abbott (2001).
In this case, upon arrival of a synaptic input at time $t_{pre}$, synaptic
weights are changed according to 
\begin{equation}
g \leftarrow g + F(\Delta t) \, g_{max} \, ,
\end{equation}
where 
\begin{equation}
F(\Delta t) = \pm A_{\pm} \, \exp\{  \pm \Delta t / \tau_{\pm} \}
\end{equation}
for $\Delta t = t_{pre} - t_{post} < 0$ and $\Delta t \geq 0$, respectively. 
Here, $t_{post}$ denotes the time of the nearest postsynaptic spike, $A_{\pm}$
quantify the maximal change of synaptic efficacy, and $\tau_{\pm}$ determine
the range of pre- to postsynaptic spike intervals in which synaptic weight
changes occur. Comparing simulation strategies at the both ends of a wide
spectrum, namely a clock-driven algorithm (see Section \ref{secClock}) and event-driven
algorithm (see Section \ref{secEvent}), we evaluate to which extent the temporal
precision of spiking events impacts on neuronal dynamics of single as well as
small networks.
These results support the argument that
the speed of neuronal simulations should not be
the sole criteria for evaluation of simulation tools, but
must complement an evaluation of their exactness.

\subsection{Neuronal systems without STDP}

In the case of a single LIF neuron with non-plastic synapses subject to a
frozen synaptic input pattern drawn from a Poisson distribution with rate
$\nu_{inp}$~=~250~Hz, differences in the discharge behavior seen in 
clock-driven simulations at different resolutions (0.1~ms, 0.01~ms, 0.001~ms)
and event-driven simulations occurred already after short periods of simulated
neural activity (Fig.~\ref{Fig_1}A). These deviations were caused by subtle differences
in the subthreshold integration of synaptic input events due to temporal
binning, and ``decayed'' with a constant which depended on the membrane
time constant. However, for a strong synaptic drive, subthreshold deviations
could accumulate and lead to marked delays in spike times, cancellation of
spikes or occurrence of additional spikes. 

Although differences at the single cell level remained widely constrained and
did not lead to changes in the statistical characterization of the discharge
activity when long periods of neural activity were considered, already
small differences in spike times of individual neurons can lead to crucial
differences in the population activity, such as synchronization (see Hansel et
al., 1998; Lee and Farhat, 2001), if neural networks are concerned. We
investigated this possibility using a small network of 15$\times$15
LIF neurons with all-to-all excitatory connectivity with fixed weights and not
distance-dependent synaptic transmission delay (0.2~ms), driven by a fixed
pattern of superthreshold random synaptic inputs to each neuron (average rate
250~Hz; weight $\Delta m$~=~0.1). In such a small network, the
activity remained primarily driven by the external inputs, i.e. the influence
of intrinsic connectivity is small. However, due to small differences in spike
times due to temporal binning could had severe effects on the occurrence of
synchronous network events where all (or most) cells discharge at the same
time. Such events could be delayed, canceled or generated depending on the
simulation strategy or temporal resolution utilized (Fig.~\ref{Fig_1}B). 

\subsection{Neuronal systems with STDP}

The above described differences in the behavior of neural systems simulated
by using different simulation strategies remain constrained to the observed
neuronal dynamics and are minor if some statistical measures, such as average
firing rates, are considered. More severe effects can be expected if
biophysical mechanism which depend on the exact times of spikes are
incorporated into the neural model. One of these mechanism is short-term
synaptic plasticity, in particular spike-timing dependent synaptic plasticity
(STDP). In this case, the self-organizing capability of the neural system
considered will yield different paths along which the systems will develop,
and, thus, possibly lead to a neural behavior which not only quantitatively
but also qualitatively may differ across various tools utilized for the
numerical simulation.

To explain why such small differences in the exact timing of events are
crucial if models with STDP are considered, consider a situation in which
multiple synaptic input events arrive in between two state updates at $t$ and
$t + dt$ in a clock-driven simulation. In the latter case, the times of
these events are assigned to the end of the interval (Fig.~\ref{Fig_2}A). In
the case these inputs drive the cell over firing threshold, the synaptic
weights of all three synaptic input channels will be facilitated by the same
amount according to the used STDP model. In contrast, if exact times are
considered, the same input pattern could cause a discharge already
after only two synaptic inputs. In this case the synaptic weights liked
to these inputs will be facilitated, whereas the weight of the input arriving
after the discharge will be depressed. 

Although the chance for the occurrence of situations such as those described
above may appear small, already one instance will push the considered neural
system onto a different path in its self-organization. The latter may lead to
systems whose qualitative behavior may, after some time, markedly vary from a
system with the same initial state but simulated by another, temporally
more or less precise simulation strategy. Such a scenario was investigated by
using a single LIF neuron ($\tau_m$~=~4.424~ms) with 1,000 plastic synapses
($A_{+}$~=~0.005, $A_{-}/A_{+}$~=~1.05, $\tau_{+}$~=~20~ms,
$\tau_{-}$~=~20~ms, $g_{max}$~=~0.4) driven by the same pattern of
Poisson-distributed random inputs (average rate 5~Hz, $\Delta m$~=0.1).
Simulating only 1,000~s neural activity led to marked differences in the
temporal development of the average rate between clock-driven simulations
with a temporal resolution of 0.1~ms and event-driven simulations
(Fig.~\ref{Fig_2}B). Considering the average firing rate over the whole
simulated window, clock-driven simulations led to an about 10~\% higher value
compared to the event-driven approach, and approached the value observed in
event-driven simulations only when the temporal resolution was increased by
two orders of magnitude. Moreover, different simulation strategies and
temporal resolutions led also to a significant difference in the synaptic
weight distribution at different times (Fig.~\ref{Fig_2}C).

Both findings show that the small differences in the precision of synaptic
events can have a severe impact even on statistically very robust measures,
such as average rate or weight distribution. Considering the temporal
development of individual synaptic weights, both depression and facilitation
were observed depending on the temporal precision of the numerical simulation
Indeed, the latter could have severe impact on the qualitative interpretation
of the temporal dynamics of structured networks, as this result suggests
that synaptic connections in otherwise identical models can be strengthened or
weakened due to the influence of the utilized simulation strategy or
simulation parameters. 

In conclusion, the results presented in this section suggest that the strategy
and temporal precision used for neural simulations can severely alter 
simulated neural dynamics. Although dependent on the neural system modeled, 
observed differences may turn out to be crucial for the qualitative
interpretation of the result of numerical simulations, in particular in 
simulations involving biophysical processes depending on the exact order
or time of spike events (e.g. as in STDP).
Thus, the search for an optimal
neural simulation tool or strategy for the numerical solution of a given
problem should be guided not only by its absolute speed and
memory consumption, but also its numerical exactness.


\clearpage
\section{Overview of simulation environments}
\label{secOverview}


\subsection{NEURON}

\subsubsection{NEURON's domain of utility}

NEURON is a simulation environment for creating and using
empirically-based models of biological neurons and neural circuits.
Initially it earned a reputation for being well-suited for
conductance-based models of cells with complex branched anatomy,
including extracellular potential near the membrane, and
biophysical properties such as multiple channel types,
inhomogeneous channel distribution, ionic accumulation and
diffusion, and second messengers. In the early 1990s, NEURON was
already being used in some laboratories for network models with
many of thousands of cells, and over the past decade it has
undergone many enhancements that make the construction and
simulation of large-scale network models easier and more efficient.

To date, more than 600 papers and books have described NEURON models
that range from a membrane patch to large scale networks with tens of
thousands of conductance-based or artificial spiking
cells\url{http://www.neuron.yale.edu/neuron/bib/usednrn.html}. In
2005, over 50 papers were published on topics such as mechanisms
underlying synaptic transmission and plasticity (Banitt et al. 
2005), modulation of synaptic integration by subthreshold active
currents (Prescott and De Koninck 2005), dendritic excitability (Day
et al. 2005), the role of gap junctions in networks (Migliore et al.
2005), effects of synaptic plasticity on the development and
operation of biological networks (Saghatelyan et al. 2005), neuronal
gain (Azouz 2005), the consequences of synaptic and channel noise for
information processing in neurons and networks (Badoual et al. 2005),
cellular and network mechanisms of temporal coding and recognition
(Kanold and Manis 2005), network states and oscillations (Wolf et al.
2005), effects of aging on neuronal function (Markaki et al. 2005),
cortical recording (Moffitt and McIntyre 2005), deep brain
stimulation (Grill et al. 2005), and epilepsy resulting from channel
mutations (Vitko et al. 2005) and brain trauma (Houweling et al.
2005).

\subsubsection{How NEURON differs from other neurosimulators}

The chief rationale for domain-specific simulators over general
purpose tools lies in the promise of improved conceptual control,
and the possibility of exploiting the structure of model equations
for the sake of computational robustness, accuracy, and efficiency.
Some of the key differences between NEURON and other
neurosimulators are embodied in the way that they approach these
goals.

{\it Conceptual control}

The cycle of hypothesis formulation, testing, and revision, which
lies at the core of all scientific research, presupposes that one
can infer the consequences of a hypothesis. The principal
motivation for computational modeling is its utility for dealing
with hypotheses whose consequences cannot be determined by unaided
intuition or analytical approaches. The value of any model as a
means for evaluating a particular hypothesis depends critically on
the existence of a close match between model and hypothesis.
Without such a match, simulation results cannot be a fair test of
the hypothesis. From the user's viewpoint, the first barrier to
computational modeling is the difficulty of achieving conceptual
control, i.e. making sure that a computational model faithfully
reflects one's hypothesis.  

NEURON has several features that facilitate conceptual control, and
it is acquiring more of them as it evolves to meet the changing
needs of computational neuroscientists. Many of these features fall
into the general category of "native syntax" specification of model
properties: that is, key attributes of biological neurons and
networks have direct counterparts in NEURON. For instance, NEURON
users specify the gating properties of voltage- and ligand-gated
ion channels with kinetic schemes or families of Hodgkin-Huxley
style differential equations. Another example is that models may
include electronic circuits constructed with the
LinearCircuitBuilder, a GUI tool whose palette includes resistors,
capacitors, voltage and current sources, and operational
amplifiers. NEURON's most striking application of native syntax may
lie in how it handles the cable properties of neurons, which is
very different from any other neurosimulator. NEURON users never
have to deal directly with compartments. Instead, cells are
represented by unbranched neurites, called sections, which can be
assembled into branched architectures (the topology of a model
cell). Each section has its own anatomical and biophysical
properties, plus a discretization parameter that specifies the
local resolution of the spatial grid. The properties of a section
can vary continuously along its length, and spatially inhomogeneous
variables are accessed in terms of normalized distance along each
section ((Hines and Carnevale 2001), chapter 5 in (Carnevale and
Hines 2006)). Once the user has specified cell topology, and the
geometry, biophysical properties, and discretization parameter for
each section, NEURON automatically sets up the internal data
structures that correspond to a family of ODEs for the model's
discretized cable equation.

{\it Computational robustness, accuracy, and efficiency}

NEURON's spatial discretization of conductance-based model neurons
uses a central difference approximation that is second order
correct in space. The discretization parameter for each section can
be specified by the user, or assigned automatically according to
the d\_lambda rule (see (Hines and Carnevale 2001), chapters 4 and
5 in (Carnevale and Hines 2006)).  

For efficiency, NEURON's computational engine uses algorithms that
are tailored to the model system equations (Hines 1984, 1989; Hines
and Carnevale 2004). To advance simulations in time, users have a
choice of built-in clock driven (fixed step backward Euler and
Crank-Nicholson) and event driven methods (global variable step and
local variable step with second order threshold detection); the
latter are based on CVODES and IDA from SUNDIALS (Hindmarsh et al.
2005). Networks of artificial spiking cells are solved analytically
by a discrete event method that is several orders of magnitude
faster than continuous system simulation (Hines and Carnevale
2004). NEURON fully supports hybrid simulations, and models can
contain any combination of conductance-based neurons and
analytically computable artificial spiking cells. Simulations of
networks that contain conductance-based neurons are second order
correct if adaptive integration is used (Lytton and Hines 2005).

Synapse and artificial cell models accept discrete events with
input stream specific state information. It is often extremely
useful for artificial cell models to send events to themselves in
order to implement refractory periods and intrinsic firing
properties; the delivery time of these "self events" can also be
adjusted in response to intervening events. Thus instantaneous and
non-instantaneous interactions of section 1.4 are supported.

Built-in synapses exploit the methods described in section 1.2.
Arbitrary delay between generation of an event at its source, and
delivery to the target (including 0 delay events), is supported by
a splay-tree queue (Sleator and Tarjan 1983) which can be replaced
at configuration time by a calendar queue. If the minimum delay
between cells is greater than 0, self events do not use the queue
and parallel network simulations are supported. For the fixed step
method, when queue handling is the rate limiting step, a bin queue
can be selected. For the fixed step method with parallel
simulations, when spike exchange is the rate limiting step,
six-fold spike compression can be selected.

\subsubsection{Creating and using models with NEURON}

Models can be created by writing programs in an interpreted
language based on hoc (Kernighan and Pike 1984), which has been
enhanced to simplify the task of representing the properties of
biological neurons and networks. Users can extend NEURON by writing
new function and biophysical mechanism specifications in the NMODL
language, which is then compiled and dynamically linked ((Hines and
Carnevale 2000), chapter 9 in (Carnevale and Hines 2006). There is
also a powerful GUI for conveniently building and using models;
this can be combined with hoc programming to exploit the strengths
of both (Fig.~\ref{FigGUI}).

The past decade has seen many enhancements to NEURON's capabilities
for network modeling. First and most important was the addition of
an event delivery system that substantially reduces the
computational burden of simulating spike-triggered synaptic
transmission, and enabled the creation of analytic
integrate-and-fire cell models which can be used in any combination
with conductance-based cells. Just in the past year the event
delivery system was extended so that NEURON can now simulate models
of networks and cells that are distributed over parallel hardware
(see NEURON in a parallel environment below).

{\it The GUI}

The GUI contains a large number of tools that can be used to
construct models, exercise simulations, and analyze results, so
that no knowledge of programming is necessary for the productive
use of NEURON. In addition, many GUI tools provide functionality
that would be quite difficult for users to replicate by writing
their own code. Some examples are: \\

\noindent {\it Model specification tools}

\begin{itemize}

\item Channel Builder--specifies voltage- and ligand-gated ion
channels in terms of ODEs (HH-style, including Borg-Graham
formulation) and/or kinetic schemes. Channel states and total
conductance can be simulated as deterministic (continuous in time),
or stochastic (countably many channels with independent state
transitions, producing abrupt conductance changes).

\item Cell Builder--manages anatomical and biophysical properties
of model cells.

\item Network Builder--prototypes small networks that can be mined
for reusable code to develop large-scale networks (chapter 11 in
(Carnevale and Hines 2006)).

\item Linear Circuit Builder--specifies models involving gap
junctions, ephaptic interactions, dual-electrode voltage clamps,
dynamic clamps, and other combinations of neurons and electrical
circuit elements.

\end{itemize}

\noindent {\it Model analysis tools}

\begin{itemize}

\item Import3D--converts detailed morphometric data (Eutectic,
Neurolucida, and SWC formats) into model cells. It automatically
fixes many common errors, and helps users identify complex problems
that require judgment.

\item Model View--automatically discovers and presents a summary of
model properties in a browsable textual and graphical form. This
aids code development and maintenance, and is increasingly
important as code sharing grows.

\item Impedance--compute and display voltage transfer ratios, input
and transfer impedance, and the electrotonic transformation.

\end{itemize}

\noindent {\it Simulation control tools}

\begin{itemize}

\item Variable Step Control--automatically adjusts the state
variable error tolerances that regulate adaptive integration.

\item Multiple Run Fitter--optimizes function and model parameters.

\end{itemize}

\subsubsection{NEURON in a parallel environment}
\label{nrn-parallel}

NEURON supports three kinds of parallel processing.

1.Multiple simulations distributed over multiple processors, each
processor executing its own simulation. Communication between
master processor and workers uses a bulletin-board method similar
to Linda (Carriero and Gelernter 1989).

2.Distributed network models with gap junctions.

3.Distributed models of individual cells (each processor handles
part of the cell). At present, setting up distributed models of
individual cells requires considerable effort; in the future it
will be made much more convenient.  

The four benchmark simulations of spiking neural networks (see
Appendix~2) were implemented under NEURON.  Figure~\ref{FigN}A
demonstrates the speedup that NEURON can achieve with distributed
network models of the four types (conductance-based, current-based,
Hodgkin-Huxley, event-based -- see Appendix~2) on a Beowulf cluster
(dashed lines are "ideal"--run time inversely proportional to
number of CPUs--and solid symbols are actual run times).
Figure~\ref{FigN}B shows that performance improvement scales with
the number of processors and the size and complexity of the
network; for this figure we ran a series of tests using a NEURON
implementation of the single column thalamocortical network model
described by Traub et al.  (2005) on the Cray XT3 at the Pittsburgh
Supercomputer Center.  Similar performance gain has been documented
in extensive tests on parallel hardware with dozens to thousands of
CPUs, using published models of networks of conductance based
neurons (Migliore et al., 2006). Speedup is linear with the number
of CPUs, or even superlinear (due to larger effective high speed
memory cache), until there are so many CPUs that each one is
solving fewer than ~100 equations.  

\subsubsection{Future plans}

NEURON undergoes a continuous cycle of improvement and revision,
much of which is devoted to aspects of the program that are not
immediately obvious to the user, e.g. improvement of computational
efficiency. More noticeable are new GUI tools, such as the recently
added Channel Builder. Many of these tools exemplify a trend toward
"form-based" model specification, which is expected to continue.
The use of form-based GUI tools increases the ability to exchange
model specifications with other simulators through the medium of
XML. With regard to network modeling, the emphasis will shift away
from developing simulation infrastructure, which is reasonably
complete, to the creation of new tools for network design and
analysis.  

\subsubsection{Software development, support, and documentation}

Michael Hines directs the NEURON project, and is responsible for
almost all code development. The other members of the development
team have varying degrees of responsibility for activities such as
documentation, courses, and user support. NEURON has benefited from
significant contributions of time and effort by members of the
community of NEURON users who have worked on specific algorithms,
written or tested new code, etc.. Since 2003, user contributions
have been facilitated by adoption of an "open source development
model" so that source code, including the latest research threads,
can be accessed from an on-line 
repository\url{http://www.neuron.yale.edu/neuron/install.html}.

Support is available by email, telephone, and consultation. Users can
also post questions and share information with other members of the
NEURON community via a mailing list and The NEURON
Forum\url{https://www.neuron.yale.edu/phpBB2/index.php}. Currently
the mailing list has more than 700 subscribers with "live" email
addresses; the Forum, which was launched in May, 2005, has already
grown to 300 registered users and 1700 posted messages.

Tutorials and reference material are
available\url{http://www.neuron.yale.edu/neuron/docs/docs.html}.  The
NEURON Book (Carnevale and Hines 2006) is the authoritative book on
NEURON.  Four books by other authors have made extensive use of
NEURON (Destexhe and Sejnowski 2001; Johnston and Wu 1995; Lytton
2002; Moore and Stuart 2000), and several of them have posted their
code online or provide it on CD with the book.

Source code for published NEURON models is available at many WWW
sites. The largest code archive is
ModelDB\url{http://senselab.med.yale.edu/senselab/ModelDB}, which
currently contains 238 models, 152 of which were implemented with
NEURON.

\subsubsection{Software availability}

NEURON runs under UNIX/Linux/OS X, MSWin 98 or later, and on parallel
hardware including Beowulf clusters, the IBM Blue Gene and Cray XT3. 
NEURON source code and installers are provided free of
charge\url{http://www.neuron.yale.edu}, and the installers do not
require "third party" software. The current standard distribution is
version 5.9.39. The alpha version can be used as a
simulator/controller in dynamic clamp experiments under real-time
Linux\url{http://rtai.org} with a National Instruments M series DAQ
card.


\clearpage
\subsection{GENESIS}

\subsubsection{GENESIS Capabilities and Design Philosophy}

GENESIS (the GEneral NEural SImulation System) was given its name because
it was designed, at the outset, be an extensible general simulation system
for the realistic modeling of neural and biological systems (Bower and
Beeman, 1998).  Typical simulations that have been performed with GENESIS
range from subcellular components and biochemical reactions (Bhalla, 2004)
to complex models of single neurons (De Schutter and Bower, 1994),
simulations of large networks (Nenadic et al., 2003), and systems-level
models (Stricanne and Bower, 1998).  Here, "realistic models" are defined
as those models that are based on the known anatomical and physiological
organization of neurons, circuits and networks (Bower, 1995).  For example,
realistic cell models typically include dendritic morphology and a large
variety of ionic conductances, whereas realistic network models attempt to
duplicate known axonal projection patterns.

Parallel GENESIS (PGENESIS) is an extension to GENESIS that runs on almost
any parallel cluster, SMP, supercomputer, or network of workstations where
MPI and/or PVM is supported, and on which serial GENESIS itself is
runnable.  It is customarily used for large network simulations involving
tens of thousands of realistic cell models (for example, see Hereld et al.,
2005).

GENESIS has a well-documented process for users themselves to extend its
capabilities by adding new user-defined GENESIS object types (classes), or
script language commands without the need to understand or modify the
GENESIS simulator code.  GENESIS comes already equipped with mechanisms to
easily create large scale network models made from single neuron models
that have been implemented with GENESIS.

While users have added, for example, the Izhikevich (2003) simplified
spiking neuron model (now built in to GENESIS), and they could also
add IF or other forms of abstract neuron models, these forms of
neurons are not realistic enough for the interests of most GENESIS
modelers.  For this reason, GENESIS is not normally provided with IF
model neurons, and no GENESIS implementations have been provided for
the IF model benchmarks (see Appendix~2).  Typical GENESIS neurons
are multicompartmental models with a variety of Hodgkin-Huxley type
voltage- and/or calcium-dependent conductances.

\subsubsection{Modeling with GENESIS}

GENESIS is an object-oriented simulation system, in which a simulation is
constructed of basic building blocks (GENESIS elements).  These elements
communicate by passing messages to each other, and each contains the
knowledge of its own variables (fields) and the methods (actions) used to
perform its calculations or other duties during a simulation.  GENESIS
elements are created as instantiations of a particular precompiled object
type that acts as a template.  Model neurons are constructed from these
basic components, such as neural compartments and variable conductance ion
channels, linked with messages.  Neurons may be linked together with
synaptic connections to form neural circuits and networks.  This
object-oriented approach is central to the generality and flexibility of
the system, as it allows modelers to easily exchange and reuse models or
model components.  Many GENESIS users base their simulation scripts on the
examples that are provided with GENESIS or in the GENESIS Neural Modeling
Tutorials package (Beeman, 2005).

GENESIS uses an interpreter and a high-level simulation language to
construct neurons and their networks.  This use of an interpreter with
pre-compiled object types, rather than a separate step to compile scripts
into binary machine code, gives the advantage of allowing the user to
interact with and modify a simulation while it is running, with no
sacrifice in simulation speed.  Commands may be issued either interactively
to a command prompt, by use of simulation scripts, or through the graphical
interface.  The 268 scripting language commands and the 125 object types
provided with GENESIS are powerful enough that only a few lines of script
are needed to specify a sophisticated simulation.  For example, the GENESIS
``cell reader'' allows one to build complex model neurons by reading their
specifications from a data file.

GENESIS provides a variety of mechanisms to model calcium diffusion and
calcium-dependent conductances, as well as synaptic plasticity.  There are
also a number of ``device objects'' that may be interfaced to a simulation
to provide various types of input to the simulation (pulse and spike
generators, voltage clamp circuitry, etc.) or measurements (peristimulus
and interspike interval histograms, spike frequency measurements, auto- and
cross-correlation histograms, etc.).  Object types are also provided for
the modeling of biochemical pathways (Bhalla and Iyengar, 1999).
A list and description of the GENESIS object types, with links to
full documentation, may be found in the {\em Objects} section of the
hypertext GENESIS Reference Manual, downloadable or viewable from the
GENESIS web site.

\subsubsection{GENESIS Graphical User Interfaces}

Very large scale simulations are often run with no GUI, with the
simulation output to either text or binary format files for later
analysis.  However, GENESIS is usually compiled to include its
graphical interface XODUS, which provides object types and
script-level commands for building elaborate graphical interfaces,
such as the one shown in Fig. \ref{FigG1} for the dual exponential
variation of the HH benchmark simulation (Benchmark~3 in Appendix~2).
GENESIS also contains graphical environments for building and running
simulations with no scripting, such as Neurokit (for single cells)
and Kinetikit (for modeling biochemical reactions).  These are
themselves created as GENESIS scripts, and can be extended or
modified.  This allows for the creation of the many educational
tutorials that are included with the GENESIS distribution (Bower and
Beeman, 1998).

\subsubsection{Obtaining GENESIS and User Support}

GENESIS and its graphical front-end XODUS are written in C and are
known to run under most Linux or UNIX-based systems with the X Window
System, as well as Mac OS/X and MS Windows with the Cygwin
environment.  The current release of GENESIS and PGENESIS (ver. 2.3,
March 17, 2006) is available from the GENESIS web
site\url{http://www.genesis-sim.org/GENESIS} under the GNU General
Public License.  The GENESIS source distribution contains full source
code and documentation, as well as a large number of tutorial and
example simulations. Documentation for these tutorials is included
along with online GENESIS help files and the hypertext GENESIS
Reference Manual. In addition to the source distribution, precompiled
binary versions are available for Linux, Mac OS/X, and Windows with
Cygwin.  The GENESIS Neural Modeling Tutorials (Beeman, 2005) are a
set of HTML tutorials intended to teach the process of constructing
biologically realistic neural models with the GENESIS simulator,
through the analysis and modification of provided example simulation
scripts.  The latest version of this package is offered as a separate
download from the GENESIS web site.

Support for GENESIS is provided through email to \texttt{genesis@genesis-sim.org},
and through the GENESIS Users Group, BABEL.  Members of BABEL receive
announcements and exchange information through a mailing list, and are
entitled to access the BABEL web page.  This serves as a repository for the
latest contributions by GENESIS users and developers, and contains
hypertext archives of postings from the mailing list.

Rallpacks are a set of benchmarks for evaluating the speed and
accuracy of neuronal simulators for the construction of single cell
models (Bhalla, et al., 1992).  However, it does not provide
benchmarks for network models.  The package contains scripts for both
GENESIS and NEURON, as well as full specifications for implementation
on other simulators.  It is included within the GENESIS distribution,
and is also available for download from the GENESIS web site.

\subsubsection{GENESIS Implementation of the HH Benchmark}

The HH benchmark network model (Benchmark 3 in Appendix~2) provides a
good example of the type of model that should probably NOT be
implemented with GENESIS.  The Vogels and Abbott (2005)
integrate-and-fire network on which it is based is an abstract model
designed to study the propagation of signals under very simplified
conditions.  The identical excitatory and inhibitory neurons have no
physical location in space, and no distance-dependent axonal
propagation delays in the connections.  The benchmark model simply
replaces the IF neurons with single-compartment cells containing fast
sodium and delayed rectifier potassium channels that fire tonically
and display no spike frequency adaptation.  Such models offer no
advantages over IF cells for the study of the situation explored by
Vogels and Abbott.

Nevertheless, it is a simple matter to implement such a model in
GENESIS, using a simplification of existing example scripts for large
network models, and the performance penalty for ``using a sledge
hammer to crack a peanut'' is not too large for a network of this
size.  The simulation script for this benchmark illustrates the power
of the GENESIS scripting commands for creating networks.  Three basic
commands are used for filling a region with copies of prototype
cells, making synaptic connections with a great deal of control over
the connectivity, and setting propagation delays.

The instantaneous rise in the synaptic conductances makes this a very
efficient model to implement with a simulator specialized for IF
networks, but such a non-biological conductance is not normally
provided by GENESIS.  Therefore, two implementations of the benchmark
have been provided.  The Dual Exponential VA HH Model script
implements synaptic conductances with a dual exponential form having
a 2 msec time-to-peak, and the specified exponential decay times of 5
msec for excitatory connections and 10 msec for inhibitory
connections.  The Instantaneous Conductance VA HH Model script uses a
user-added {\em isynchan} object type that can be compiled and linked
into GENESIS to provide the the specified conductances with an
instantaneous rise time.  There is little difference in the behavior
of the two versions of the simulation, although the Instantaneous
Conductance model executes somewhat faster.

Figure \ref{FigG1} shows the implementation of the Dual Exponential VA HH
Model with a GUI that was created by making small changes to the example
{\em RSnet.g}, {\em protodefs.g}, and {\em graphics.g} scripts, which are
provided in the GENESIS Modeling Tutorial (Beeman, 2005) section {\em
Creating Large Networks with GENESIS}.

These scripts and the tutorial specify a rectangular grid of excitatory
neurons.  An exercise suggests adding an additional layer of inhibitory
neurons.  The GENESIS implementations of the HH benchmark use a
layer of $64 \times 50$ excitatory neurons and a layer of $32 \times 25$
inhibitory neurons.  A change of one line in the example {\em RSnet.g}
script allows the change from the nearest-neighbor connectivity of the
model to the required infinite-range connectivity with 2\% probability.

The identical excitatory and inhibitory neurons used in the network
are implemented as specified in App. 2.  For both versions of the model,
Poisson-distributed random spike inputs with a mean frequency of 70 Hz were
applied to the excitatory synapses of the all excitatory neurons.  The
the simulation was run for 0.05 sec, the random input was removed, and
it was then run for an additional 4.95 sec.

The Control Panel at the left is used to run the simulation and to set
parameters such as maximal synaptic conductances, synaptic weight scaling,
and propagation delays.  There are options to provide current injection
pulses, as well as random synaptic activation.  The plots in the middle
show the membrane potentials of three excitatory neurons (0, 1536, and
1567), and inhibitory neuron 0.  The netview displays at the right show the
membrane potentials of the excitatory neurons (top) and inhibitory neurons
(bottom).  With no propagation delays, the positions of the neurons on the
grid are irrelevant.  Nevertheless, this two-dimensional representation of
the network layers makes it easy to visualize the number of cells firing at
any time during the simulation.

Figure \ref{FigG2} shows the plots for the membrane potential of the same
neurons as those displayed in Fig. \ref{FigG1}, but produced by the
Instantaneous Conductance VA HH Model script.  The plot at the right shows
a zoom of the interval between 3.2 and 3.4 sec.

In both figures, excitatory neuron 1536 has the lowest ratio of excitatory
to inhibitory inputs of the four neurons plotted.  It fires only rarely,
whereas excitatory neuron 0, which has the highest ratio, fires most
frequently.

\subsubsection{Future Plans for GENESIS}

The GENESIS simulator is now undergoing a major redevelopment effort,
which will result in GENESIS 3.  The core simulator functionality is
being reimplemented in C++ using an improved scheme for messaging
between GENESIS objects, and with a platform-independent and
browser-friendly Java-based GUI.  This will result in not only
improved performance and portability to MS Windows and non-UNIX
platforms, but will also allow the use of alternate script parsers
and user interfaces, as well as the ability to communicate with other
modeling programs and environments.  The GENESIS development team is
participating in the NeuroML (Goddard et al., 2001; Crook et al.,
2005) project\url{http://www.neuroml.org}, along with the developers
of NEURON.  This will enable GENESIS 3 to export and import model
descriptions in a common simulator-independent XML format. 
Development versions of GENESIS are available from the Sourceforge
GENESIS development
site\url{http://sourceforge.net/projects/genesis-sim}.


\clearpage
\subsection{NEST}

\subsubsection{The NEST initiative\label{sec:The-NEST-initiative}}

The problem of simulating neuronal networks of biologically realistic
size and complexity has long been underestimated. This is reflected
in the limited number of publications on suitable algorithms and data
structures in high-level journals. The lack of awareness of researchers
and funding agencies of the need for progress in simulation technology
and sustainability of the investments may partially originate from
the fact that a mathematically correct simulator for a particular
neuronal network model can be implemented by an individual in a few
days. However, this has routinely resulted in a cycle of unscalable
and unmaintainable code being rewritten in unmaintainable fashion
by novices, with little progress in the theoretical foundations.

Due to the increased availability of computational resources, simulation
studies are becoming ever more ambitious and popular. Indeed, many
neuroscientific questions are presently only accessible through simulation.
An unfortunate consequence of this trend is that it is becoming ever
harder to reproduce and verify the results of these studies. The ad
hoc simulation tools of the past cannot provide us with the appropriate
degree of comprehensibility. Instead we require carefully crafted,
validated, documented and expressive neuronal network simulators with
a wide user community. Moreover, the current progress towards more
realistic models demands correspondingly more efficient simulations.
This holds especially for the nascent field of studies on large-scale
network models incorporating plasticity. This research is entirely
infeasible without parallel simulators with excellent scaling properties,
which is outside the scope of ad hoc solutions. Finally, to be useful
to a wide scientific audience over a long time, simulators must be
easy to maintain and to extend.

On the basis of these considerations, the NEST initiative was founded
as a long term collaborative project to support the development of
technology for neural systems simulations (Diesmann et al., 2002). 
The NEST simulation tool is the reference implementation of this
initiative.  The software is provided to the scientific community
under an open source license through the NEST initiative's
website\url{http://www.nest-initiative.org}.  The license requests
researchers to give reference to the initiative in work derived from
the original code and, more importantly, in scientific results
obtained with the software. The website also provides references to
material relevant to neuronal network simulations in general and is
meant to become a scientific resource of network simulation
information.  Support is provided through the NEST website and a
mailing list. At present NEST is used in teaching at international
summer schools and in regular courses at the University of Freiburg.

\subsubsection{The NEST simulation tool}

In the following we give a brief overview of the NEST simulation tool
and its capabilities.

{\it Domain and design goals}

The domain of NEST is large neuronal networks with biologically realistic
connectivity. The software easily copes with the threshold network
size of $10^{5}$ neurons (Morrison et al., 2005) at which each neuron
can be supplied with the natural number of synapses and simultaneously
a realistic sparse connectivity can be maintained. Typical neuron
models in NEST have one or a small number of compartments. The simulator
supports heterogeneity in neuron and synapse types. In networks of
realistic connectivity the memory consumption and work load is dominated
by the number of synapses. Therefore, much emphasis is placed on the
efficient representation and update of synapses. In many applications
network construction has the same computational costs as the integration
of the dynamics. Consequently, NEST parallelizes both. NEST is designed
to guarantee strict reproducibility: the same network is required
to generate the same results independent of the number of machines
participating in the simulation. It is considered an important principle
of the project that the development work is carried out by neuroscientists
operating on a joint code base. No developments are made without the
code being directly tested in neuroscientific research projects. This
implements an incremental and iterative development cycle. Extensibility
and long-term maintainability are explicit design goals.

{\it Infrastructure}

The primary user interface is a simulation language interpreter which
processes a rather high level expressive language with an extremely
simple syntax which incorporates heterogeneous arrays, dictionaries,
and pure (i.e. unnamed) functions and is thus suited for interactive
work. There is no built-in graphical user interface as it would not
be particularly helpful in NEST's domain: network specification is
procedural, and data analysis is generally performed off-line for
reasons of convenience and efficiency. The simulation language is
used for data pre- and post-processing, specification of parameters,
and for the compact description of the network structure and the protocol
of the virtual experiment. The neuron models and synapse types are
not expressed in the simulation language as this would result in a
slower performance. They are implemented as derived classes on the
C++ level such that all models provide the same minimal functionality
and are thus easily interchangeable on the simulation language level.
A mechanism for error handling propagates errors messages through
all levels of the software. Connections between nodes (i.e. neurons,
generators and recording devices) are checked for consistency at the
time of creation. User level documentation is provided in a browsable
format (the ``helpdesk'') and is generated directly from source code. 

The code of NEST is modularized to facilitate the development of new
neuron models that can be loaded at run time and to decouple the development
of extensions from a specific NEST release. In the framework of the
FACETS project a Python interface and a {}``facetsmodule'' has been
created. In addition to providing an interface between user-defined
modules and the core code, NEST can interface with other software
- for example, in order to provide a graphical user interface. The
primary strategy used is interpreter-interpreter interaction, whereby
each interpreter emits code that the other interpreter accepts as
its native language. This approach minimizes the need to define protocols
and the dependency of NEST on foreign libraries.

{\it Kernel}

There is a common perception that event-driven algorithms are exact
and time-driven algorithms are approximate. We have recently shown
that both parts of this perception are generally false; it
depends on the dynamics of the neuron model whether an event-driven
algorithm can find an exact solution, just as it does for time-driven
algorithms (Morrison et al., 2006b). NEST is designed for large scale
simulations where performance is a critical issue. We have therefore
argued that when comparing different integration strategies, one
should evaluate the efficiency, i.e. the simulation time required to
achieve a given integration error, rather than the plain simulation
time (Morrison et al., 2006b).  This philosophy is reflected in the
simulation kernel of NEST. Although it implements a globally
time-driven algorithm with respect to the ordering of neuron updates
and the delivery of events, spike times are not necessarily
constrained to the discrete time grid. Neuron implementations
treating incoming and outgoing spikes in continuous time are
seamlessly integrated into the time-driven infrastructure with no
need for a central event queue. This permits a great flexibility in
the range of neuron models which can be represented, including
exactly solvable continuous time neuron models, models requiring
approximation techniques to locate threshold passing and models with
grid-constrained dynamics and spike times. 

The simulation kernel of NEST supports parallelization by
multi-threading and message passing, which allows distribution of a
simulation over multiple processors of an SMP machine or over
multiple machines in a cluster. Communication overhead is minimized
by only communicating in intervals of the minimum propagation delay
between neurons, and communication bulk is minimized by storing
synapses on the machine where the post-synaptic neuron is located
(Morrison et al., 2005). This results in supra-linear speed-up in
distributed simulations; scaling in multi-threaded simulations is
reasonable, but more research is required to understand and overcome
present constraints. The user only needs to provide a serial script,
as the distribution is performed automatically. Interactive usage of
the simulator is presently only possible in purely multi-threaded
operation. Reproducibility of results independent of the number of
machines/processors is achieved by dividing a simulation task into a
fixed number of abstract (virtual) processes which are distributed
amongst the actual machines used (Morrison et al., 2005).

\subsubsection{Performance}

The supplementary material contains simulation scripts for all of the
benchmarks specified in Appendix~2. Considering the domain of NEST,
the benchmarks can only demonstrate NEST's capabilities in a limited
way. Therefore, a fifth benchmark is included which is not only
significantly larger than the other benchmarks (three times as many
neurons and forty times as many synapses), but also incorporates
spike-timing dependent plasticity in its excitatory-excitatory
synapses. The neuron model for this benchmark is the same as for
benchmark $2$. All the benchmarks were simulated on a Sun Fire V40z
equipped with four dual core AMD Opteron $875$ processors at $2.2$
GHz and $32$ Gbytes RAM running Ubuntu 6.06.1 LTS with kernel
2.6.15-26-amd64-server. Simulation jobs were bound to specific
cores using the \texttt{taskset} command. The simulations were
performed with a synaptic propagation delay of $0.1$ ms and a
computation time step of $0.1$ ms unless otherwise stated.

Fig.~\ref{cap:nest_benchmarks}A shows the simulation time for one biological
second of Benchmarks $1-3$. To compare the benchmarks fairly despite their
different firing rates, the spiking was suppressed in all three benchmarks by
removing the initial stimulus, and in the case of Benchmark~$2$, the intrinsic
firing was suppressed by setting the resting potential to be lower than the
threshold. For networks of integrate-and-fire neuons of this size and
activity, the delivery of spikes does not contribute significantly to the
simulation times, which are dominated by the neuron updates. If the spiking is
not suppressed, the simulation times for Benchmarks $1$ and $2$ are less than
$10\%$ longer. The simulation time for Benchmark~$3$ is about $15\%$ longer
because of the computational cost associated with the integration of the
action potential. Benchmark~$2$ (current-based integrate-and-fire neuron
model) is significantly faster than the other two as its linear subthreshold
dynamics permits the use of exact integration techniques (see Rotter and
Diesmann, 1999).  The non-linear dynamics of the conductance based
integrate-and-fire neuron model in Benchmark~$1$ and the Hodgkin-Huxley neuron
in benchmark~$3$ are propagated by one global computation time step by one or
more function calls to the standard adaptive time stepping method of the GNU
Scientific Library (GSL; Galassi et al., 2001) with a required accuracy of
$1\:\mathrm{\mu}V$ . The ODE-solver used is the embedded Runge-Kutta-Fehlberg
$(4,5)$ provided by the GSL , but this is not a constraint of NEST - a neuron
model may employ any method for propagating its dynamics. In a distributed
simulation, processes must communicate in intervals of the minimum synaptic
delay in order to preserve causality (Morrison et al., 2005). It is therefore
more efficient to simulate with realistic synaptic delays than with
unrealistically short delays, as can be seen in
Fig.~\ref{cap:nest_benchmarks}A. The simulation times for the benchmark
networks incorporating a synaptic delay of $1.5$ ms are in all cases
significantly shorter than the simulation times for the networks if the
synaptic delay is assumed to be $0.1$ ms.

Benchmark~$4$ (integrate-and-fire neuron model with voltage jump
synapses) is ideal for an event-driven simulation, as all spike times
can be calculated analytically - they occur either when an excitatory
spike is received, or due to the relaxation of the membrane potential
to the resting potential, which is above the threshold. Therefore the
size of the time steps in which NEST updates the neuron dynamics
plays no role in determining the accuracy of the simulation. The 
primary constraint on the step
size is that it must be less than or equal to the
minimum synaptic delay between the neurons in the network. 
Fig.~\ref{cap:nest_benchmarks}B shows the simulation time for one
biological second of Benchmark~$4$ on two processors as a function of
the minimum synaptic delay. Clearly, the simulation time is strongly
dependent on the minimum delay in this system. At a realistic value
of $1$ ms, the network simulation is approximately a factor of $1.3$
slower than real time; at a delay of $0.125$ ms the simulation is
approximately $7.3$ times slower than real time. In the case of
neuron models where the synaptic time course is not invertible, the
computational time step determines the accuracy of the calculation of
the threshold crossing. For a discussion of this case and the
relevant quantitative benchmarks, see Morrison et al.\ (2006b).

Fig.~\ref{cap:nest_benchmarks}C shows the scaling of an application which lies
in the domain of neural systems for which NEST is primarily designed. The
simulated network contains $11250$ neurons, of which $9000$ are excitatory and
$2250$ inhibitory. Each neuron receives $900$ inputs randomly chosen from the
population of excitatory neurons and $225$ inputs randomly chosen from the
inhibitory population. The scaling is shown for the case that all the synapses
are static, and for the case that the excitatory-excitatory synapses implement
multiplicative spike-timing dependent plasticity with an all-to-all spike
pairing scheme (Rubin et al., 2001). For implementation details of the STDP,
see Morrison et al.\ (2006a), for further network parameters, see the
supplementary material. The network activity is in the asynchronous irregular
regime at $10$ Hz. Both applications scale supra-linearly due to the
exploitation of fast cache memory.  When using eight processors, the static
network is a factor of $6.5$ slower than real time and the plastic network is
a factor of $14$ slower. Compared to Benchmark~$2$, the network contains $3$
times as many neurons, $40$ times as many synapses and the firing rate is
increased by a factor of $2$.  However, using the same number of processors
($2$), the static network simulation is only a factor of $17$ slower, and the
plastic network simulation is only a factor of $32$ slower. This demonstrates
that NEST is capable of simulating large, high-connectivity networks with
computationally expensive synaptic dynamics with a speed suitable for
interactive work. Although for this network the presence of the STDP synapses
increases the simulation time by a factor of two, this factor generally
depends on the number of synapses and the activity.

\subsubsection{Perspectives}

Future work on NEST will focus on an interactive mode for distributed
computing, an improvement of performance with respect to modern
multi-core computer clusters, and a rigorous test and validation
suite. Further information on NEST and the current release can be
found at the NEST web site\url{http://www.nest-initiative.org}.


\clearpage
\subsection{NCS}

The NeoCortical Simulator (NCS), as its name suggests, is optimized
to model the horizontally dispersed, vertically layered
distribution of neurons characteristic of the mammalian neocortex.
NCS development began in 1997, a time at which fascinating details
of synaptic plasticity and connectivity were being discovered
(Markram et al., 1997a, 1997b) yet available simulators such as
GENESIS and NEURON did not offer parallel architectures nor the
degree of neuronal compartmental simplification required for
reasonable performance times. Also emerging at the time were
inexpensive clusters-of-workstations, also known as Beowulf
clusters, operating under the LINUX operating system. Following a
1997 neuroscience fellowship with Rodney Douglas and Kevan Martin
at the Institute for Neuroinformatics in Z\"urich, Philip Goodman
programmed the first NCS using Matlab in collaboration with Henry
Markram (then at the Weizmann Institute, now at the Swiss EPFL) and
Thomas McKenna, Neural Computation Program Officer at the U.S.
Office of Naval Research. Preliminary results led to ONR funding
(award N000140010420) in 1999, which facilitated the subsequent
collaboration with UNR computer scientists Sushil Louis and
Frederick Harris, Jr. This led to a C++ implementation of NCS using
LINUX MPI on a Beowulf cluster. NCS was first made available to
outside investigators beginning in 2000, with further development
targeting the following specifications:

\begin{enumerate}
	\item \underline{Compartments:}  sampling frequency and  membrane compartmental realism sufficient 
	to capture biological  response properties, arbitrary voltage- and ion-sensitive channel behaviors, 
	and multicompartmental models  distributed in 3-D (dendritic, somatic, and axonal systems);

	\item \underline{Synapses:} short-term depression and facilitation (Markram et al., 1998a), 
	augmentation (Wang et al., 2006) and Hebbian spike-timing dependent plasticity (Markram et al., 1997b);

	\item \underline{3-D Connectionism:} a layout to easily allocate neurons into subnetwork groupings, 
	layers, column, and sheets separated by real micron- or millimeter spacings, with realistic propagation
	distances and axonal conduction speeds;

	\item \underline{Parallelism:} an inherently parallel, efficient method of passing messages of 
	synaptic events among neurons;

	\item \underline{Reporting:} an efficient way to collect, sample and analyze selected compartmental 
	and neuronal behaviors;

	\item \underline{Stimulation:} ability to (a) specify fixed, standard neurophysiological stimulation 
	protocols, (b) port signals from an external device, and (c) export neuronal responses and await 
	subsequent replies from external systems (e.g., dynamic clamps, in vitro or in vivo preparations, 
	robotic emulations);

	\item \underline{Freeze/resume system state:} the ability to stop a simulation and hibernate all 
	hardware and software parameters into a binary blob, for unpacking and resuming in later experiments;

	\item \underline{Command files:} simplicity in generating and modifying scripts.

\end{enumerate}

As of 2005, NCS developers achieved all the objectives above, using
an ASCII file based command input file to define a hierarchy of
reusable brain objects (Figure~\ref{fig:NCS}A). NCS uses a
clock-based integrate and fire neurons whose compartments contain
conductance-based synaptic dynamics and Hodgkin-Huxley formulations
of ionic channel gating
particles\url{http://brain.unr.edu/publications/thesis.ecw01.pdf}. 
Although a user-specified active spike template is usually used for
our large simulations, Hodgkin-Huxley channel equations can be
specified for the rapid sodium and delayed rectifier spike behavior. 
No nonlinear simplifications, such as the Izhikevich formulation, are
supported.  Compartments are allocated in 3-D space, and are
connected by forward and reverse conductances without detailed cable
equations.  Synapses are conductance-based, with phenomenological
modeling of depression, facilitation, augmentation, and STDP.  

NCS runs on any LINUX cluster. We run NCS on our 200-CPU  hybrid of
Pentium and AMD processors, and also on the 8,000-CPU Swiss EPFL
IBM Blue Brain. NCS can run in single-PC mode under LINUX or LINUX
emulation (e.g., Cygwin) and on the new Pentium-based Macintosh.  

Although NCS was motivated by the need to model the complexity of
the neocortex and hippocampus, limbic and other structures can be
modeled by variably collapsing layers and specifying the relevant
3-D layouts. Large-scale models often require repetitive patterns
of interconnecting brain objects, which can be tedious using only
the basic ASCII command file. We therefore developed a suite of
efficient Python-based scripting tools called Brainlab (Drewes,
2005). An Internet-based library and control system was also
developed (Waikul et al., 2002).  

NCS delivers reports on any fraction of neuronal cell groups, at any
specified interval. Reports include membrane voltage (current clamp
mode), current (voltage clamp), spike-event-only timings
(event-triggered), calcium concentrations, synaptic dynamics
parameter states, and any Hodgkin-Huxley channel parameter.  Although
NCS does not provide any direct visualization software, report files
are straightforward to view in any graphics environment. Two such
Matlab-based tools are available for download from the lab's web
site\url{http://brain.unr.edu/publications/neuroplot.m;
http://brain.unr.edu/publications/EVALCELLTRACINGS.zip}.

Benchmark. We ran the Vogels and Abbot (2005) benchmark under the
conditions specified for the conductance-based integrate and fire
model (see Benchmark~1 in Appendix~2), and obtained the expected
irregularly-bursting sustained pattern (first second shown in
Figure~\ref{fig:NCS}B).  At the default 10:1 ratio of inhibitory to
excitatory synaptic conductances, the overall mean firing rate was
15.9 Hz.  

The largest simulations to-date have been on the order of a million
single-compartment neurons using membrane AHP, M, A-type channels.
Neurons were connected by 1 trillion synapses using short-term and
STDP dynamics; this required about 30 minutes on 120 CPUs to
simulate one biological second
(Ripplinger et al., 2004). Intermediate-complexity simulations have
examined multimodal sensory integration and information transfer
\url{http://brain.unr.edu/publications/Maciokas\_Dissertation\_final.zip},
and genetic algorithm search for parameter sets which support
learning of visual patterns (Drewes et al., 2004). Detailed work
included evaluation of interneuronal membrane channels (Maciokas et
al., 2005) underlying the spectrum of observed firing behaviors
(Gupta et al., 2000), and potential roles in speech recognition
(Blake and Goodman, 2002) and neuropathology (Kellogg et al., 1999;
Wills et al., 1999; Wiebers et al., 2003; Opitz and Goodman, 2005).
Recent developments focus on IP port-based real time input-output
of the "brain" to remotely behaving and learning 
robots\url{http://brain.unr.edu/publications/jcm.hierarch\_robotics.unr\_ms\_thesis03.pdf; 
http://brain.unr.edu/publications/JGKingThesis.pdf
(Macera-Rios et al., 2004)}.  

The UNR Brain Computation Laboratory is presenting collaborating
with the Brain Mind Institute of the Swiss EPFL. Their 8,000-CPU
Blue Brain cluster\url{http://bluebrainproject.epfl.ch} currently
runs NCS alone or as in a hybrid configuration as an efficient
synaptic messaging system with CPU-resident instances of NEURON.
The Reno and Swiss teams are exploring ways to better calibrate
simulated to living microcircuits, and to effect real-time robotic
behaviors. Under continuing ONR support, the investigators and two
graduate students provide part-time assistance to external users at
no cost through e-mail and online documentation. User manual and
programmer specifications with examples are 
available\url{http://brain.unr.edu/ncsDocs}.


\clearpage
\subsection{CSIM}

\newcommand{\mypar}[1]{\emph{#1}:}

\subsubsection{Feature overview}

The \emph{\emph{C}ircuit \emph{SIM}ulator} CSIM is a tool for

simulating heterogeneous networks composed of (spike emitting) point
neurons. CSIM is intended to simulate networks containing a few
neurons, up to networks with a few thousand neurons and on the order
of 100000 synapses. It was written to do modeling at the network level
in order to analyze the computational effects which can not be
observed at the single cell level. To study single cell computations
in detail we give the advice to use simulators like GENESIS or NEURON.

\mypar{Easy to use Matlab interface} The core of CSIM is written in C++
which is controlled by means of Matlab (there is no standalone version
of CSIM). We have chosen Matlab since it provides very powerful
graphics and analysis capabilities and is a widely used programming
language in the scientific community. Hence it is not necessary to
learn yet another script language to set up and run simulations with
CSIM. Furthermore the results of a simulation are directly returned as
Matlab arrays and hence any plotting and analysis tools available in
Matlab can easily be applied.

Until now CSIM does not provide a GUI. However one can easily use Matlab
powerful GUI builder to make a GUI for a specific application based on CSIM.
  
\mypar{Object oriented design} We adopted an object oriented design for CSIM
which is similar to the approaches taken in GENESIS and NEURON. That is there
are objects (e.g. a \texttt{LifNeuron} object implements the standard
leaky-integrate-and-fire model) which are interconnected by means of well
defined signal channels. The creation of objects, the connection of objects
and the setting of parameters of the objects is controlled at the level of
Matlab whereas the actual simulation is done in the C++ core.

\mypar{Fast C++ core} Since CSIM is implemented in C++ and is not as general
as e.g. GENESIS simulations are performed quite fast.  We also implemented
some ideas from event driven simulators which result in a considerable speedup
(up to a factor of three for low firing rates; see the subsection about
implementation aspects below).
  
\mypar{Runs on Windows and Linux (Unix)} CSIM is developed on Linux
(Matlab 6.5 and 7.2, gcc 4.0.2). From the site 
\verb+www.lsm.tugraz.at/csm+ precompiled versions for Linux and Windows
are available.  Since CSIM is pure C++ it should not be hard to port
it to other platforms for which Matlab is available.
  
\mypar{Different levels of modeling} By providing different neuron models CSIM
allows to investigate networks at different levels of abstraction: sigmoidal
neurons with analog output, linear and non-linear leaky-integrate-and-fire
neurons and compartmental based (point) neurons with spiking output. A broad
range of synaptic models is also available for both spiking and non-spiking
neuron models: starting from simple static synapses ranging over synapses with
short-term plasticity to synapse models which implement different models for
long-term plasticity.

\subsubsection{Built-in models}

\mypar{Neuron models} CSIM provides two different classes of neurons: neurons
with analog output and neurons with spiking output. Neurons with analog output
are useful for analyzing population responses in larger circuits. For example
CSIM provides a sigmoidal neuron with leaky integration. However, there are
much more different objects available to build models of spiking neurons:
\begin{itemize}
  
\item Standard (linear) leaky-integrate-and-fire neurons
  
\item Non-linear leaky-integrate-and-fire neurons based on the models
  of Izhikevich
  
\item Conductance based point neurons with and without a spike template. There
  are general conductance based neurons where the user can insert any number
  of available ion-channel models to build the neuron model. On the other hand
  there is a rich set of predefined point neurons available used in several
  studies.

\end{itemize}

\mypar{Spiking Synapses} As for the neurons CSIM also implements synapses
which transmit analog values and spike transmitting synapses.  Two types of
synapses are implemented: static and dynamic synapses. While for static
synapses the amplitude of each postsynaptic response (current of conductance
change) is the same, the amplitude of an postsynaptic response in the case of
a dynamic synapse depends on the spike train that it has seen so far, i.e.
dynamic synapses implement a form of short term plasticity (depression,
facilitation). For synapses transmitting spikes the time course of a
postsynaptic response is modeled by $A \times \mathrm{exp}(-t/\tau_{syn})$,
where $\tau_{syn}$ is the synaptic time constant and $A$ is the synaptic
strength which is constant for static synapses and given by the model
described in (Markram~et~al.,~1998b) for dynamic synapses.

Note that static as well as dynamic synapses are available as current
supplying or conductance based models.

\mypar{Analog Synapses} For synapses transmitting analog values, such as the
output of a sigmoidal neuron, static synapses are simply defined by their
strength (weight), whereas for dynamic synapses we implemented a continuous
version of the dynamic synapse model for spiking neurons
(Tsodyks~et~al.~1998).

\mypar{Synaptic plasticity} CSIM also supports spike time dependent
plasticity, STDP, applying a similar model as in (Song~et~al.,~2000). STDP can
be modeled most easily by making the assumption that each pre- and
postsynaptic spike pair contributes to synaptic modification independently and
in a similar manner. Depending on the time difference $\Delta t =
t_{pre}-t_{post}$ between pre- and postsynaptic spike the absolute synaptic
strength is changed by an amount $L(\Delta t)$. The typical shape for the
function $L(\Delta t)$ as found for synapses in neocortex layer 5
(Markram~et~al.,~1997) is implemented. Synaptic strengthening and weakening
are subject to constraints so that the synaptic strength does not go below
zero or above a certain maximum value. Furthermore additional variants as
suggested in (Froemke and Dan, 2002) and (G{\"u}tig~et~al.~2003) are also
implemented.

\subsubsection{Implementation aspects}

\mypar{Network input and output} There are two forms of inputs which can be
supplied to the simulated neural microcircuit: spike trains and analog
signals. To record the output of the simulated model special objects called
\texttt{Recorder} are used. A recorder can be connected to any object to
record any field of that object.

\mypar{Simulation Strategy} CSIM employees a clock based simulation strategy
with a fixed simulation step width $dt$. Typically the exponential Euler
integration method is used.  A spike which occurs during a simulation time
step is assumed to occur at the end of that time step. That implies that
spikes can only occur at multiples of $dt$.

\mypar{Efficient processing of spikes} In a typical simulation of a neural
circuit based on simple neuron models the CPU time spent in advancing
\emph{all} the synapses may by larger then the time needed to integrate the
neuron equations. However if one considers the fact that synapses are
actually ``idle'' most of the time (at least in low firing rate scenarios) it
makes sense to update during one time step only those synapses whose
postsynaptic response is not zero, i.e. are active. CSIM implements this idea
by dividing synapses into a list of idle and a list of active synapses where
only the latter is updated during a simulation time step. A synapse becomes
active (i.e. is moved from the idle list to the active list) if a spike
arrives. After its postsynaptic response has vanished the synapse becomes idle
again (i.e. is moved back from the active list to the idle list). This trick
can result in considerable speed up for low firing rate scenarios.

\subsubsection{Further information}

CSIM ins distributed under the GNU General Public License and is
available for download\url{http://www.lsm.tugraz.at/csim}. Support
for CSIM (and its related tools) can be obtained by writing email to
\verb+lsm@igi.tu-graz.ac.at+

At the site \verb+http://www.lsm.tugraz.at+ one can find besides the
download area for CSIM (including the user manual and an object
reference manual) a list of publications which used CSIM (and its
related tools) and also the code of published models.

\mypar{Related tools} Furthermore the site
\verb+http://www.lsm.tugraz.at+ provides two sets of Matlab scripts
and objects which heavily build on CSIM.  The \emph{circuit tool}
supports the construction of multi-column circuits by providing
functionality to connect pools of neurons to pools of neurons. The
\emph{learning tool} was developed to analyze neural circuits in the
spirit of the liquid state machine (LSM) approach (Maass et al.,
2002) and therefore contains several machine learning methods (see
(Natschl\"{a}ger~et~al.,~2003) for more information about this tools).
  
As of this writing resources are devoted to develop a parallel
version of CSIM called PCSIM which allows distributed simulation of
large scale networks.  PCSIM will have a python interface which
allows an easy implementation of the upcoming PyNN application
programming interface (see appendix 1).  The current development
version of PCSIM can be obtained from the SourceForge 
site\url{http://sourceforge.net/projects/pcsim}.

\subsubsection{CSIM implementations of the benchmark simulations}

We implemented the benchmark networks 1 to 3 as specified in
Appendix~2.

The integrate-and-fire benchmark networks (Benchmark 1 and 2) are
well suited to be simulated with CSIM and can be implemented by only
using built-in objects: \texttt{CbNeuron} and
\texttt{StaticSpikingCbSynapse} as the neuron and synapse model for
the COBA network and \texttt{LifNeuron} and
\texttt{StaticSpikingSynapse} as neuron and synapse model for the
CUBA network.

To implement Benchmark 3 (HH network) it is necessary to add the
desired channel dynamics to CSIM by implementing it at the C++ level.
The user defined neuron model (\texttt{TraubsHHNeuron}) is easily
implemented in C++ (see the files \verb+traubs_hh_channels.[cpp|h]+
and \verb+TraubsHHNeuron.[cpp|h]+).  After these files are compiled
and linked to CSIM they are available for use in the simulation. We
refer the user to the CSIM manual for details on how to add user
defined models at C++ level to CSIM.

For each benchmark network we provide two implementations: the first
implementation uses the plain CSIM interface only while the second
implementation makes use of the \emph{circuit tool} mentioned in the
previous subsection (filename suffix \verb+*_circuit.m+).

To provide the initial stimulation during the first 50\,ms of the
simulation we set up a pool of input neurons
(\texttt{SpikingInputNeuron} objects) which provide random spikes to
the network.

Results of CSIM simulations of all implemented benchmarks are
depicted in Figure~\ref{fig:CSIM1}. This figures were produced by the
simulation scripts provided for each benchmark using Matlab's
powerful graphics capabilities (see the file \verb+make_figures.m+)
and illustrate the sustained irregular activity described by Vogels
and Abbott (2005) for such networks.

The current development version of PCSIM has been used to perform
scalability tests based on the CUBA benchmark (Benchmark~2). The
results are summarized in Figure~\ref{fig:PCSIM}.  For the small 4000
neuron network the speedup for more than four machines vanishes while
for the larger networks a more than expected speedup occurs up to six
machines. This shows that PCSIM is scalable with regard to the
problem size and the number of available machines.  The development
version of PCSIM together with the python script for the CUBA
benchmark can be obtained from the SourceForge
site\url{http://sourceforge.net/projects/pcsim}.


\clearpage
\subsection{XPPAUT}

\newcommand{\XPP}{{\em XPPAUT \,}}

 \XPP is a general numerical tool for simulating, animating, and
analyzing dynamical systems.  These can range from discrete finite
state models (McCulloch-Pitts) to stochastic Markov models, to
discretization of partial differential and integrodifferential
equations.  \XPP was not specifically developed for neural
simulations but because of its ability to provide a complete
numerical analysis of the dependence of solutions on parameters
(``bifurcation diagrams'') it is widely used by the community of
computational and theoretical neuroscientists.  There are many online
tutorials many of which are geared to neuroscience.  While it can be
used for modest sized networks, it is not specifically designed for
this purpose and due to its history, there are limits on the size of
problems which can be solved (about 2000 differential equations is
the current limit). The benchmarks were not performed due to this
limitation in size, however, a reduced version is included.  Rather
than a pure simulator, \XPP is a tool for understanding the equations
and the results of simulating the equations.  \XPP uses a highly
optimized parser to produce a pseudocode which is interpreted and
runs very fast -- at about half the speed of directly compiled code. 
Since no compiler is required, \XPP is a stand alone program and runs
on all platforms which have an X-windows interface available (UNIX,
MAC OSX, Windows, etc.)  The program is open source and available as
source and various binary versions.  

\XPP can be run interactively (the preferred method) but can also be
run in batch mode with no GUI with the results dumped to one or more
files. Graphical output in  postscript, GIF, PBM, and animated GIF is
possible. (There are codecs available for AVI format but these are not
generally included in the compiled versions.) Numerous packages for
controlling \XPP have been written, some stand-alone such as JigCell
and others using Matlab or PERL.  Data from simulations can be saved
for other types of analysis and or plotting with other packages. The
``state'' of the program can be saved as well so that users can come
back where they let off.  

There are no limits as far as the form of the equations is concerned
since the actual equations that you desire to solve are written down
like you would write them in a paper. For example the voltage equation
for a conductance-based model would be written as:
\begin{verbatim}
dv/dt = (-gl*(v-el) - gna*m^3*h*(v-ena)-gk*n^4*(v-ek))/cm
\end{verbatim}
There is a method for writing indexed networks as well, so that one
does not have to write every equation. Special operators exist for
speeding up network functions like discrete convolutions and
implementation of the stochastic Gillespie algorithm. 
Furthermore, the
user can link the right-hand sides of differential equations to
external C libraries to solve complex equations (for example,
equation-free firing rate models, Laing JCNS 2006).  Because it is a
general purpose solver, the user can mix different types of equations
for example stochastic discrete time events with continuous
ODEs. Event driven simulations are also possible and can be performed
in such as way that output occurs only when an event happens. There
are many ways to display the results of simulations including
color-coded plots showing space-time behavior, a built-in animation
language, and one- two- and three-dimensional phase-space plots.   

 \XPP provides a variety of
numerical methods for solving differential equations, stochastic
systems, delay equations, Volterra integral equations, and
boundary-value problems (BVP).  The numerical integrators are very robust
and vary from the simple Euler method to the standard method for
solving stiff differential equations, CVODE. The latter allows the
user to specify whether the system is banded and thus can improve
calculation speed by up to two orders of magnitude.  The use of BVP
solvers is rare in neuroscience applications but they can be used to
solve, for example, the steady-state behavior of Fokker-Planck
equations for noisy neurons and to find the speed of traveling waves
in spatially distributed models. 

 Tools for analysis dynamical properties such as equilibria,
basins of attraction, Lyapunov exponents, Poincare maps, embedding,
and temporal averaging are all available via menus. Some
statistical analysis of simulations is possible such as power spectra,
mean and variance, correlation analysis and histograms are also
included in the package. There is a very robust parameter fitting
algorithm (Marquardt-Levenburg) which allows the user to find
parameters and initial conditions which best approximate specified
data.

One part of \XPP which makes it very popular is the inclusion of the
continuation package, AUTO.  This package allows the user to track
equilibria, limit cycles, and solutions to boundary-value problems as
parameters vary. The stability of the solutions is irrelevant so that
users can track the entire qualitative behavior of a differential
equation.  \XPP provides a simple to use GUI for AUTO which allows the
user to seamlessly switch back and forth between simulation and
analysis. 

\XPP is used in many different courses and workshops including the
Methods in Computational Neuroscience course at the Marine Biological
Laboratory (where it was developed 15 years ago), various European CNS
courses as well as in classroom settings.  Since equations are written
for the software as you would write them on paper, it is easy to teach
students how to use \XPP for their own problems.  There are many
features for the qualitative analysis of differential equations such
as direction fields, nullclines and color coding of solutions by some
property (such as energy or speed).  

\XPP can be considered a stable mature package. It is developed and
maintained by the author. While a list of users is not maintained, a
recent Google search revealed 38500 hits and a search on Google
Scholar showed over 250 papers citing the software. In the future, the
parser will be rewritten so that there will be no limit to the number
of equations and methods for implementing large spatially distributed
systems will also be incorporated.  Parts of the analysis code in \XPP
may possible be included in NEURON in the near future. A book has been written on the use of the program (Ermentrout, 2004) and it comes with 120 pages of documentation and dozens of examples.


\clearpage
\subsection{SPLIT}
\label{split-sec}

\subsubsection{Parallel simulators} The development of parallel
simulation in computational neuroscience has been relatively slow. 
Today there are a few publicly available parallel simulators, but
they are far from as general, flexible, and documented as commonly
used serial simulators such as Neuron (Hines and Carnevale, 1997) and
Genesis ([Bower and Beeman, 1998).  For Genesis there is PGENESIS and
the development of a parallel version of Neuron has started. In
addition there exists simulators like
NCS\url{http://brain.cse.unr.edu/ncsdocs} (see Frye, 2005), 
NEST (Morrison et~al., 2005),
and our own parallelizing simulator SPLIT (Hammarlund and Ekeberg,
1998). However, they are in many ways still on the experimental and
developmental stage.

\subsubsection{The simulator}

SPLIT is a tool specialized for efficiently simulating large-scale
multicompartmental models based on Hodgkin-Huxley formalism. It
should be regarded as experimental software for demonstrating the
possibility and usefulness of very large scale biophysically detailed
neuronal network simulations.  Recently, this tool was used for one
of the largest cortex simulations ever performed (Djurfeldt et~al.,
2005).  It supports massive parallelism on cluster computers using
MPI. The model is specified by a C++ program written by the SPLIT
user. This program is then linked with the SPLIT library to obtain
the simulator executable.  Currently, there is no supported graphical
interface, although an experimental Java/QT-based graphical interface
has been developed.  There is no built-in support for analysis of
results.  Rather, SPLIT should be regarded as a pure, generic, neural
simulation kernel with the user program adapting it into a simulator
specific to a certain model.  Although this approach is in some sense
``raw'', this means that the model specification benefits from the
full power of a general purpose programming language.

SPLIT provides conductance-based synaptic interactions with short-term
plasticity (facilitation and depression).  Long-term plasticity (such
as STDP) and integrate-and-fire formalism have not yet been
implemented, although this is planned for the future.

The user program specifies the model through the SPLIT API which is
provided by the class \texttt{split}.  The user program is serial and
parallelism is hidden from the user.  The program can be linked with
either a serial or parallel version of SPLIT.  In the parallel case,
some or all parts of the program run in a master node on the cluster
while SPLIT internally sets up parallel execution on a set of slave
nodes.  As an option, parts of the user program can execute
distributed onto each slave via a callback interface.  However, SPLIT
provides a set of tools which ensures that also such distributed code
can be written without explicit reference to parallelism.

The SPLIT API provides methods to dynamically inject spikes to an
arbitrary subset of cells during a simulation.  Results of a
simulation are logged to file.  Most state variables can be
logged. This data can be collected into one file at the master node or
written down at each slave node.  In the latter case, a separate
program might be used to collect the files at each node after the
simulation terminates.

\subsubsection{Large scale simulations} Recently,
Djurfeldt et~al., 2005 have described an effort to optimize SPLIT for
the Blue Gene/L supercomputer.  BG/L (Gara et~al., 2005) represents a
new breed of cluster computers where the number of processors,
instead of the computational performance of individual processors, is
the key to higher total performance.  By using a lower clock
frequency, the amount of heat generated decreases dramatically. 
Therefore, CPU chips can be mounted more densely and need less
cooling equipment.  A node in the BG/L cluster is a true ``system on
a chip'' with two processor cores, 512 MiB of on chip memory and
integrated network logic.  A BG/L system can contain up to 65536
processing nodes.

During this work, simulations of a neuronal network model of layers
II/III of the neocortex were performed using conductance-based
multicompartmental model neurons based on Hodgkin-Huxley formalism.
These simulations comprised up to 8 million neurons and 4 billion
synapses.  After a series of optimization steps, performance
measurements showed linear scaling behavior both on the Blue Gene/L
supercomputer (see Figure 1) and on a more conventional cluster
computer.  Optimizations included parallelization of model setup and
domain decomposition of connectivity meta data.  Computation time was
dominated by the synapses which allows for a ``free'' increase of cell
model complexity. Furthermore, communication time was hidden by
computation.

\subsubsection{Implementation aspects}

SPLIT has so far been used to model neocortical networks (Frans{\'e}n
and Lansner, 1998; Lundqvist et~al., 2006), the Lamprey spinal cord
(Kozlov et~al., 2003; 2006) and the olfactory cortex (Sandstr{\"o}m
et~al., 2006).

The library exploits data locality for better cache-based
performance.  In order to gain performance on vector architectures,
state variables are stored as sequences. It uses techniques such as
adjacency lists for compact representation of projections and AER
(Address Event Representation; Bailey and Hammerstrom, 1988) for
efficient communication of spike events.

Perhaps the most interesting concept in SPLIT is its asynchronous
design: On a parallel architecture, each slave process has its own
simulation clock which runs asynchronously with other slaves. Any pair
of slaves only need to communicate at intervals determined by the
smallest axonal delay in connections crossing from one slave to the
other.

The neurons in the model can be distributed arbitrarily over the set
of slaves. This gives great freedom in optimizing communication so
that densely connected neurons reside on the same CPU and so that
axonal delays between neurons simulated on different slaves are
maximized. The asynchronous design, where a slave process does not
need to communicate with all other slaves at each time step, gives two
benefits: 1. By communicating more seldom, the communication overhead
is reduced. 2.  By allowing slave processes to run out of phase, to a
degree determined by the mutually smallest axonal delay, the waiting
time for communication is decreased.

\subsubsection{Benchmark}

The SPLIT implementation of the HH benchmark (Benchmark~3 in
Appendix~2) consists of a C++ program which specifies what entities
are to be part of the simulation (cell populations, projections,
noise-generators, plots), makes a call which distributes these
objects onto the cluster slaves (in the parallel case), sets the
parameters of the simulation objects, initializes, and simulates. 
While writing the code, close attention needs to be payed to which
parameters are scalar and which are vectorized over the sets of cells
or axons.  Channel equations are pre-compiled into the library, and a
choice of which set of equations to use needs to be made.  Parameters
are specified using SI units.

The Benchmark~3 simulation (4000 cells, 5 s of simulated time) took
386 s on a 2~GHz Pentium M machine (Dell D810).  Outputs are written
in files on disk and can easily be displayed using \texttt{gnuplot}. 
Figure \ref{fig:raster} shows a raster of spiking activity in 100
cells during the first second of activity.  Figure \ref{fig:3cells}
shows membrane potential traces of 3 of the cells during 5 s (left)
and 100 ms (right).

\subsubsection{Future plans}

Ongoing and possible future developments of SPLIT include:

\begin{itemize}
\item a revision of the simulation kernel API
\item the addition of a Python interpreter interface
\item compatibility with channel models used in popular simulators such as
  Neuron and Genesis, enabling easy transfer of neuron models
\item gap junctions
\item graded transmitter release
\item better documentation and examples
\end{itemize}

Currently, SPLIT is developed, in part time, by two people.  There
exists some limited documentation and e-mail support.


\clearpage
\subsection{Mvaspike}

\subsubsection{Modelling with events}

It has been argued many times that action potentials as produced by
many types of neurones can be considered as {\em events}: they
consist of stereotypical impulses that appear superimposed on the
internal voltage dynamics of the neurons. As a result, many models of
neurons offer ways of defining event times associated with each
emitted action potential, often through the definition of a firing
threshold\footnote{The firing threshold here has to be taken in a
very broad sense, from a simple spike detection threshold in a
continuous model (e.g. Hodgkin-Huxley) to an active threshold that is
uses in the mathematical expression of the dynamics
(integrate-and-fire model)}. Neural simulation tools have taken
advantage of this for a long time, through the use of {\em event
driven algorithms} (see section \ref{secAlgorithms}). Indeed, when
one speaks of {\em events} in the context of simulation of neural
networks, {\em event-driven} algorithms come to mind and it it the
author impression that the use of events upstream, during the
modeling stage, is often understated.  

Mvaspike was designed as an event-based modeling and simulation
framework. It is grounded on a well established set-theoretic
modeling approach (DEVS: Discrete EVent system Specification (Zeigler
and Vahie, 1993; Zeigler et al., 2000). Target models are discrete
events systems: their dynamics can be described by changes of state
variables at arbitrary moments in time \footnote{as opposed to
discrete time systems, in which state changes occurs periodically,
and continuous systems where state changes continuously.}. One aspect
of Mvaspike is to bridge the gap between the more familiar expression
of continuous dynamics, generally in use in the neuroscience
community, and the event-centric use of models in the simulator (see
figure \ref{figRochel1}). This is conveniently easy for many simple
models that represent the models of choice in Mvaspike (mostly
integrate-and-fire or phase models, and SRMs). Watts (1994) already
noted that many neuronal properties can be explicitly and easily
represented in discrete event systems. Think of absolute refractory
{\em periods}, rising {\em time} of PSPs, axonal propagation {\em
delays}, these are notions directly related to time intervals (and
therefore, events) that are useful to describe many aspects of the
neuronal dynamics. This being obviously quite far from the well
established, more electro-physiologically correct conductance based
models, another aim of Mvaspike is therefore to take into account as
much as possible of these more complex models, through the explicit
support of discrete-time events, and, possibly, state space

discretization for the integration of continuous or hybrid dynamics.

The DEVS formalism makes also possible the modeling of large,
hierarchical or modular systems (e.g. networks of coupled populations
of neurons, or micro-circuits, cortical columns etc.), through a
well-defined coupling and composition system. This helps modeling
large and complex networks, but also favor code reusability,
prototyping, and the use of different levels of modeling. Additional
tools have been implemented in Mvaspike to take into account e.g.
synaptic or axonal propagation delays, the description of structured
or randomly connected networks in an efficient way, through the use
of generic iterators to describe the connectivity (Rochel and
Martinez, 2003).

\subsubsection{The simulator}

The core simulation engine in Mvaspike is event-driven, meaning that
is is aimed at simulating networks of neurons where event-times can
be computed efficiently. Firing times will then be calculated exactly
(in fact, to the precision of the machine). This does not mean
however that it is restricted to models that offer analytical
expressions of the firing times, as numerical approximations can be
used in many situations.  

Mvaspike consists of a core C++ library, implementing a few generic
classes to describe networks, neurons and additional input/output
systems. It has been designed to be easy to access from other
programming languages (high level or scripting languages, e.g.
Python) and extensible. Well established simulation algorithms are
provided, based on state of the art priority queue data structures.
They have been found to be sufficiently efficient on average;
however, the object-oriented approach has been designed to permit the
use of dedicated, optimized sub-simulators when possible.

On top of the core engine lies a library that includes a few common
models of neurons, including linear or quadratic integrate-and-fire
(or SRM) neurons, with Dirac synaptic interactions, or various forms
of piecewise linear and exponential PSPs. Other available ingredients
include plasticity mechanisms (STDP), refractory periods, input spike
trains generation (Poisson). Some connectivity patterns (e.g.
all-to-all, ring, etc) are also included.

There is no graphical user interface, nor pre- and post-processing
tools included, as these are elements of the modeling and simulation
work-flow that we believe to be easy  to handle using third-party
environments or high level languages, tailored to the needs and
habits of the user.

\subsubsection{Benchmarks}

The simplest model available in Mvaspike corresponds to the one
defined for Benchmark~4 (see Appendix~2). A straightforward
implementation of the corresponding network can be done using only
available objects from the library.  

The typical output of a Mvaspike simulation is a list of events,
corresponding e.g. to spikes emitted (or received) by the neurons. In
particular, the membrane potential is not available directly. In
order to obtain the voltage trace presented in figure
\ref{figRochel2}, a simple post-processing stage was necessary in
order to obtain values for the membrane potential at different
instants between the event times. To this aim, the differential
equation governing the dynamics between events is used (in a
integrated form), together with the values already available at each
event times, to find new intermediary values. Here, this is as simple
as computing the effect of the leak (exponential) and the refractory
period. As this only has to be done between events, each neuron can
be treated independently of the others. In a sense, this illustrates
how the hybrid formalism (as presented in section \ref{hybrid}) is
handled in Mvaspike: the flow of discrete events is the main point of
interest, continuous dynamics come second.

\subsubsection{Current status and further perspectives}

Mvaspike is currently usable for the modeling of medium to large
scale networks of spiking neurons. It is released under the GPL
license, maintained and supported by its main author and various
contributors.

It has been used to model networks of integrate-and-fire neurons, for
e.g. modeling the early stages of the visual system (see eg. Hugues
et al., 2002; Wohrer et al., 2006), and more theoretical research on
computing paradigms offered by spiking neurons (for instance, Rochel
and Cohen, 2005; Rochel and Vieville, 2006).
A partial parallel implementation was developed and successfully
tested on small clusters of PCs and parallel machines (16 processors
max), and should be completed to take into account all aspects of the
framework and more ambitious hardware platforms.

Work is ongoing to improve the interface of the simulator regarding
input and output data formatting, through the use of structured data
language (XML). While a proof-of-concept XML extension has already
been developed, this is not a trivial task, and further work is
needed in the context of existing initiatives (such as NeuroML).

Meanwhile, it is expected that the range of models available to the
user will be extended, for instance through the inclusion of models
of stochastic point processes, and generic implementation of state
space discretization methods.


\clearpage
\section{Discussion}

We have presented here an overview of different strategies and
algorithms for simulating spiking neural networks, as well as an
overview of most of the presently available simulation environment to
implement such simulations.  We also have conceived a set of
benchmark simulations of spiking neural networks (Appendix~2) and
provide as supplementary material (linked to ModelDB) the codes for
implementing the benchmarks in the different simulators.  We believe
this should constitute a very useful resource, especially for new
researchers in the field of computational neuroscience.

We voluntarily did not approach the difficult problem of simulation
speed and comparison of different simulators in this respect.  In
Table~1 we have tried to enumerate the features of every simulator,
in particular regarding the models that are implemented, the
possibility of distributed simulation and the simulation environment.
In summary, we can classify the simulators presented in Section~3
into four categories according to their most relevant range of
application: 1) single-compartment models: CSIM, NEST and NCS, 2)
multi-compartment models: NEURON, GENESIS, SPLIT, 3) event-driven
simulation: MVASPIKE, 4) dynamical system analysis: XPP. The
simulators NEST, NCS, PCSIM (the new parallel version of CSIM) and
SPLIT are specifically designed for distributed simulations of very
large networks. Three simulators (NEURON, GENESIS and XPP) constitute
a complete simulation environment which includes a graphical
interface and sophisticated tools for representation of model
structure and analysis of the results, as well as a complete book for
documentation.  In other simulators, analysis and graphical interface
are obtained through the use of an external front-end (such as MATLAB
or Python).

It is interesting to note that the different simulation environments
are often able to simulate the same models, but unfortunately the
codes are not compatible with each-other.  This underlines the need
for a more transparent communication channel between simulators.
Related to this, the present efforts with simulator-independent
codes (such as NeuroML, see Appendix~1) constitutes the main advance
for a future inter-operability.  We illustrated here that, using a
Python-based interface,  one of the benchmarks can be run in either
NEURON or NEST using the same code (see Fig.~\ref{interop} and
Appendix~1).

Thus, future work should focus on obtaining a full compatibility
between simulation environments and XML-based specifications. 
Importing and exporting XML should enable to convert simulation codes
between simulators, and thereby provide very efficient means of
combining existing models.  A second direction for future
investigations is to adapt simulation environments to current hardware
constraints, such as parallel computations on clusters.  Finally,
more work is also needed to clarify the differences between
simulation strategies and integration algorithms, which may
considerably differ for cases where the timing of spikes is important
(Fig.~\ref{Fig_1}).


\section*{Acknowledgments}

Research supported by the European Community (FACETS project, IST
15879), NIH (NS11613), CNRS and HFSP.  We are also grateful for the
feedback and suggestions from users that have led to improvements of
the simulators reviewed here.


\clearpage

\section*{Appendix 1: Simulator-independent model specification}

As we have seen, there are many freely-available, open-source and
well-documented tools for simulation of networks of spiking neurons. There is
considerable overlap in the classes of network that each is able to simulate,
but each strikes a different balance between efficiency, flexibility,
scalability and user-friendliness, and the different simulators encompass a
range of simulation strategies. This makes the choice of which tool to use for a
particular project a difficult one.  Moreover, we argue that using just one
simulator is an undesirable state of affairs. This follows from the general
principle that scientific results must be reproducible, and that any given
instrument may have flaws or introduce a systematic bias.  The simulators
described here are complex software packages, and may have hidden bugs or
unexamined assumptions that may only be apparent in particular circumstances.
Therefore it is desirable that any given model should be simulated using at
least two different simulators and the results cross-checked.

This is, however, more easily said than done. The configuration files, scripting
languages or graphical interfaces used for specifying model structure are very
different for the different simulators, and this, together with subtle
differences in the implementation of conceptually-identical ideas, makes the
conversion of a model from one simulation environment to another an extremely
non-trivial task; as such it is rarely undertaken.

We believe that the field of computational neuroscience has much to gain from
the ability to easily simulate a model with multiple simulators. First, it would
greatly reduce implementation-dependent bugs, and possible subtle systematic
biases due to use of an inappropriate simulation strategy. Second, it would
facilitate communication between investigators and reduce the current
segregation into simulator-specific communities; this, coupled with a
willingness to publish actual simulation code in addition to a
model description, would perhaps lead to reduced fragmentation of research
effort and an increased tendency to build on existing models rather than
redevelop them de novo. Third, it would lead to a general improvement in
simulator technology since bugs could be more easily identified, benchmarking
greatly simplified, and hence best-practice more rapidly propagated.

This goal of simulator independent model specification is some way off, but some
small steps have been taken. 
There are two possible approaches (which will probably prove to be
complementary) to developing simulator-independent model specification, which
mirror the two approaches taken to model specification by individual simulators:
declarative and programmatic. 
Declarative model specification is exemplified by the use of configuration
files, as used for example by NCS. 
Here there is a fixed library of neuron models, synapse
types, plasticity mechanisms, connectivity patterns, etc., and a
particular model is specified by choosing from this library. 
This has the advantages of simplicity in setting up a model, and of well-defined
behaviors for individual components, but has less flexibility than the
alternative, programmatic model specification. 
Most simulators reviewed here use a more or less general purpose programming
language, usually an interpreted one, which has neuroscience specific functions
and classes together with more general control and data structures. 
As noted, this gives the flexibility to generate new structures beyond those
found in the simulator's standard library, but at the expense of the very
complexity that we identified above as the major roadblock in converting models
between simulators.

\subsection*{Declarative model specification using NeuroML}

The NeuroML project\url{http://www.neuroml.org (Crook et al., 2005)}
is an open-source 
collaboration\url{http://sourceforge.net/projects/neuroml}
whose stated aims are:
\begin{enumerate}
\item To support the use of declarative specifications for models in
neuroscience using XML.
\item To foster the development of XML standards for particular areas of
computational neuroscience modeling.
\end{enumerate}
The following standards have so far been developed:
\begin{itemize}
 \item \textbf{MorphML}: specification of neuroanatomy (i.e. neuronal
morphology)
 \item \textbf{ChannelML}: specification of models of ion channels and receptors
(see Figure~\ref{fig:NeuroML} for an example)
 \item \textbf{Biophysics}: specification of compartmental cell models, building
on MorphML and ChannelML
 \item \textbf{NetworkML}: specification of cell positions and connections in a
network.
\end{itemize}

The common syntax of these specifications is XML (Extensible Markup
Language\url{http://www.w3.org/XML}). This has the
advantages of being both human- and machine-readable, and standardized by an
international organization, which in turn has led to wide uptake and developer
participation.

Other XML-based specifications that have been developed in
neuroscience and in biology more generally  include
BrainML\url{http://brainml.org} for exchanging neuroscience data,
CellML\url{http://www.cellml.org} for models of cellular and
subcellular processes and SBML\url{http://sbml.org} for representing
models of biochemical reaction networks.

Although XML has become the most widely used technology for the
electronic communication of hierarchically structured information,
the real standardization effort is orthogonal to the underlying
technology, and concerns the structuring of domain-specific
knowledge, i.e. a listing of the objects and concepts of interest in
the domain and of the relationships between them, using a
standardized terminology. To achieve this, NeuroML uses the XML
Schema Language\url{http://www.w3.org/XML/Schema} to define the
allowed elements and structure of a NeuroML document. The validity of
a NeuroML document may be checked with reference to the schema
definitions. The NeuroML Validation
service\url{http://morphml.org:8080/NeuroMLValidator} provides a
convenient way to do this.

\subsubsection*{Using NeuroML for specifying network models}

In order to use NeuroML to specify spiking neuronal network models we require
detailed
descriptions of \begin{enumerate}
 \item point spiking neurons (integrate and fire neurons and generalizations
thereof),
 \item compartmental models with Hodgkin-Huxley-like biophysics,
 \item large networks with structured internal connectivity related to  a
network topology (e.g.: full-connectivity, 1D or 2D map with local 
connectivity, synfire chains patterns, with/without randomness)
 and structured map to map connectivity (e.g., point-to-point, point-to-many, etc.).
\end{enumerate}

At the time of writing, NeuroML supports the second and third items, but not the
first. However, an extension to support specification of integrate-and-fire-type
neuron models is currently being developed, and will hopefully be incorporated
into the NeuroML standard in the near future.

Specification of Hodgkin-Huxley-type models uses the MorphML, ChannelML and
Biophysics standards of NeuroML (see Fig.~\ref{fig:NeuroML} for an example. We
focus here only on specification of networks, using the NetworkML standard.

A key point is that a set of neurons and network connectivity may be defined
either by {\em extension} (providing the list of all neurons, parameters and
connections), for example:
\footnotesize
\begin{verbatim}
          <population name="PopulationA">
            <cell_type>CellA</cell_type>
            <instances>
              <instance id="0"><location x="0" y="0" z="0"/></instance>
              <instance id="1"><location x="0" y="10" z="0"/></instance>
              <instance id="2"><location x="0" y="20" z="0"/></instance>
              . . .
            </instances>
          </population>
\end{verbatim}
\normalsize
(note that \texttt{CellA} is a cell model described earlier in the NeuroML
document), or by {\em specification}, i.e. an implicit enumeration, for example:
\footnotesize
\begin{verbatim}
          <population name="PopulationA">
            <cell_type>CellA</cell_type>
            <pop_location> 
              <random_arrangement>
                <population_size>200</population_size>
                <spherical_location>
                  <meta:center x="0" y="0" z="0" diameter="100"/>
                </spherical_location>
              </random_arrangement>
            </pop_location>
          </population>
\end{verbatim}
\normalsize
Similarly, for connectivity, one may define an explicit list of connections,
\footnotesize
\begin{verbatim}
          <projection name="NetworkConnection1">
            <source>PopulationA</source>
            <target>PopulationB</target>
            <connections>
              <connection id="0">
                <pre cell_id="0" segment_id = "0"/>
                <post cell_id="1" segment_id = "1"/>
              </connection>
              <connection id="1">
                <pre cell_id="2" segment_id = "0"/>
                <post cell_id="1" segment_id = "0"/>
              </connection>
              . . .
            </connections>
          </projection>
\end{verbatim}
\normalsize
or specify an algorithm to determine the connections:
\footnotesize
\begin{verbatim}
          <projection name="NetworkConnection1">
            <source>PopulationA</source>
            <target>PopulationB</target>
            <connectivity_pattern>
              <num_per_source>3</num_per_source>
              <max_per_target>2</max_per_target>
            </connectivity_pattern>
          </projection>
\end{verbatim}
\normalsize

\subsubsection*{Using NeuroML with a specific simulator}
One very interesting feature of XML is that any language such as NeuroML is not
fixed for ever:
\begin{itemize}
 \item it may be adapted to your own\footnote{{\bf Pragmatic generic
coding-rules.} There are always several ways to represent information as a
logical-structure. Here are a few key ideas to make such choices:
  \\ * {\em  Maximizing atomicity.} i.e. structure the data with a maximal
decomposition (e.g. atomic values must only contain ``words'' else there is
still a ``structure'' and is thus to be decomposed itself in terms of elements).
  \\ * {\em  Maximizing factorization.} i.e. prohibit data redundancy, but use
references to index a data fragment from another part of the data. This saves
place and time, but also avoid data inconsistency.
  \\ * {\em  Maximizing flat representation.} i.e. avoid complex tree
structures, when the data can be represented as uniform lists of data, i.e.
tables with simple records, such as a field-set.
  \\ * {\em  Maximizing generic description.} i.e. abstract representation,
without any reference to file format or operating-system syntax: independent of
how the data is going to be used.
  \\ * {\em  Maximizing parameterization of functionality.} i.e. specify, as much
as possible, the properties (i.e. characteristics / parameters / options) of a
software module or a function as a static set of data (instead of
``putting-it-in-the-code'').} way of presenting data and models (e.g. words may
be written in your own native language) as soon as the related logical-structure
can be translated to/from standard NeuroML
 \item add-ons are always easily defined, as soon as they are compatible with
the original NeuroML specifications.
\end{itemize}

Then using NeuroML simply means editing such data-structures using a suitable
XML editor, validating them (i.e.\ verify that the related logical-structures
are well-formed and valid with respect to the specification, conditions, etc.)
and normalizing them (i.e.\ translate it to an equivalent logical-structure but
without redundancy, while some factorization simplifies subsequent
manipulation).

Translation from this validated normalized form is efficient and
safe.  Translation can be achieved by one of two methods: Either a
simulator may accept a NeuroML document as input, and translation
from NeuroML elements to native simulator objects is performed by the
simulator, or the XSL Transformation
language\url{http://www.w3.org/TR/xslt} may be used to generate
native simulator code (e.g. \texttt{hoc} or \texttt{NMODL} in the
case of NEURON). For example, the NeuroML Validator service provides
translation of ChannelML and MorphML files to NEURON and GENESIS
formats.  

The process of editing, validating, normalizing and translating NeuroML
data-structures is summarized in
Figure~\ref{fig:facetsml-use}.

\subsubsection*{Future extensions}

The NetworkML standard is at an early stage of development. Desirable future
extensions include:
\begin{itemize}
\item specification of point spiking models, such as the integrate-and-fire
model.
\item more flexible specification of numerical parameters. Numerical parameter
values are not simple ``numbers''
 but satisfy certain standard conditions (parameter values are physical
quantities with a unit, may take a default value, have values bounded within a
certain range with minimal/maximal values and are defined
   up to a certain precision) or specific conditions defined by a boolean
expression, and may have their default value not simply defined by a constant
but from an algebraic expression. In the current NeuroML standards all numerical
parameters are simple numbers, and all units must be consistent with either a
``physiological units'' system or the SI system (they may not be mixed in a
single NeuroML document).
\item specifying parameter values as being drawn from a defined random
distribution.
\end{itemize}

\subsection*{Programmatic model specification using Python}

For network simulations, we may well require more flexibility than can easily be
obtained using a declarative model specification, but we still wish to obtain
simple conversion between simulators, i.e. to be able to write the simulation
code for a model only once, then run the same code on multiple simulators. This
requires first the definition of an API (Application Programming Interface) or
meta-language, a set of functions/classes which provides a superset of the
capabilities of the simulators we wish to run on\footnote{Note that since we
choose a superset, the system must emit a warning/error if the underlying
simulator engine does not support a particular feature.}. Having defined an API,
there are two possible next stages: (i) each simulator implements a parser which
can interpret the meta-language; (ii) a separate program either translates the
meta-language into simulator-specific code or controls the simulator directly,
giving simulator-specific function calls.

In our opinion, the second of these possibilities is the better one, since 
\begin{enumerate}
\item it avoids replication of effort in writing parsers,
\item we can then use a general purpose, state-of-the-art interpreted
programming language, such as Python or Ruby, rather than a simulator-specific
language, and thus leverage the effort of outside developers in areas that are
not neuroscience specific, such as data analysis and visualization\footnote{For
Python, examples include efficient data storage and transfer (HDF5, ROOT), data
analysis (SciPy), parallelization (MPI), GUI toolkits (GTK, QT).}
\end{enumerate}

The PyNN project\footnote{pronounced `pine'} has begun to develop both the API
and the binding to individual simulation engines, for both purposes using the
Python programming language. The API has two parts, a low-level, procedural API
(functions \texttt{create()}, \texttt{connect()}, \texttt{set()},
\texttt{record()}), and a high-level, object-oriented API (classes
\texttt{Population} and \texttt{Projection}, which have methods like
\texttt{set()}, \texttt{record()}, \texttt{setWeights()}, etc.). The low-level
API is good for small networks, and perhaps gives more flexibility. The
high-level API is good for hiding the details and the book-keeping, and is
intended to have a one-to-one mapping with NeuroML, i.e. a \texttt{population}
element in NeuroML will correspond to a \texttt{Population} object in PyNN.

The other thing that is required to write a model once and run it on multiple
simulators is standard cell models. PyNN translates standard cell-model names
and parameter names into simulator-specific names, e.g. standard model
\texttt{IF\_curr\_alpha} is \texttt{iaf\_neuron} in NEST and \texttt{StandardIF}
in NEURON, while \texttt{SpikeSourcePoisson} is a \texttt{poisson\_generator} in
NEST and a \texttt{NetStim} in NEURON.

An example of the use of the API to specify a simple network is given in
Figure~\ref{fig:PyNN_example}.

Python bindings currently exist to control NEST (PyNEST\footnote{a Python
interface to NEST}) and Mvaspike, and Python can be used as an alternative
interpreter for NEURON (nrnpython), although the level of integration (how easy
it is to access the native functionality) is variable. Currently PyNN supports
PyNEST and NEURON (via nrnpython), and there are plans to add support for other
simulators with Python bindings, initially Mvaspike and CSIM, and to add support
for the distributed simulation capabilities of NEURON and NEST.

\subsubsection*{Example simulations}

Benchmarks 1 and 2 (see Appendix~2) have been coded in PyNN and run
using both NEURON and NEST (Fig.~\ref{interop}).  The results for
the two simulators are not identical, since we used different
random number sequences when determining connectivity, but the
distributions of inter-spike intervals (ISIs) and of the
coefficient of variation of ISI are almost indistinguishable. All
the cell and synapse types used in the benchmarks are standard
models in PyNN. Where these models do not come as standard in
NEURON or NEST, the model code is distributed with PyNN (in the
case of NEURON) or with PyNEST (in the case of NEST). We do not
report simulation times, as PyNN has not been optimized for either
simulator.


\clearpage
\section*{Appendix 2: Benchmark simulations}

In this appendix, we present a series of ``benchmark'' network
simulations using both integrate-and-fire (IF) or Hodgkin-Huxley (HH)
type neurons.  They were chosen such that at least one of the
benchmark can be implemented in the different simulators (the code
corresponding to these implementations will be provided in the
ModelDB database\url{http://senselab.med.yale.edu/senselab/ModelDB}.

The models chosen were networks of excitatory and inhibitory neurons
inspired from a recent study (Vogels and Abbott, 2005).  This paper
considered two types of networks of leaky IF neurons, one with
current-based synaptic interactions (CUBA model), and another one
with conductance-based synaptic interactions (CUBA model; see below).
We also introduce here a HH-based version of the COBA model, as well
as a fourth model consisting of IF neurons interacting through
voltage deflections (``voltage-jump'' synapses).

\subsection*{Network structure}

Each model consisted of 4,000 IF neurons, which were separated into
two populations of excitatory and inhibitory neurons, forming 80\%
and 20\% of the neurons, respectively.  All neurons were connected
randomly using a connection probability of 2\%.

\subsection*{Passive properties}

The membrane equation of all models was given by:

\begin{equation}
 C_m \ {dV \over dt} \ = \ -g_L (V-E_L) \ + \ S(t) \ + \ G(t) ~ ,
\end{equation}
where $C_m$ = 1~$\mu$F/cm$^2$ is the specific capacitance, $V$ is the
membrane potential, $g_L$ = 5$\times$10$^{-5}$~S/cm$^2$ is the leak
conductance density and $E_L$ = -60~mV is the leak reversal
potential.  Together with a cell area of 20,000~$\mu$m$^2$, these
parameters give a resting membrane time constant of 20~ms and an
input resistance at rest of 100~M$\Omega$.  The function $S(t)$
represents the spiking mechanism and $G(t)$ stands for synaptic 
interactions (see below).

\subsection*{Spiking mechanisms}

\subsubsection*{IF neurons}

In addition to passive membrane properties, IF neurons had a firing
threshold of -50~mV.  Once the Vm reaches threshold, a spike is
emitted and the membrane potential is reset to -60~mV and remains at
that value for a refractory period of 5~ms.

\subsubsection*{HH neurons}

HH neurons were modified from Traub and Miles (1991) and were
described by the following equations:

\begin{eqnarray}
 C_m \ {dV \over dt} & = & -g_L (V-E_L) \ 
    - \bar{g}_{Na} \ m^3h \ (V-E_{Na}) 
    - \bar{g}_{Kd} \ n^4 \ (V-E_K) \ + \ G(t) \\ 
 {dm \over dt} & = & \alpha_m(V) \ (1-m) - \beta_m(V) \ m \nonumber \\
 {dh \over dt} & = & \alpha_h(V) \ (1-h) - \beta_h(V) \ h \nonumber \\
 {dn \over dt} & = & \alpha_n(V) \ (1-n) - \beta_n(V) \ n \nonumber
 ~ ,
\end{eqnarray}
where $\bar{g}_{Na} = 100~mS/cm^2$ and $\bar{g}_{Kd} = 30~mS/cm^2$
are the maximal conductances of the sodium current and delayed
rectifier with reversal potentials of $E_{Na} = 50~mV$ and $E_K =
-90~mV$.  $m$, $h$, and $n$ are the activation variables which time
evolution depends on the voltage-dependent rate constants $\alpha_m$,
$\beta_m$, $\alpha_h$, $\beta_h$, $\alpha_n$ and $\beta_n$.  The
voltage-dependent expressions of the rate constants were modified
from the model described by Traub and Miles (1991):

\begin{eqnarray}
  \alpha_m & = & 0.32 * (13-V+V_T) / [ \exp((13-V+V_T)/4) - 1]	\nonumber \\
  \beta_m  & = & 0.28 * (V-V_T-40) / [ \exp((V-V_T-40)/5) - 1]	\nonumber \\
  \alpha_h & = & 0.128 * \exp((17-V+V_T)/18) 			\nonumber \\
  \beta_h  & = & 4 / [ 1 + \exp((40-V+V_T)/5) ] 		\nonumber \\
  \alpha_n & = & 0.032 * (15-V+V_T) / [ \exp((15-V+V_T)/5) - 1] \nonumber \\
  \beta_n  & = & 0.5 * \exp((10-V+V_T)/40) ~ ,			\nonumber
\end{eqnarray}
where $V_T$ = -63~mV adjusts the threshold (which was around -50~mV
for the above parameters).

\subsection*{Synaptic interactions}

\subsubsection*{Conductance-based synapses}

For conductance-based synaptic interactions, the membrane equation of
neuron $i$ was given by:

\begin{equation}
 C_m \ {dV_i \over dt} \ = \ -g_L (V_i-E_L) \ + \ S(t) 
   \ - \ \sum_{j} g_{ji}(t) (V_i-E_j) ~ ,	\label{gcond}
\end{equation}
where $V_i$ is the membrane potential of neuron $i$, $g_{ji}(t)$ is
the synaptic conductance of the synapse from neuron $j$ to neuron $i$,
and $E_j$ is the reversal potential of that synapse.  $E_j$ was of 
0~mV for excitatory synapses, or -80~mV for inhibitory synapses.  

Synaptic interactions were implemented as follows: when a spike occurred 
in neuron $j$, the synaptic conductance $g_{ji}$ was instantaneously 
incremented by a quantum value (6~nS and 67~nS for excitatory and
inhibitory synapses, respectively) and decayed exponentially with a
time constant of 5~ms and 10~ms for excitation and inhibition,
respectively.

\subsubsection*{Current-based synapses}

For implementing current-based synaptic interactions, the following 
equation was used:

\begin{equation}
 C_m \ {dV_i \over dt} \ = \ -g_L (V_i-E_L) \ + \ S(t) 
   \ - \ \sum_{j} g_{ji}(t) (\bar{V}-E_j) ~ ,	\label{gcurr}
\end{equation}
where $\bar{V}$ = -60~mV is the mean membrane potential.  The 
conductance quanta were of 0.27~nS and 4.5~nS for excitatory and
inhibitory synapses, respectively.  The other parameters are the 
same as for conductance-based interactions.

\subsubsection*{Voltage-jump synapses}

For implementing voltage-jump type of synaptic interactions, the 
membrane potential was abruptly increased by a value of 0.25~mV 
for each excitatory event, and it was decreased by 2.25~mV for
each inhibitory event.  

\subsection*{Benchmarks}

Based on the above models, the following four benchmarks were
implemented.

\begin{quote}

\item[{\it Benchmark 1:}] {\it Conductance-based IF network}.  This
benchmark consists of a network of IF neurons connected with
conductance-based synapses, according to the parameters above.  It
is equivalent to the COBA model described in Vogels and Abbott
(2005).

\item[{\it Benchmark 2:}] {\it Current-based IF network}.  This
second benchmark simulates a network of IF neurons connected with
current-based synapses, which is equivalent to the CUBA model
described in Vogels and Abbott (2005).  It has the same parameters
as above, except that every cell needs to be depolarized by about
10~mV, which was implemented by setting $E_L$ = -49~mV (see Vogels
and Abbott, 2005).

\item[{\it Benchmark 3:}] {\it Conductance-based HH network}.  This
benchmark is equivalent to Benchmark~1, except that the HH model
was used.

\item[{\it Benchmark 4:}] {\it IF network with voltage-jump
synapses}.  This fourth benchmark used voltage-jump synapses, and
has a membrane equation which is analytically solvable, and can be
implemented using
event-driven
simulation strategies.

\end{quote}

For all four benchmarks, the models simulate a self-sustained
irregular state of activity, which is easy to identify: all cells
fire irregularly and are characterized by important subthreshold
voltage fluctuations.  The neurons must be randomly stimulated
during the first 50~ms in order to set the network in the active
state.

\subsection*{Supplementary material}

The supplementary material to the paper contains the codes for
implementing those benchmarks in the different simulators reviewed
here (see Section~3 for details on specific implementations).  We
provide the codes for those benchmarks, implemented in each
simulator, and this code is made available in the ModelDB 
database\url{https://senselab.med.yale.edu/senselab/modeldb/ShowModel.asp?model=83319 
(if necessary, use "reviewme" as password).}.

In addition, we provide a clock-driven implementation of Benchmarks 1
and 2 with Scilab, a free vector-based scientific software.   In this
case, Benchmark~1 is integrated with Euler method, second order
Runge-Kutta and Euler with spike timing interpolation (Hansel et al,
1998), while Benchmark~2 is integrated exactly (with spike timings
aligned to the time grid).  The event-driven implementation
(Benchmark~4) is also possible with Scilab but very inefficient
because the programming language is interpreted, and since the
algorithms are asynchronous, the operations cannot be vectorized. 
Finally, we also provide a C++ implementation of Benchmark~2 and of a
modified version of the COBA model (Benchmark~1, with identical
synaptic time constants for excitation and inhibition).



\clearpage

\begin{table}

\footnotesize

\begin{tabular}{lllllllllll}
Question        &NEURON &GENESIS & NEST & NCS & CSIM &  XPP &   SPLIT & Mvaspike \\
\ \\
\hline 
HH		&B.I.	&B.I.	&YES	&B.I.	&B.I.	&YES	&B.I.	&POSS	\\
leaky IF	&B.I.	&POSS	&YES	&B.I.	&B.I.	&YES	&POSS**	&B.I.	\\
Izhikevich IF	&YES	&B.I.	&YES	&NO	&B.I.	&YES	&POSS**	&POSS**	\\
Cable eqs	&B.I.	&B.I.	&NO	&NO	&NO	&YES	&B.I.	&NO	\\
ST plasticity	&YES	&B.I.	&YES	&B.I.	&B.I.	&YES	&B.I.	&YES	\\
LT Plasticity	&YES	&YES	&YES	&B.I.	&B.I.	&YES	&NO**	&YES	\\
Event-based	&B.I.	&NO	&YES	&NO	&NO	&YES	&NO	&YES	\\
  exact		&B.I.	&-	&YES	&-	&-	&NO	&-	&YES	\\
Clock-based	&B.I.	&B.I.	&YES	&B.I.	&YES	&YES	&YES	&POSS**	\\
  interpolated	&B.I.	&NO	&YES	&NO	&NO	&YES	&B.I.	&POSS	\\
G synapses	&B.I.	&B.I.	&YES	&B.I.	&B.I.	&YES	&B.I.	&POSS**	\\
parallel	&B.I.	&YES	&B.I.	&B.I.	&NO**	&NO	&B.I.	&NO**	\\
graphics	&B.I.	&B.I.	&NO(*)	&NO(*)	&NO(*)	&YES	&NO	&NO	\\
simple analysis	&B.I.	&B.I.	&YES	&NO(*)	&NO(*)	&YES	&NO	&NO	\\
complx analysis	&B.I.	&YES	&NO(*)	&NO(*)	&NO(*)	&YES	&NO	&NO	\\
development	&YES	&YES	&YES	&YES	&YES	&YES	&YES	&YES	\\
  how many p.	&3	&2-3	&4	&2-3	&2	&1	&2	&1	\\
support		&YES	&YES	&YES	&YES	&YES	&YES	&YES	&YES	\\
  type		&e,p,c	&e	&e	&e	&e	&e	&e	&e	\\
user forum	&YES	&YES	&YES	&NO	&NO	&YES	&YES	&NO	\\
publ list	&YES	&YES	&YES	&YES	&YES	&NO	&NO	&NO	\\
codes		&YES	&YES	&YES	&YES	&YES	&YES	&NO	&NO	\\
online manual	&YES	&YES	&YES	&YES	&YES	&YES	&YES	&YES	\\
book		&YES	&YES	&NO	&NO	&NO	&YES	&NO	&NO	\\
XML import	&NO**	&POSS	&NO**	&NO**	&NO	&YES	&NO	&NO**	\\
XML export	&B.I.	&NO**	&NO**	&NO**	&NO	&NO	&NO	&NO**	\\
web site	&YES	&YES	&YES	&YES	&YES	&YES	&YES	&YES	\\
LINUX		&YES	&YES	&YES	&YES	&YES	&YES	&YES	&YES	\\
Windows		&YES	&YES	&YES	&YES	&YES	&YES	&NO	&NO	\\
Mac-Os		&YES	&YES	&YES	&NO	&NO	&YES	&NO	&NO	\\
Interface	&B.I.	&B.I.	&POSS	&B.I	&YES	&POSS	&POSS	&POSS	\\
Save option	&B.I.	&YES	&NO**	&B.I.	&NO	&NO	&NO	&NO	\\
\ \\
\hline 
\end{tabular}

\caption{\label{table} Comparison of features of the different
simulators.}

\end{table}

\clearpage
\subsection*{Table caption}

\centerline{Table 1: Comparison of features of the different
simulators}

\

Different questions were asked (see below), and for each question, the answer
is either: \\
B.I. = Built-in feature, incorporated in the simulator without need
       to load additional mechanisms; \\
YES  = feature very easy to simulate or implement (ie., a few
       minutes of programming); \\
POSS = feature possible to implement, but requires a bit of user
       programming;  \\
NO   = feature not implemented, would require modifying the code; \\
**   = feature planned to be implemented in a future version of the 
simulator;  \\
(*) graphical interface and analysis possible via front-ends like 
Python or MATLAB.  \\

The list of questions were:  \\
HH:		can it simulate HH models?; \\
leaky IF:	can it simulate leaky IF models?; \\
Izhikevich IF:	can it simulate multivariable IF models, for example Izhikevich type?; \\
Cable eqs:	can it simulate compartmental models with dendrites?; \\
ST plasticity:	can it simulate short-term synaptic plasticity? (facilitation, depression); \\
LT Plasticity:  can it simulate long-term synaptic plasticity? (LTP, LTD, STDP); \\
Event-based:	can it simulate event-based strategies?; \\
  exact:	in this case, is the integration scheme exact?; \\
Clock-based:	can it simulate clock-based strategies? (e.g., Runge-Kutta); \\
  interpolated:	in this case, does it use interpolation for spike times?; \\
G synapses:	can it simulate conductance-based synaptic interactions?; \\
parallel:	does it support parallel processing?; \\
graphics:	does it have a graphical interface?; \\
simple analysis: is it possible to use the interface for simple analysis?
		(spike count, correlations, etc); \\
complx analysis: can more complex analysis be done? (parameter fitting, fft,
		matrix operations, ...); \\
development:	is it currently developed?; \\
  how many p.:	if yes, how many developers are working on it?; \\
support:	is it supported? (help for users); \\
  type:		what type of support (email, phone, consultation?); \\
user forum:	is there a forum of users or mailing list?; \\
publ list:	is there a list of publications of articles that used it?; \\
codes:		are there codes available on the web of published models?; \\
online manual:	are there tutorials and reference material available on the web?; \\
book:		are there published books on the simulator? \\
XML import:	can it import model specifications in XML? \\
XML export:   	can it export model specifications in XML? \\
web site:       is there a web site of the simulator where all can be found?
		(including help and source codes) \\
LINUX:		does it run on LINUX? \\
Windows:	does it run on Windows? (98, 2K, XP) \\
Mac-Os:		does it run on Mac-OS X? \\
Interface:	Is there a possibility to interface the simulator to outside
		signals ?  (such as a camera, or a real neuron) \\
Save option:    Does it have a "save option", (different than ctrl-z), allowing 
		the user to interrupt a simulation, and continue it later on ?
                (this feature is important on a cluster when simulations must
		be interrupted)


\clearpage 

\begin{figure}
\begin{center}
\begin{boxedminipage}{4in}
\begin{tabular}{ll}
\verb|t=0|&\\
\verb|while t<duration|&\\
\verb|  for every neuron|&\emph{State updates}\\
\verb|    process incoming spikes|&\\
\verb|    advance neuron dynamics by dt|&\\
\verb|  end|&\\
\verb||&\\
\verb|  for every neuron|&\\
\verb|    if vm>threshold|&\\
\verb|      reset neuron|&\emph{Propagation}\\
\verb|      for every connection|&\emph{of spikes}\\
\verb|        send spike|&\\
\verb|      end|&\\
\verb|    end|&\\
\verb|  end|&\\
\verb||&\\
\verb|  t=t+dt|&\\
\verb|end|&
\end{tabular}
\end{boxedminipage}
\end{center}
\caption{A basic clock-driven algorithm}
\label{figclockdriven}
\end{figure}

\begin{figure}
\begin{center}
\begin{boxedminipage}{5in}
\begin{tabular}{ll}
\verb|while queue not empty and t<duration|&\\
\verb|  extract event with lowest timing|&\emph{Process event}\\
\verb|  (= timing t, target i, weight w)|&\\
\verb|  compute state of neuron i at time t|&\\
\verb|  update state of neuron i (+w)|&\\
\verb|  if vm>threshold|&\\
\verb|    for each connection i->j|&\emph{Propagate spike}\\
\verb|      insert event in the queue|&\\
\verb|    end|&\\
\verb|    reset neuron i|&\\
\verb|  end|&\\
\verb|end|&\\
\end{tabular}
\end{boxedminipage}
\end{center}
\caption{A basic event-driven algorithm with instantaneous synaptic interactions}
\label{figeventdriven1}
\end{figure}

\begin{figure}
\begin{center}
\begin{boxedminipage}{5in}
\begin{tabular}{ll}
\verb|for every neuron i|&\\
\verb|  compute timing of next spike|&\emph{Initialization}\\
\verb|  insert event in priority queue|&\\
\verb|end|&\\
\verb||&\\
\verb|while queue not empty and t<duration|&\\
\verb|  extract event with lowest timing|&\emph{Process spike}\\
\verb|  (event = timing t, neuron i)|&\\
\verb|  compute state of neuron i at time t|&\\
\verb|  reset membrane potential|&\\
\verb|  compute timing of next spike|&\\
\verb|  insert event in queue|&\\
\verb||&\\
\verb|  for every connection i->j|&\emph{Communicate spike}\\
\verb|    compute state of neuron j at time t|&\\
\verb|    change state with weight w(i,j)|&\\
\verb|    compute timing of next spike|&\\
\verb|    insert/change/suppress event in queue|&\\
\verb|  end|&\\
\verb|end|&\\
\end{tabular}
\end{boxedminipage}
\end{center}
\caption{A basic event-driven algorithm with non-instantaneous synaptic interactions}
\label{figeventdriven2}
\end{figure}

\clearpage 

\begin{figure}
\begin{center}
\psfig{figure=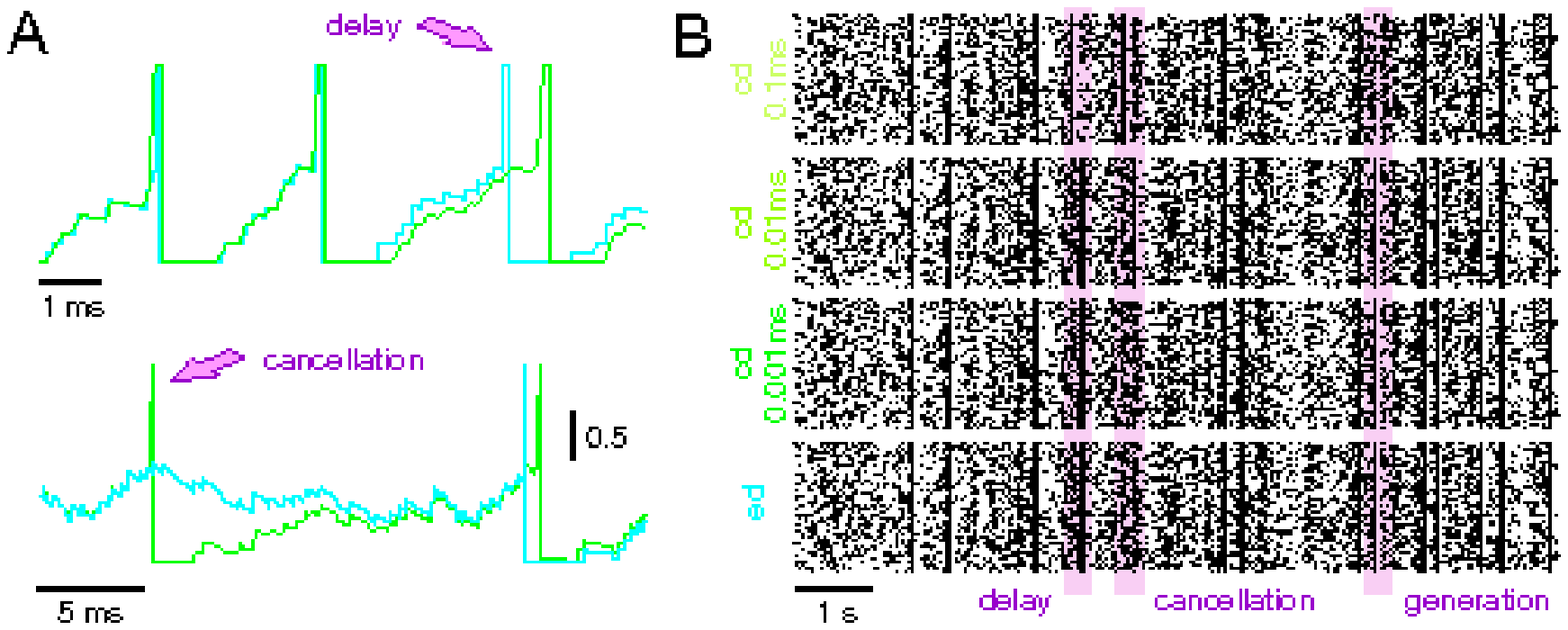,width=16cm}
\end{center}

\caption{\label{Fig_1} Modelling strategies and dynamics in neuronal
systems without STDP.} A: Small differences in spike times can
accumulate and lead to severe delays or even cancellation (see
arrows) of spikes, depending on the simulation strategy utilized or
the temporal resolution within clock-driven strategies used. B:
Rasterplots of spike events in a small neuronal network of LIF
neurons simulated with event-driven and clock-driven approaches with
different temporal resolutions. Observed differences in neural
network dynamics include delays, cancellation or generation of
synchronous network events (Figure modified from Rudolph \& Destexhe,
2007).

\end{figure}

\begin{figure}
\begin{center}
\psfig{figure=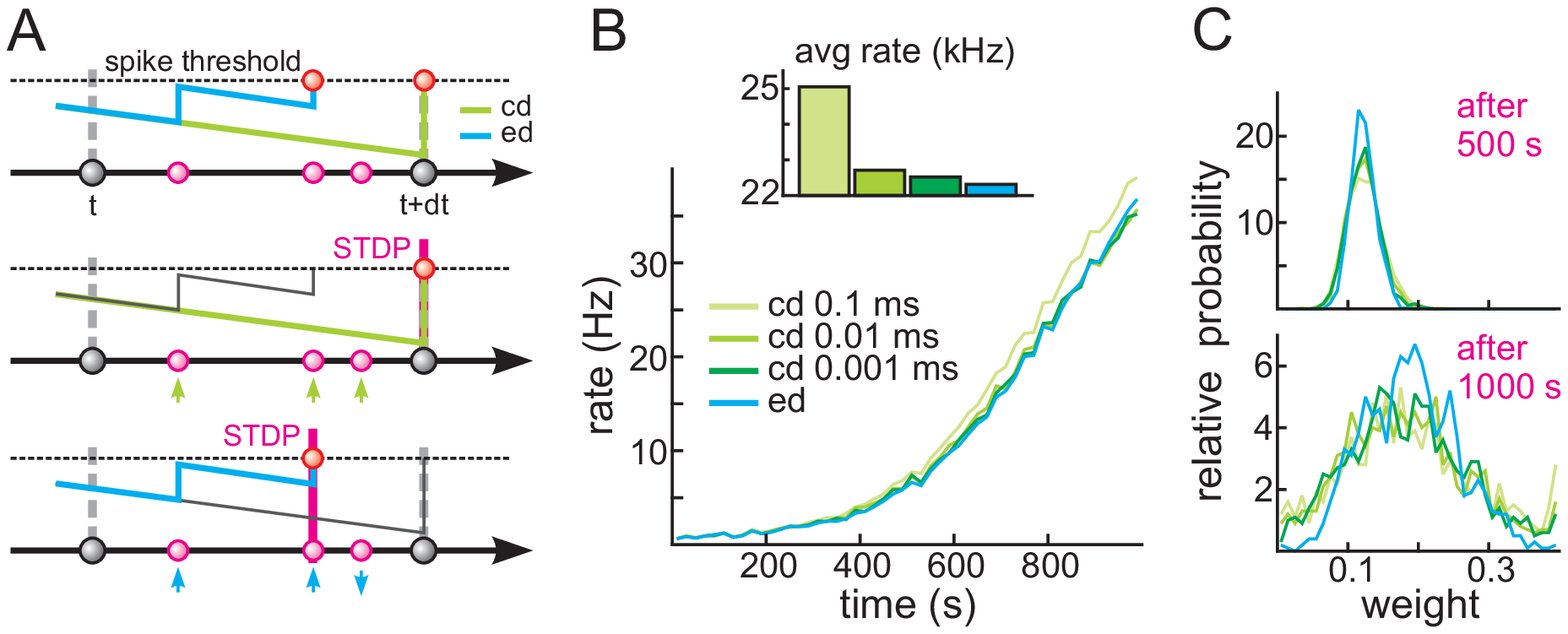,width=16cm}
\end{center}

\caption{\label{Fig_2} Dynamics in neuronal systems with STDP.} A:
Impact of the simulation strategy (clock-driven: {\it cd};
event-driven: {\it ed}) on the facilitation and depression of
synapses. B: Time course and average rate (inset) in a LIF model with
multiple synaptic input channels for different simulation strategies
and temporal resolution. C: Synaptic weight distribution after 500~s
and 1,000~s (Figure modified from Rudolph \& Destexhe, 2007).

\end{figure}

\clearpage 

\begin{figure}[ht]
\begin{center}
\psfig{file=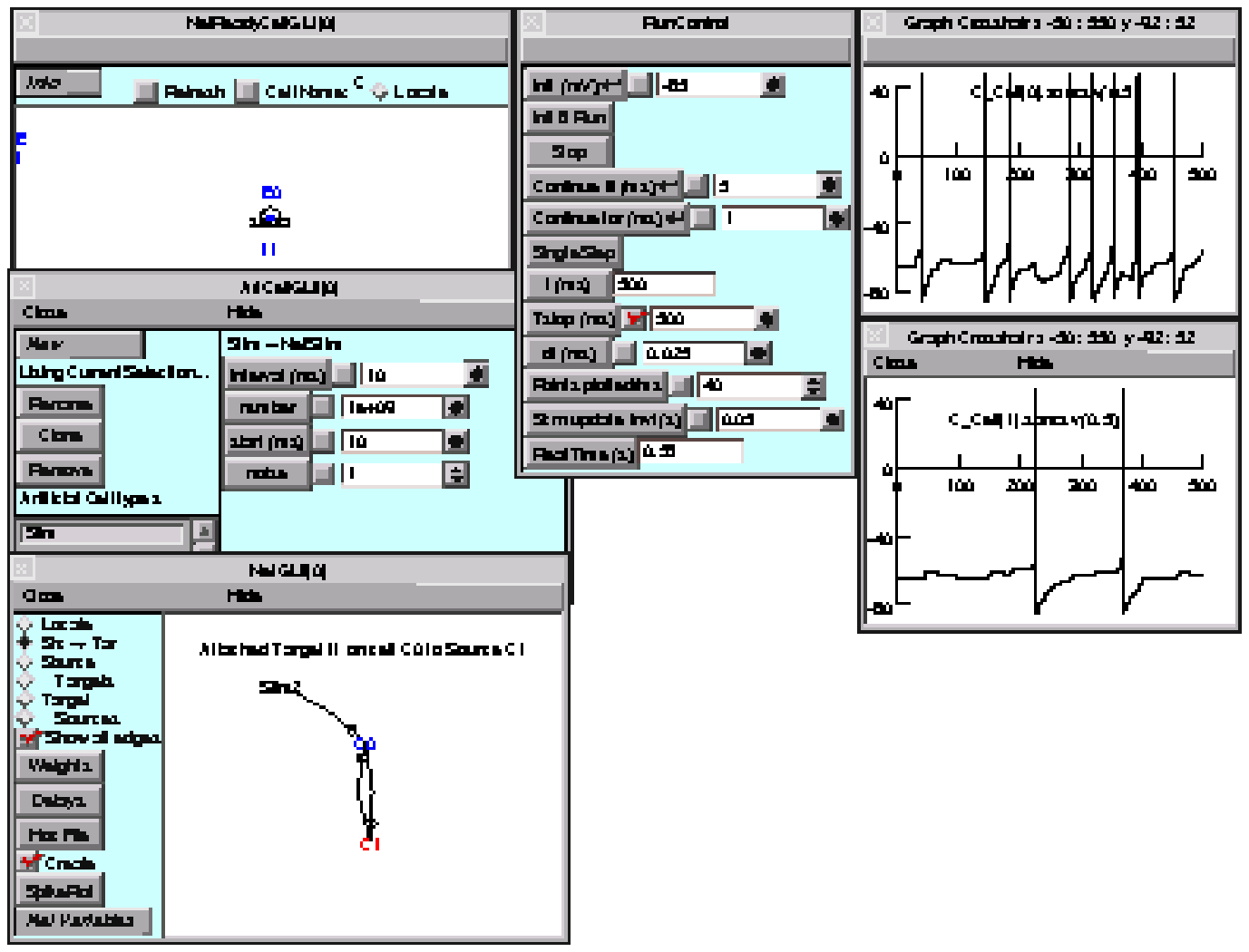,width=6in}
\end{center}

\caption{NEURON graphical user interface.} In developing large scale
networks, it is helpful to start by debugging small prototype nets. 
NEURON's GUI, especially its Network Builder (shown here), can
simplify this task.  Also, at the click of a button the Network
Builder generates hoc code that can be reused as the building blocks
for large scale nets (see chapter 11, {\it Modeling networks} in
Carnevale and Hines 2006).

\label{FigGUI}
\end{figure}

\begin{figure}[ht]
\begin{center}
\psfig{file=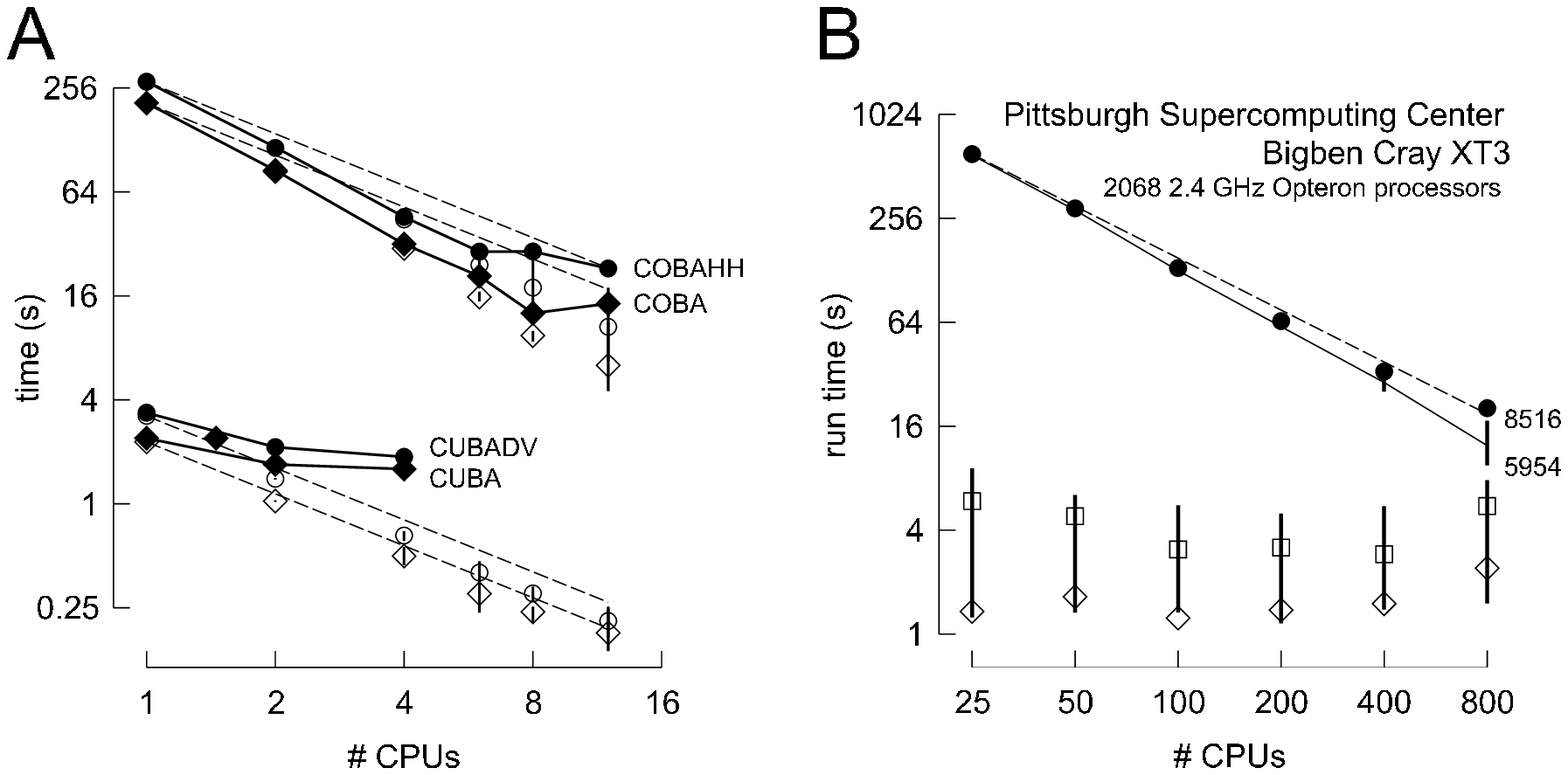,width=6in}
\end{center}

\caption{Parallel simulations using NEURON.} A. Four benchmark
network models were simulated on 1, 2, 4, 6, 8, and 12 CPUs of a
Beowulf cluster (6 nodes, dual CPU, 64-bit 3.2 GHz Intel Xeon with
1024 KB cache).  Dashed lines indicate "ideal speedup" (run time
inversely proportional to number of CPUs). Solid symbols are run
time, open symbols are average computation time per CPU, and
vertical bars indicate variation of computation time. The CUBA and
CUBADV models execute so quickly that little is gained by
parallelizing them. The CUBA model is faster than the more
efficient CUBADV because the latter generates twice as many spikes
(spike counts are COBAHH 92,219, COBA 62,349, CUBADV 39,280, CUBA
15,371).  B. The Pittsburgh Supercomputing Center's Cray XT3 (2.4
GHz Opteron processors) was used to simulate a NEURON
implementation of the thalamocortical network model of Traub et al.
(2005). This model has 3,560 cells in 14 types, 3,500 gap
junctions, 5,596,810 equations, and 1,122,520 connections and
synapses, and 100 ms of model time it generates 73,465 spikes and
19,844,187 delivered spikes. The dashed line indicates "ideal
speedup" and solid circles are the actual run times. The solid
black line is the average computation time, and the intersecting
vertical lines mark the range of computation times for each CPU. 
Neither the number of cell classes nor the number of cells in each
class were multiples of the number of processors, so load balance
was not perfect. When 800 CPUs were used, the number of equations
per CPU ranged from 5954 to 8516. Open diamonds are average spike
exchange times. Open squares mark average voltage exchange times
for the gap junctions, which must be done at every time step; these
lie on vertical bars that indicate the range of voltage exchange
times. This range is large primarily because of synchronization
time due to computation time variation across CPUs. The minimum
value is the actual exchange time.

\label{FigN}
\end{figure}

\begin{figure}[ht]
\begin{center}
\psfig{file=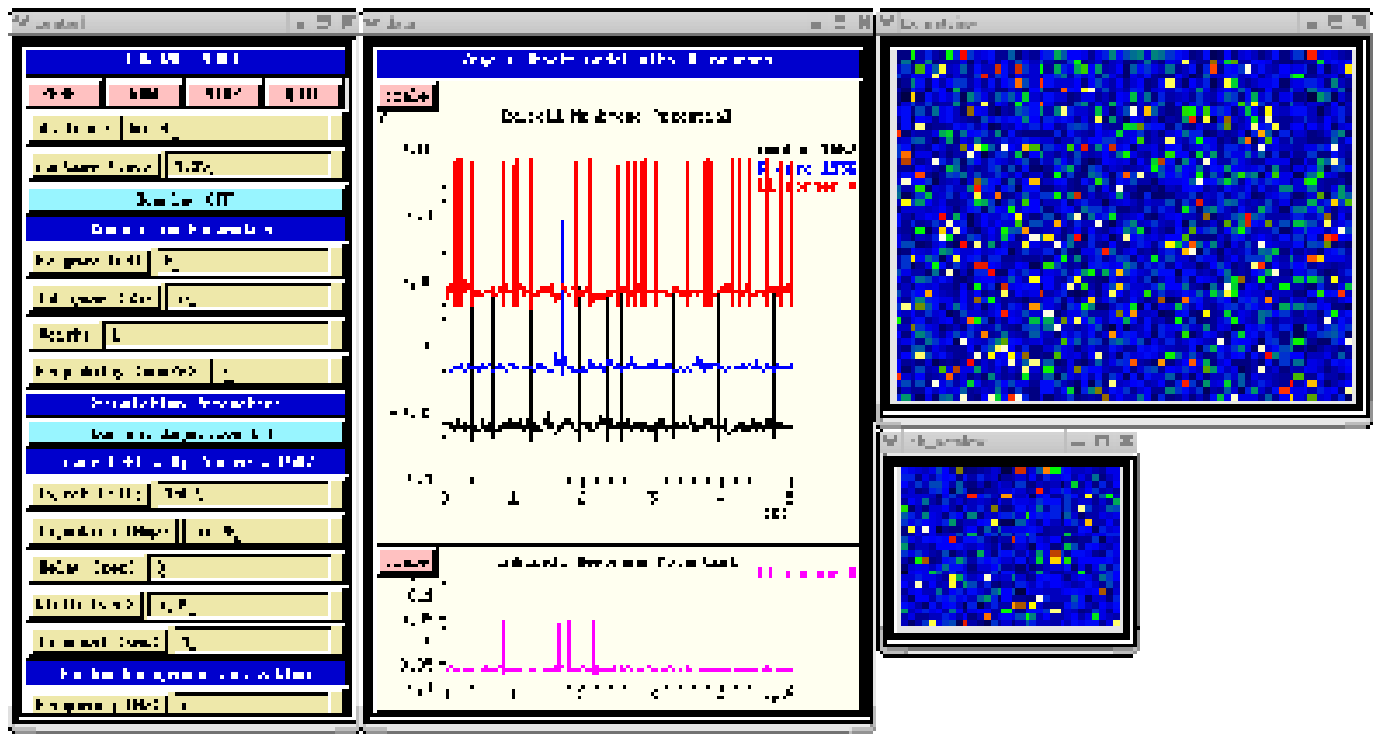,width=6.25in}
\end{center}

\caption{The GUI for the GENESIS implementation of the HH benchmark,
using the dual-exponential form of synaptic conductance.}

\label{FigG1}
\end{figure}

\begin{figure}[ht]
\begin{center}
\psfig{file=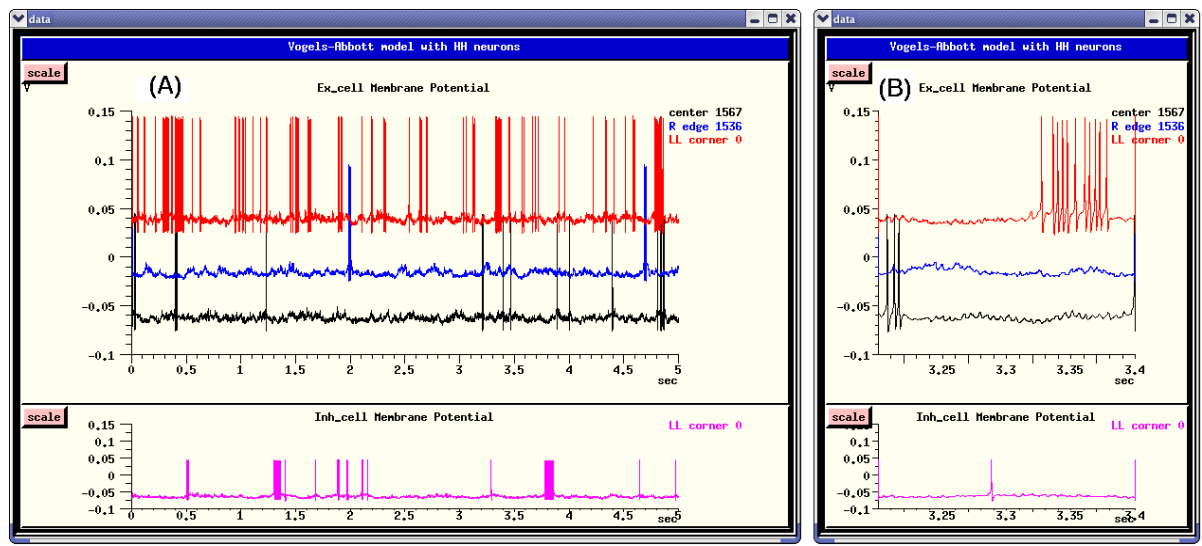,width=6.25in}
\end{center}

\caption{Membrane potentials for four selected neurons of the
Instantaneous Conductance VA HH Model in GENESIS.}  A. The entire
5 seconds of the simulation.  B. Detail of the interval 3.2--3.4 sec.

\label{FigG2}
\end{figure}

\begin{figure}
\begin{center}
\psfig{figure=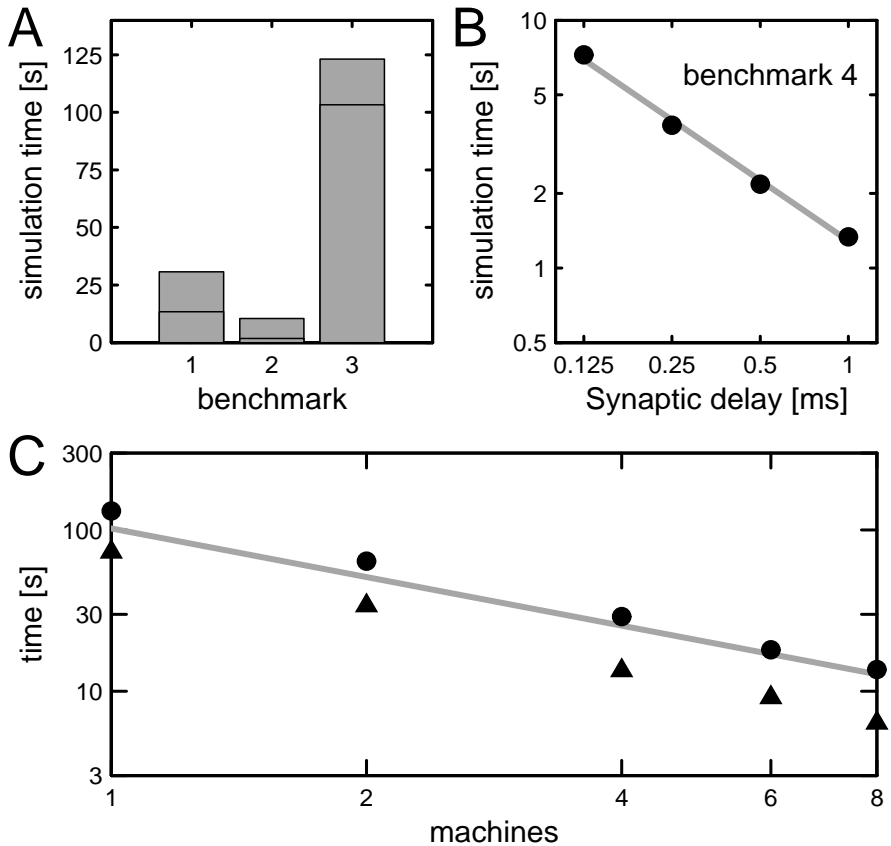,width=12cm}
\end{center}

\caption{\label{cap:nest_benchmarks}Performance of NEST on Benchmarks
$1$-$4$ and an additional benchmark (5) with spike timing dependent
plasticity (STDP).} A. Simulation time for one biological second of
Benchmarks $1$-$3$ distributed over two processors, spiking
supressed, with a synaptic delay of $0.1$ ms. The horizontal lines
indicate the simulation times for the benchmarks with the synaptic
delay increased to $1.5$ ms.  B. Simulation time for one biological
second of Benchmark $4$ as a function of the minimum synaptic delay
in double logarithmic representation. The gray line indicates a
linear fit to the data (slope$-0.8$).  C. Simulation time for one
biological second of Benchmark 5, a network of $11250$ neurons and
connection probability of $0.1$ (total number of synapses:
$12.7\times10^{6}$) as a function of the number of processors in
double logarithmic representation.  All synapses static, triangles;
excitatory-excitatory synapses implementing multiplicative STDP with
an all-to-all spike pairing scheme, circles.  The gray line indicates
a linear speed-up.

\end{figure}

\begin{figure}
\begin{center}
\psfig{figure=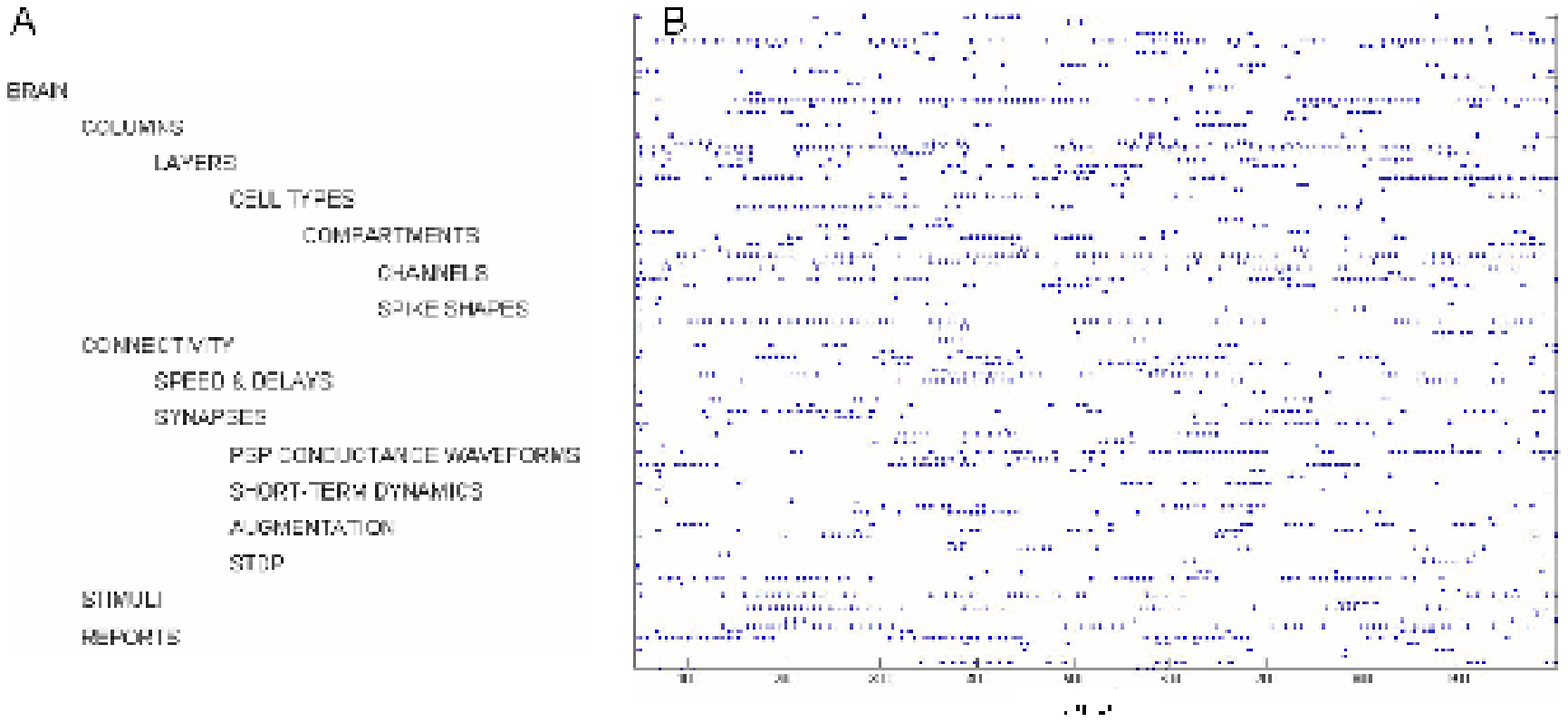,width=16cm}
\end{center}

\caption{NCS file specifications and example of simulation.}
A. Hierarchy of the NCS Command File Objects. The file is
ASCII-based with simple object delimiters. Brainlab scripting tools
are available for repetitive structures (Drewes, 2005). B. 1-second
spike rastergram of 100 arbitrarily selected neurons in the benchmark
simulation.\label{fig:NCS}

\end{figure}

\begin{figure}
\begin{center}
\begin{tabular}{cc}
\psfig{figure=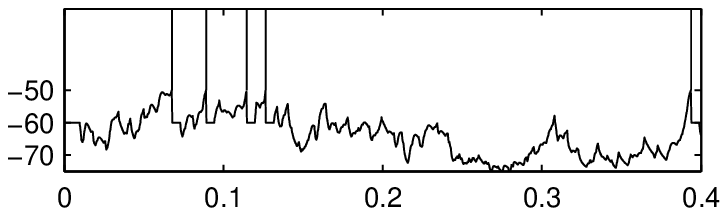,width=76mm}     & 
\psfig{figure=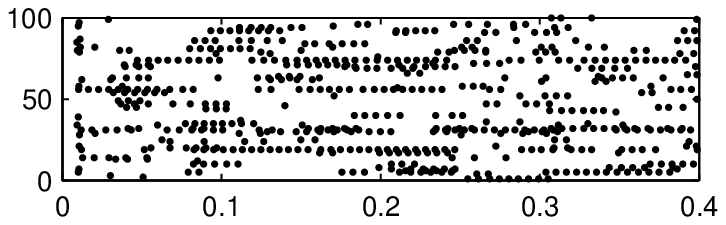,width=76mm} \\
\psfig{figure=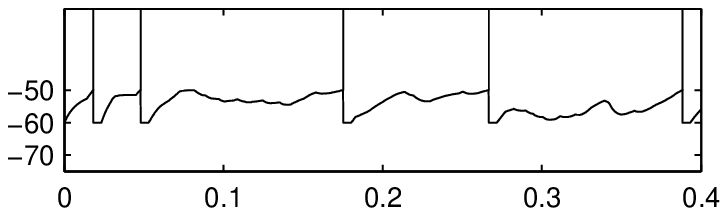,width=76mm}     & 
\psfig{figure=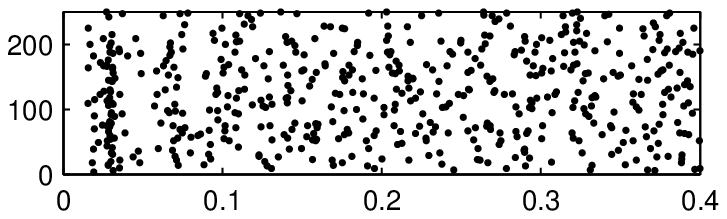,width=76mm} \\
\psfig{figure=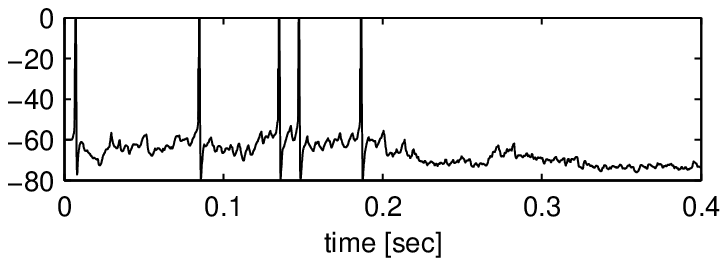,width=76mm}       & 
\psfig{figure=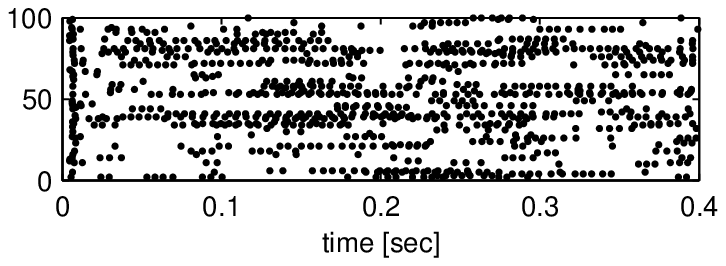,width=76mm}
\end{tabular}
\end{center}

\caption{\label{fig:CSIM1} Results of CSIM simulations of the
benchmarks 1 to 3 (top to bottom).} The left panels show the voltage
traces (in mV) of a selected neuron. For Benchmark~1 (COBA) and
Benchmark~2 (CUBA) models (top two rows), the spikes superimposed as
vertical lines.  The right panels show the spike raster for randomly
selected neurons for each of the three benchmarks.

\end{figure}

\begin{figure}
\begin{center}
\psfig{figure=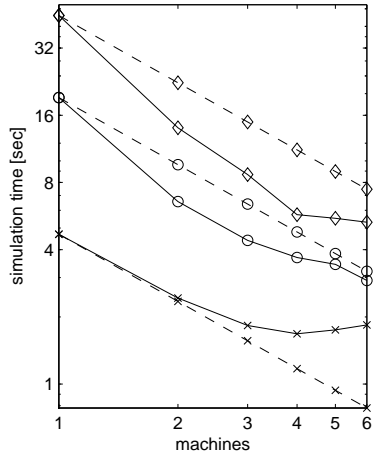,width=50mm}
\end{center}

\caption[]{\label{fig:PCSIM} Performance of PCSIM.} The time needed
to simulate the Benchmark~2 (CUBA) network (1ms synaptic delay, 0.1ms
time step) for one second of biological time (solid line) as well as
the expected times (dashed line) are plotted against the number of
machines (Intel Xeon, 3.4 Ghz, 2 Mb cache). The CUBA model was
simulated for three different sizes: 4000 neurons and $3.2 \times
10^5$ synapses (stars), 10000 neurons and $2 \times 10^6$ synapses
(circles), and 20000 neurons and $20 \times 10^6$ synapses (diamonds).

\end{figure}

\begin{figure}
\begin{center}
\psfig{file=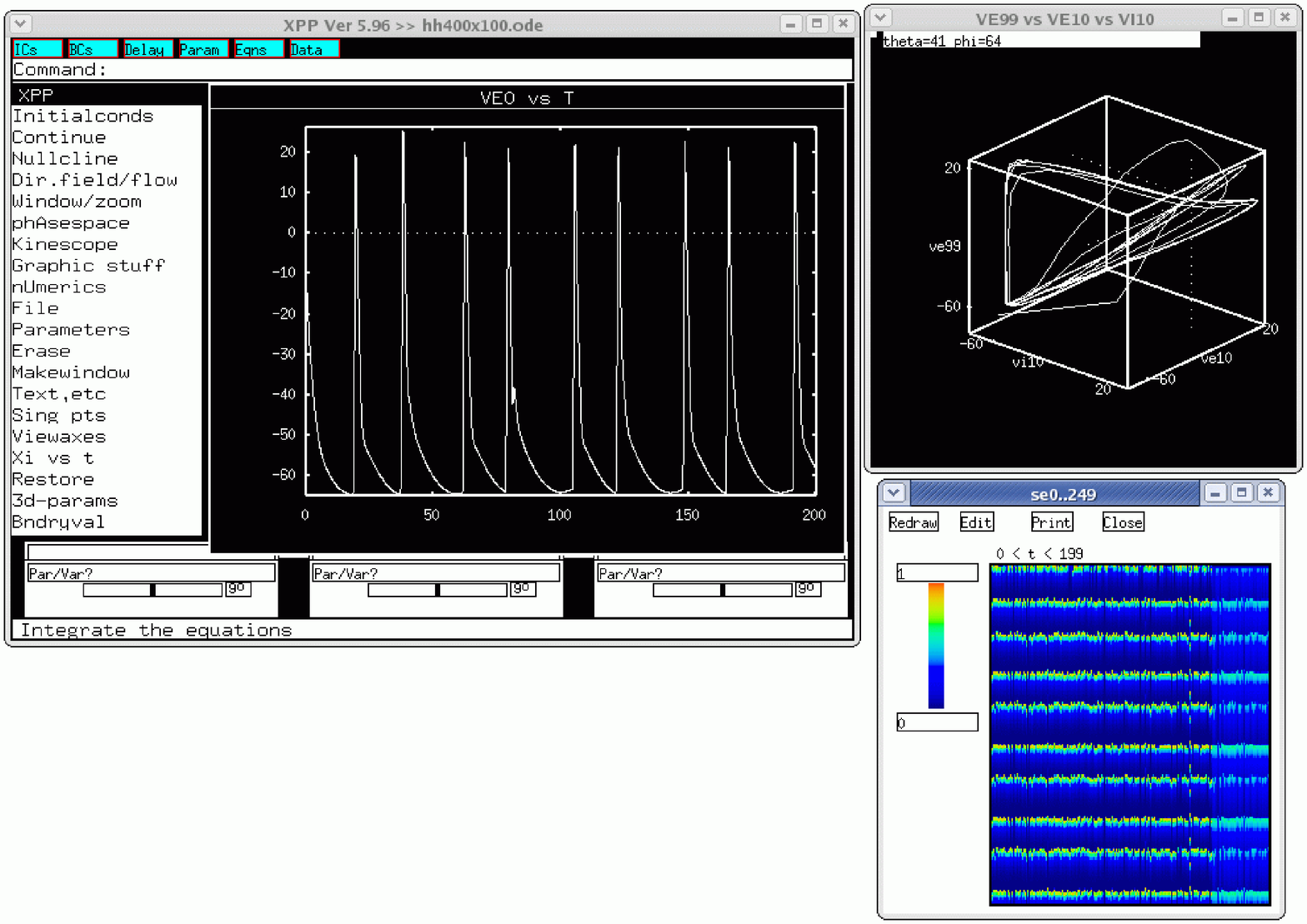,width=5in}
\end{center}

\caption{\XPP interface for a network of 200 excitatory and 50
inhibitory Hodgkin-Huxley neurons with random connectivity,
conductance-based dynamical synapses.}  Each neuron is also given a
random drive. Main window, a three-dimensional phase plot, and an
array plot are shown.

\end{figure}

\begin{figure}
\begin{center}
\psfig{file=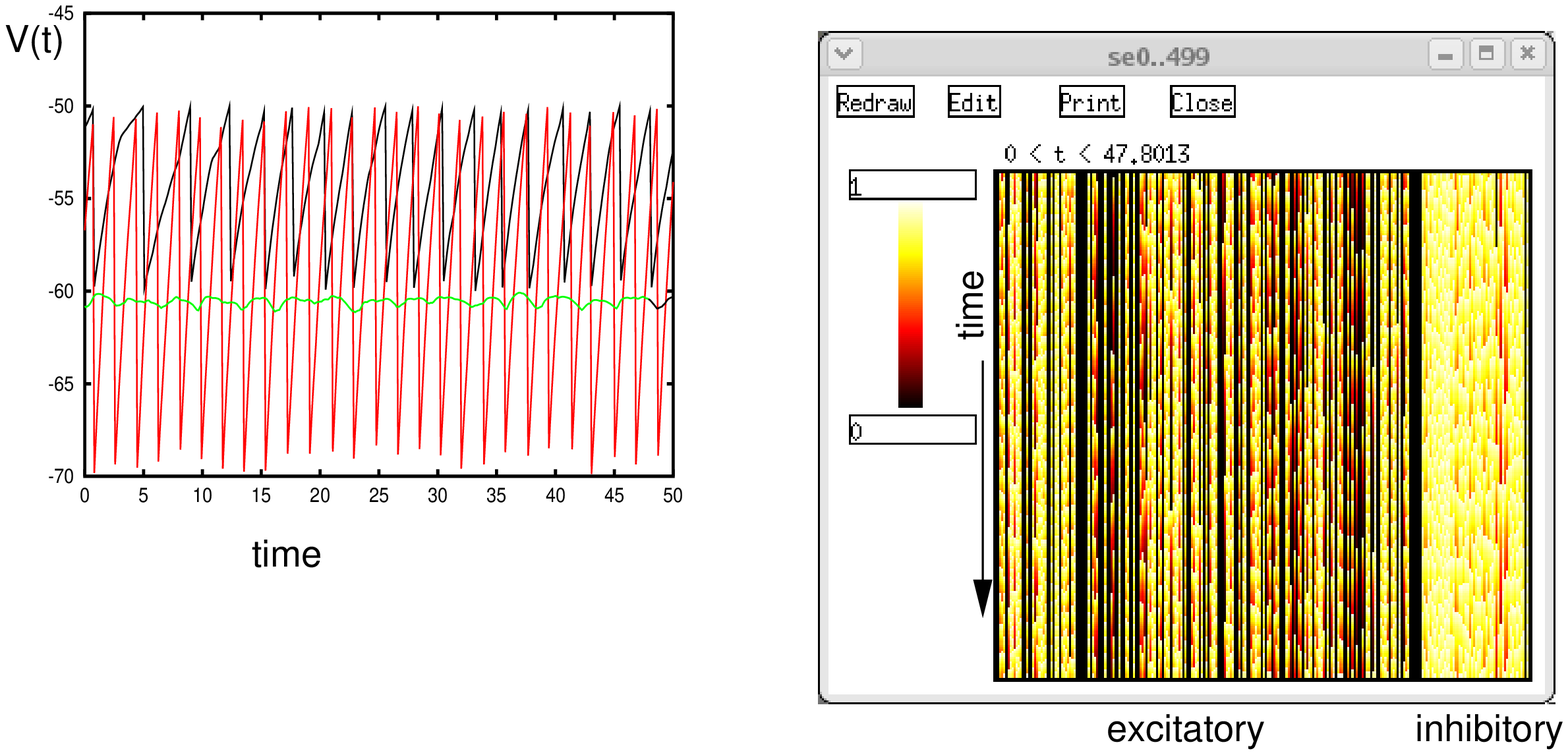,width=5in}
\end{center}

\caption{Persistent state in an integrate-and-fire network with 400
excitatory and 100 inhibitory cell.} \XPP simulation with exponential
conductance-based synapses, sparse coupling and random
drive.Excitatory and inhibitory synapses are shown as well as
voltages traces from 3 neurons.

\end{figure}

\begin{figure*}
\begin{center}
\psfig{file=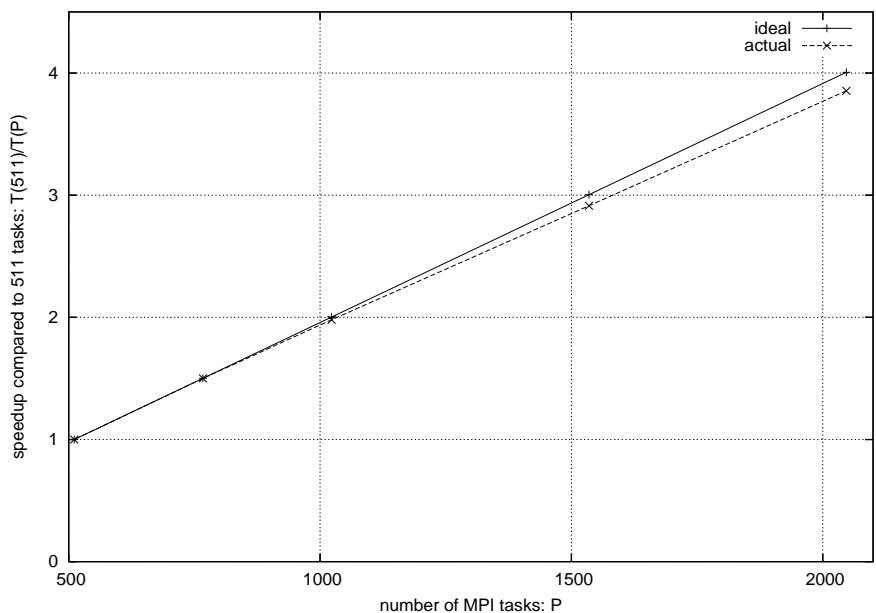,width=120mm}
\end{center}

\caption{Speedup for model with 4 million cells and 2 billion
synapses simulated with SPLIT on BG/L (from Djurfeldt et~al., 2005).}

\label{fig:speedup}
\end{figure*}

\begin{figure}
\begin{center}
\psfig{file=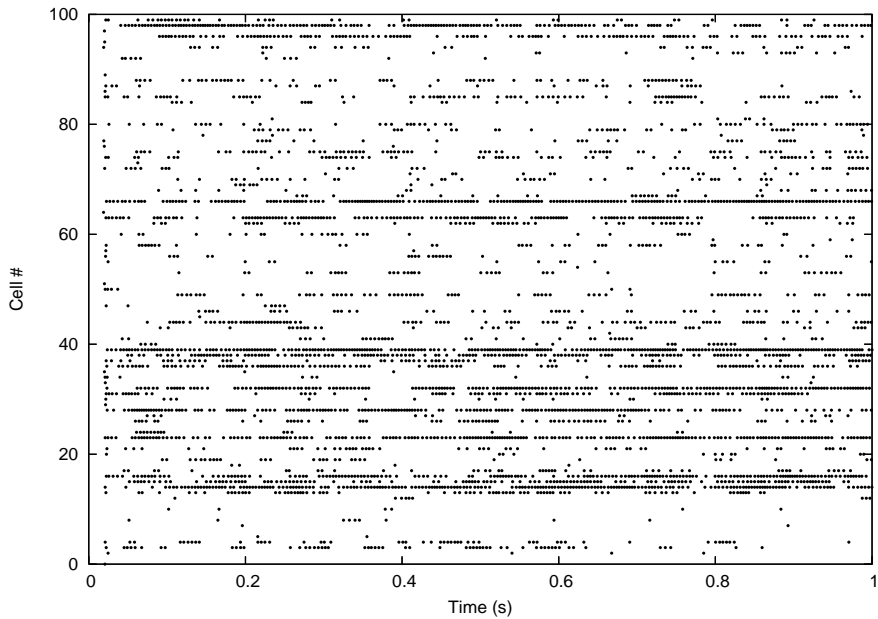,width=120mm}
\end{center}

\caption{Raster plot showing spikes of 100 cells during the first
  second of activity (SPLIT simulation of Benchmark~3).}

\label{fig:raster}
\end{figure}

\begin{figure}
\begin{center}
\centerline{\psfig{file=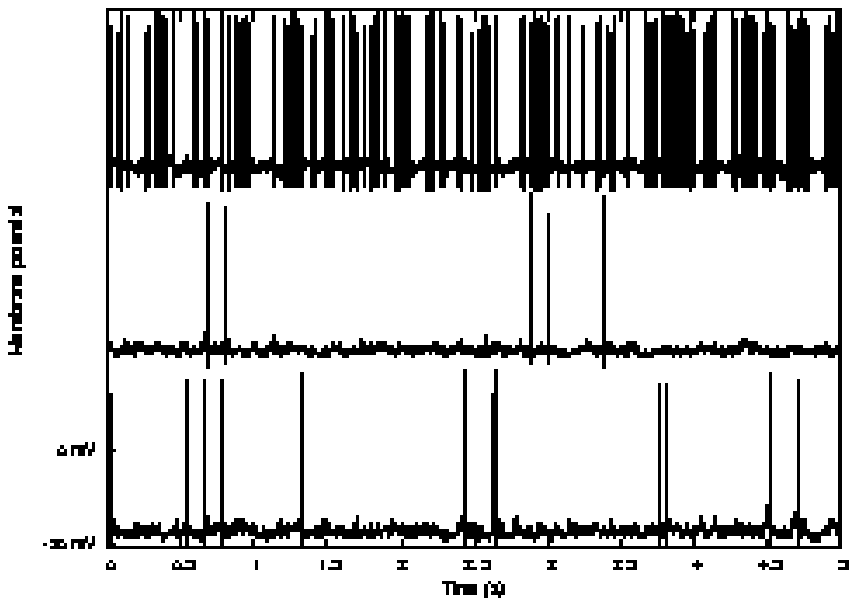,width=80mm}
\psfig{file=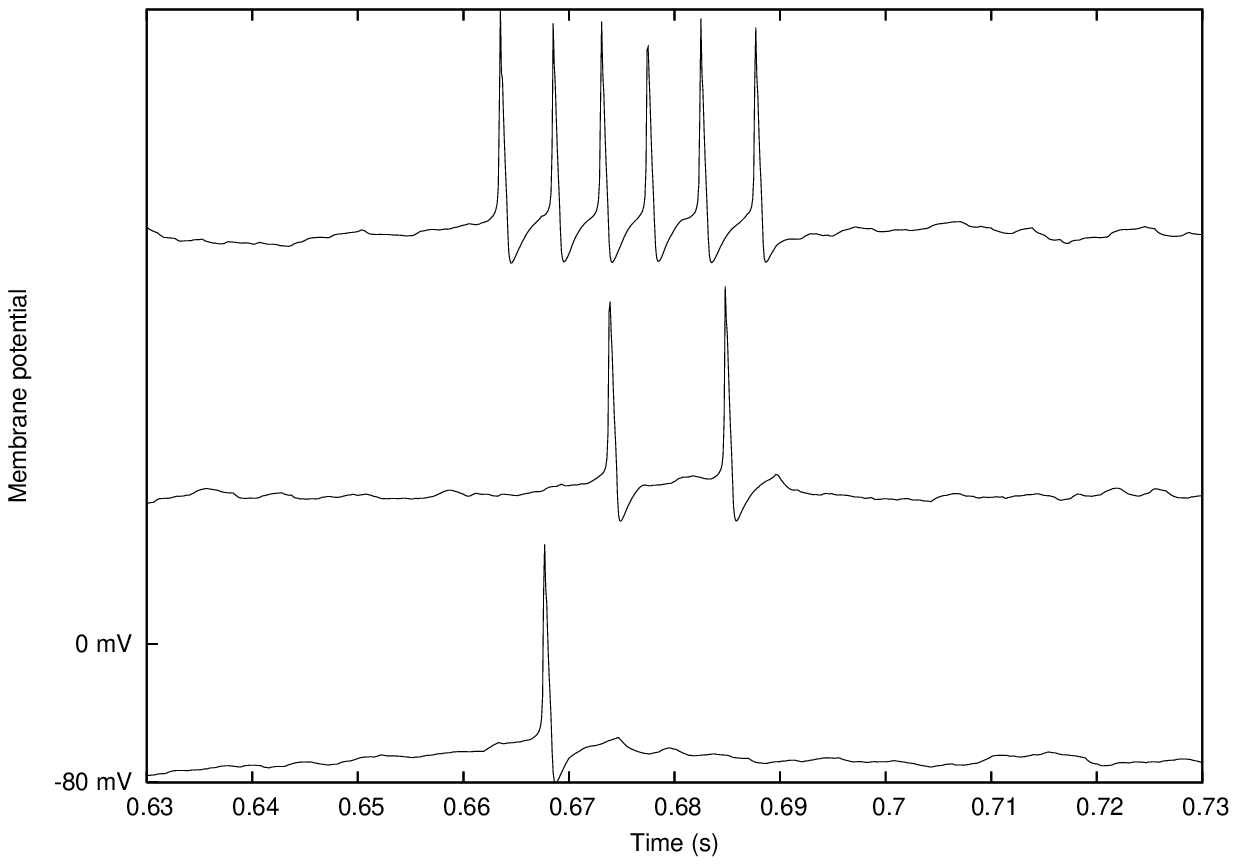,width=80mm}}
\end{center}

\caption{Plots of the membrane potential for 3 of the 4000 cells. }
The right plot shows a subset of the data in the left plot, with
higher time resolution (SPLIT simulation of Benchmark~3).

\label{fig:3cells}
\end{figure}

\begin{figure}
   \begin{center}
       \psfig{file=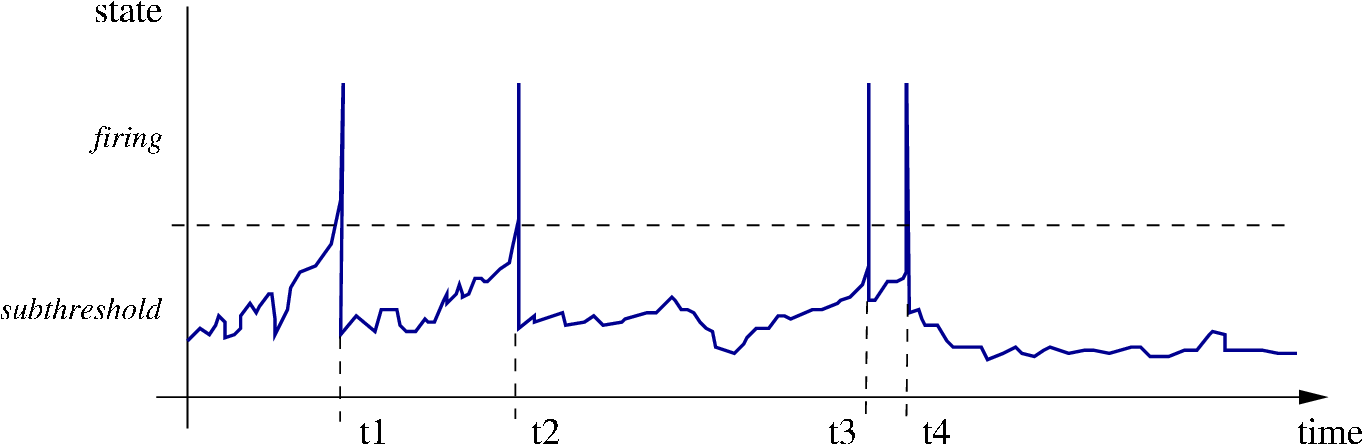,width=12cm}
   \end{center}

   \caption{Neuronal dynamics from a discrete-event dynamical systems
perspective.} Events (t1-t4), corresponding to the state variable
switching from the sub-threshold to the firing dynamics, can occur at
any arbitrary point in time. They correspond here to change of the
neuron output that can be passed to the rest of the systems (e.g.
other neurons). Internal changes (e.g. end of the refractory period)
can also be described in a similar way.

   \label{figRochel1}
\end{figure}

\begin{figure}
   \begin{center}
       \psfig{file=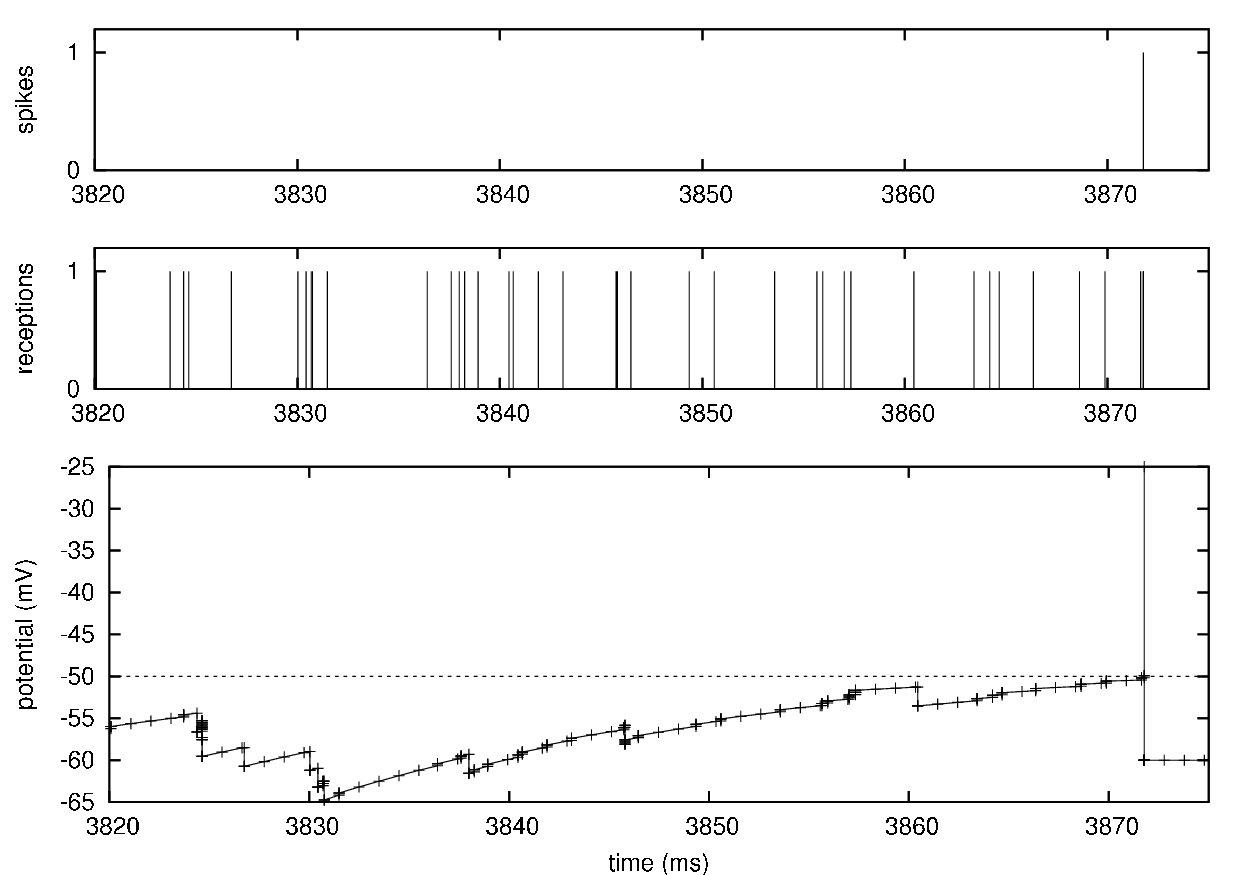,width=12cm}
   \end{center}

   \caption{Membrane potential of a single neuron, from a Mvaspike
implementation of Benchmark 4.} Top: membrane potential dynamics
(impulses have been superimposed at firing time to make them more
apparent). Bottom: Mvaspike simulation result typically consists of
lists of events (here, spiking and reception time, top and middle
panels) and the corresponding state variables at these instants (not
shown). In order to obtain the full voltage dynamics, a
post-processing stage is used to add new intermediary values between
events (bottom trace).

   \label{figRochel2}
\end{figure}

\begin{figure}
\begin{center}
\begin{scriptsize}
\begin{verbatim}
<?xml version="1.0" encoding="UTF-8"?>

<channelml xmlns="http://morphml.org/channelml/schema" 
  xmlns:xsi="http://www.w3.org/2001/XMLSchema-instance" 
  xmlns:meta="http://morphml.org/metadata/schema" 
  xsi:schemaLocation="http://morphml.org/channelml/schema 
    ../../Schemata/v1.1/Level2/ChannelML_v1.1.xsd"
  units="Physiological Units">

  <ion name="k"  default_erev="-77.0" charge="1"/>  <!-- phys units: mV -->

  <channel_type name="KChannel" density="yes">

    <meta:notes>Simple example of K conductance in squid giant axon. 
        Based on channel from Hodgkin and Huxley 1952</meta:notes>

    <current_voltage_relation>
      <ohmic ion="k">
        <conductance default_gmax="36">  <!-- phys units: mS/cm2-->
          <gate power="4">
            <state name="n" fraction="1">
              <transition>
                <voltage_gate>
                  <alpha>
                    <parameterised_hh type="linoid" expr="A*(k*(v-d))/(1 - exp(-(k*(v-d))))">
                      <parameter name="A" value="0.1"/>
                      <parameter name="k" value="0.1"/>
                      <parameter name="d" value="-55"/>
                    </parameterised_hh>
                  </alpha>
                  <beta>
                    <parameterised_hh type="exponential" expr="A*exp(k*(v-d))">
                      <parameter name="A" value="0.125"/>
                      <parameter name="k" value="-0.0125"/>
                      <parameter name="d" value="-65"/>
                    </parameterised_hh>
                  </beta>
                </voltage_gate>
              </transition>
            </state>
          </gate>
        </conductance>
      </ohmic>
    </current_voltage_relation>
  </channel_type>
</channelml>
\end{verbatim}
\end{scriptsize}
\end{center}
\caption{Example of Hodgkin-Huxley K$^{+}$ conductance specified in ChannelML, a
component of NeuroML.}\label{fig:NeuroML}
\end{figure}

\clearpage

\begin{figure}
\begin{center}
\psfig{file=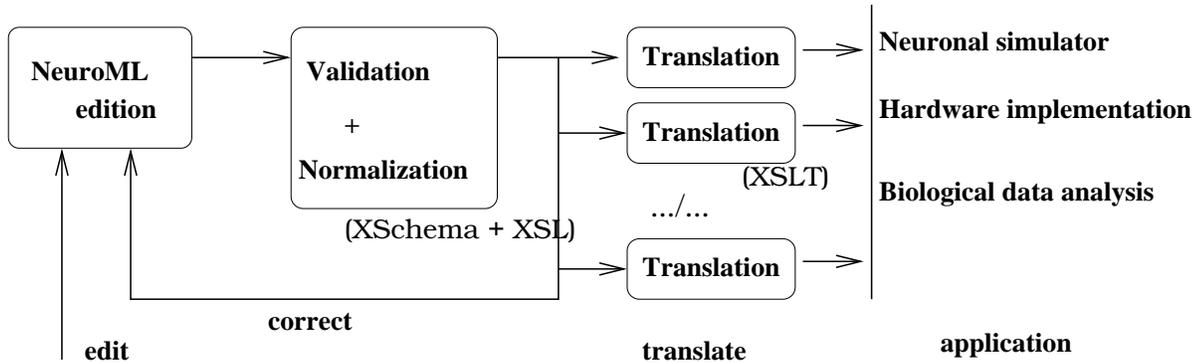,width=16cm}
\end{center}

\caption{From NeuroML to simulator}
\label{fig:facetsml-use}
\end{figure}

\begin{figure}
\centering
\begin{scriptsize}
\begin{verbatim}
          cell_params = { 'tau_m' : 20.0,  'tau_syn' : 2.0,   'tau_refrac': 1.0,
                          'v_rest': -65.0, 'v_thresh': -50.0, 'cm': 1.0}
          
          populationA = Population((10,), "IF_curr_alpha", cell_params)
          populationB = Population((5,5), "IF_curr_alpha", cell_params)
          populationA.randomInit('uniform', v_reset, v_thresh)
          
          connAtoB = Projection(populationA, populationB, 'fixedProbability', 0.2)
          connAtoA = Projection(populationA, populationA, 'distanceDependentProbability', "exp(-abs(d))")
          connBtoA = Projection(populationB, populationA, 'allToAll')
          
          connAtoB.setWeights(w_AB)
          connAtoA.setWeights(w_AA)
          connBtoA.setWeights(w_BA)
          
          populationA.record()
          populationB.record()
          
          run(1000.0)
          
          populationA.printSpikes("populationA.spiketimes")
          populationB.printSpikes("populationA.spiketimes")
\end{verbatim}
\end{scriptsize}
\caption{Example of the use of the PyNN API to specify a network that can then
be run on multiple simulators.}\label{fig:PyNN_example}
\end{figure}

\clearpage

\begin{figure}
\begin{center}
\psfig{figure=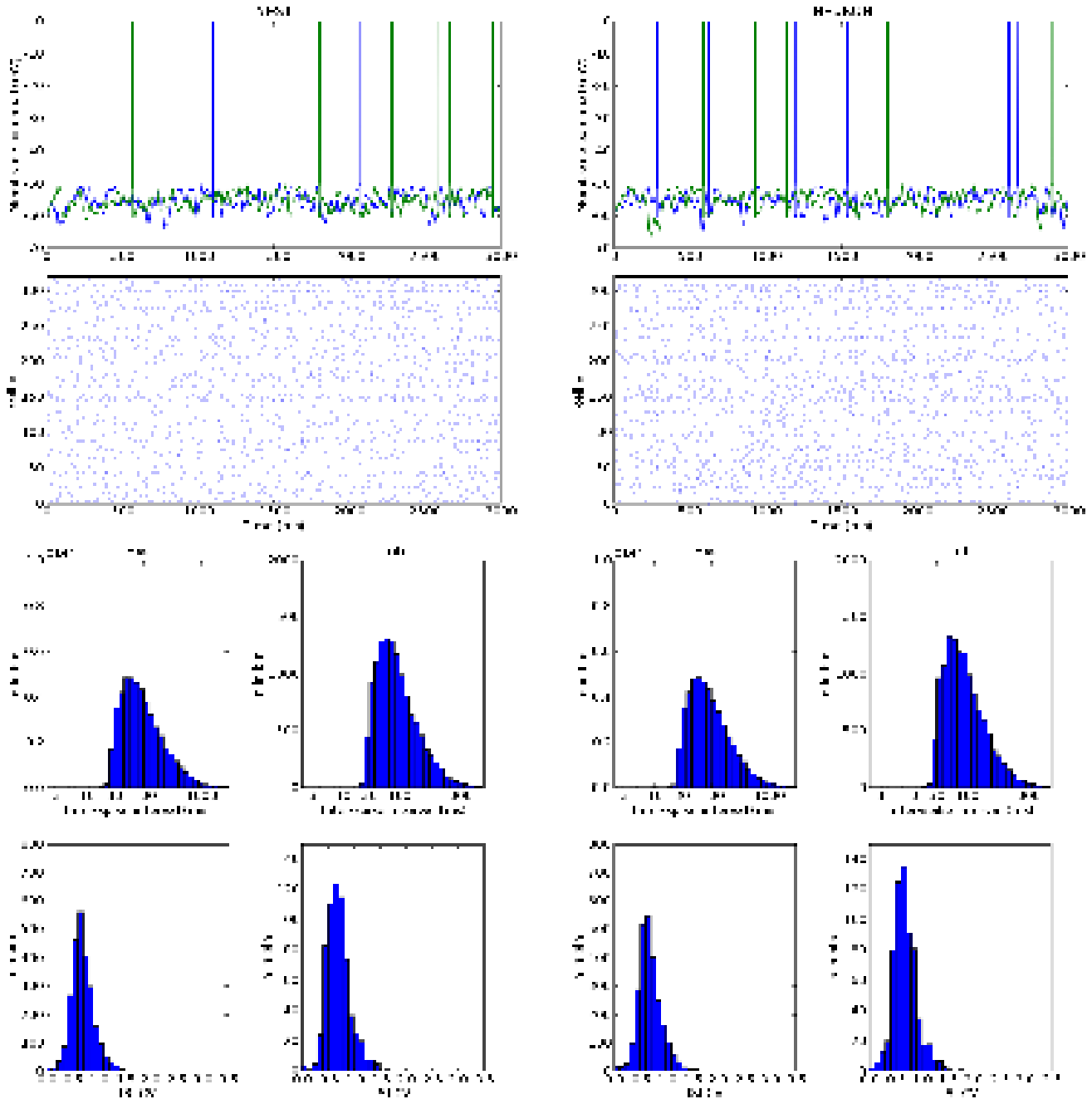,width=16cm}
\end{center}

\caption{\label{interop}Same network model run on two different
simulators using the same source code.} The model considered was the
Vogels-Abbott integrate-and-fire network with current-based synapses
and displaying self-sustained irregular activity states (benchmark 2
in Appendix~2). This network implemented with the PyNN
simulator-independent network modelling API, and simulated using NEST
(left column) and NEURON (right column) as the simulation engines.
The same sequence of random numbers was used for each simulator, so
the connectivity patterns were rigorously identical. The membrane
potential trajectories of individual neurons simulated in different
simulators rapidly diverge, as small numerical differences are
rapidly amplified by the large degree of recurrency of the circuit,
but the interspike interval (ISI) statistics of the populations are
almost identical for the two simulators. (Top row) Voltage traces for
two cells chosen at random from the population.  (Second row) Spike
raster plots for the first 320 neurons in the population. (Third row)
Histograms of ISIs for the excitatory and inhibitory cell
populations. (Bottom row) Histograms of the coefficient of variation
(CV) of the ISIs.

\end{figure}

\end{document}